\documentclass[aps,twocolumn,prb,showpacs,10pt,floatfix,superscriptaddress,floatfix]{revtex4-1}
\usepackage[dvips]{graphics}
\usepackage{makecell}
\usepackage{bbold}
\usepackage{xstring}
\usepackage{boldline}
\usepackage{graphicx}
\usepackage[utf8]{inputenc}
\usepackage{amssymb}
\usepackage{epstopdf}
\usepackage{bm}
\usepackage{dcolumn,multirow}
\usepackage{epsfig}
\usepackage{latexsym}
\usepackage{amsmath,mathtools}
\usepackage{color}
\usepackage{array}
\usepackage{bbm}
\usepackage{enumerate,dsfont}
\usepackage{array}
\usepackage{booktabs}
\newcommand{\mm}{m}

\def\XXint#1#2#3{{\setbox0=\hbox{$#1{#2#3}{\int}$ }
\vcenter{\hbox{$#2#3$ }}\kern-.5\wd0}}

\newcommand*{\IsInteger}[3]{%
    \IfStrEq{#1}{ }{%
        #3% is a blank string
    }{%
        \IfInteger{#1}{#2}{#3}%
    }%
}%
\newcommand{\ZZ}{\mathbb{Z}}
\newcommand{\rM}{{\rm M}}
\newcommand{\pf}{{\rm Pf}}
\newcommand{\cM}{{\hat\rho}}
\newcommand{\K}[1][1]{K^{\IfInteger{#1}{\ifcase#1\relax\or\prime\or\prime\prime\or\prime\prime\prime\else(#1)\fi}{(#1)}}}
\newcommand{\ki}[2][\mathrm{a}]{\mathcal{K}_{#1}^{\IfInteger{#2}{\ifcase#2\relax\or\prime\or\prime\prime\or\prime\prime\prime\else(#2)\fi}{(#2)}}}
\newcommand{\ks}[2][\mathrm{a}]{\mathcal{D}_{#1}^{\IfInteger{#2}{\ifcase#2\relax\or\prime\or\prime\prime\or\prime\prime\prime\else(#2)\fi}{(#2)}}}
\newcommand{\bki}[2][\mathrm{a}]{\mathcal{\bar K}_{#1}^{\IfInteger{#2}{\ifcase#2\relax\or\prime\or\prime\prime\or\prime\prime\prime\else(#2)\fi}{(#2)}}}
\newcommand{\ke}[2][1]{\mathcal{K}_{#1}^{\IfInteger{#2}{\ifcase#2\relax\or\prime\or\prime\prime\or\prime\prime\prime\else(#2)\fi}{(#2)}}}
\newcommand{\kd}[2][\mathrm{d}]{\mathcal{K}_{#1}^{\IfInteger{#2}{\ifcase#2\relax\or\prime\or\prime\prime\or\prime\prime\prime\else(#2)\fi}{(#2)}}}
\newcommand{\vk}{\vec k}
\newcommand{\omap}{\omega}

\newcommand{\vn}{\vec n}

\renewcommand{\vr}{\vec \varphi}

\newcommand{\sign}{{\rm sign}\,}

\renewcommand{\vec}[1]{\bm{#1}}

\newcommand{\img}{\mathrm{img}\,}

\newcommand{\cR}[1]{\mathcal{M}_{#1}}
\newcommand{\cO}{\mathcal{O}}

\makeatletter
\newcommand*{\rom}[1]{\expandafter\@slowromancap\romannumeral #1@}
\makeatother

\begin{document}
 
\title{Higher-order bulk-boundary correspondence for topological crystalline phases}

\author{Luka Trifunovic and Piet W. Brouwer}
\affiliation{Dahlem Center for Complex Quantum Systems and Physics Department, Freie Universit\"at Berlin, Arnimallee 14, 14195 Berlin, Germany}
\date{\today}
\begin{abstract}
We study the bulk-boundary correspondence for topological crystalline phases, where
the crystalline symmetry is an order-two (anti)symmetry, unitary or
antiunitary. We obtain a formulation of the bulk-boundary correspondence in
terms of a subgroup sequence of the bulk classifying groups, which uniquely
determines the topological classification of the boundary states. This
formulation naturally includes higher-order topological phases as well as
topologically nontrivial bulk systems without topologically protected boundary
states. The complete bulk and boundary classification of higher-order
topological phases with an additional order-two symmetry or antisymmetry is
contained in this work.
\end{abstract}

\maketitle
\section{Introduction}
A central paradigm in the field of topological insulators and superconductors
is the bulk-boundary correspondence: A nontrivial topology of the bulk band
structure uniquely manifests itself through an anomalous gapless, topologically
nontrivial boundary, irrespective of the orientation of the boundary or the
lattice termination.\cite{hasan2010,bernevig2013,qi2011} On the other hand, for
topological crystalline phases, which are protected by an additional non-local
crystalline
symmetry,~\cite{fu2011,fang2012,turner2012,fang2013,chiu2013,morimoto2013,slager2013,jadaun2013,liu2014b,alexandradinata2014,dong2016,chiu2016,trifunovic2017,kruthoff2017,po2017,bradlyn2017,khalaf2018,shiozaki2014,shiozaki2016,shiozaki2017,shiozaki2018,thorngren2018,rhim2018}
the existence of gapless boundary states for a nontrivial bulk topology is
guaranteed only if the boundary is invariant under the crystalline symmetry. 

\begin{figure}
	\includegraphics[width=0.8\columnwidth]{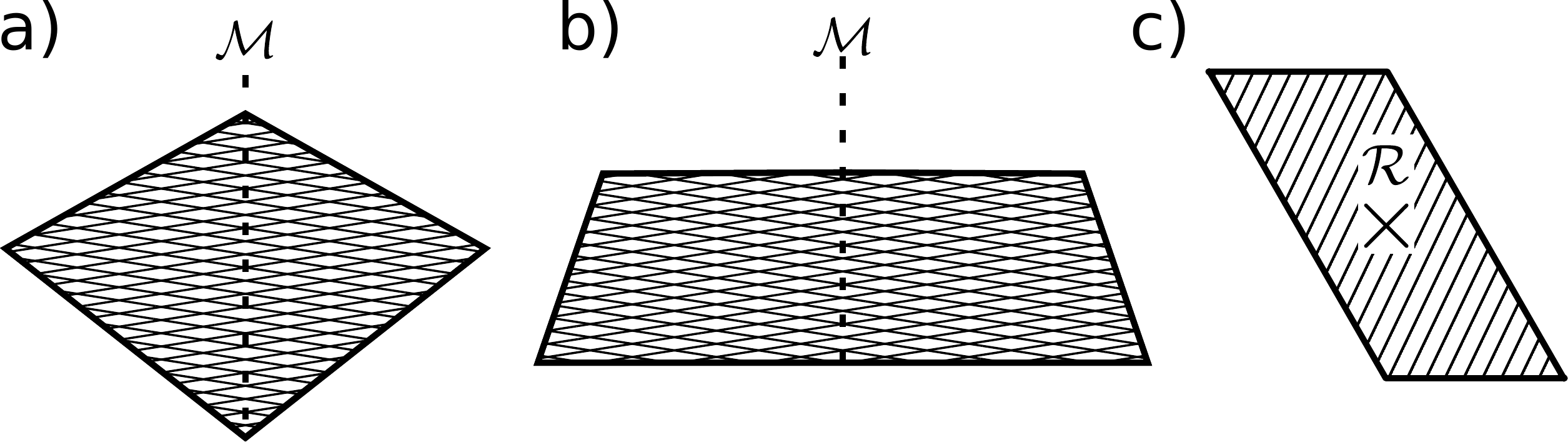}
	\caption{\label{fig:1} Schematic pictures of a two-dimensional crystal
      for which the shape is compatible with mirror symmetry (a and b) and with
    twofold rotation symmetry (c). The crystal in (b) has a boundary that is invariant
  under the mirror symmetry, whereas the boundaries of the crystals in (a) and
(c) appear in symmetry-related pairs. The special situation of a crystal with
mirror symmetric boundary, as shown in panel (b), is excluded from the
definition of the higher-order topological phases.}
\end{figure}

Recently, it was realized that a nontrivial crystalline topology of a
$d$-dimensional crystal may also manifest itself through protected boundary
states of dimension less than
$d-1$.\cite{parameswaran2017,schindler2018,peng2017,langbehn2017,song2017,benalcazar2017,benalcazar2017b,fang2018,ezawa2018,shapourian2017,zhu2018,yan2018,wang2018,wang2018b}
A topological phase with such lower-dimensional boundary states is called a
``higher-order topological phase'', where the order $n$ of the topological
phase corresponds to the codimension of the boundary
states.\cite{schindler2018} [According to this definition, a topological
insulator or superconductor with the conventional $(d-1)$-dimensional boundary
states is a first-order topological phase.] The condition that guarantees the
protection of such higher-order boundary states is that the orientation of the
crystal faces and the lattice termination be compatible with the crystalline
symmetry --- {\em i.e.}, the crystal faces and the corresponding lattice
termination must be related to each other by the crystalline symmetry operation. This is a much weaker condition than the condition that the crystal
boundary be invariant under the symmetry operation (compare with
Fig.~\ref{fig:1}). For example, whereas inversion symmetry leaves no crystal
faces invariant, compatibility with inversion symmetry merely requires that
crystal faces appear in inversion-related pairs (see Fig.~\ref{fig:1}c).
Topological crystalline insulators with second-order boundary states were
theoretically predicted for models with certain magnetic
symmetries,\cite{schindler2018} mirror
symmetry,~\cite{schindler2018,langbehn2017} and rotation and inversion
symmetries.~\cite{peng2017,song2017,khalaf2018,geier2018,vanmiert2018b,calugaru2018}
The latter two symmetries are relevant for the semimetal Bi, which shows
boundary states reminiscent of that of a second-order topological
insulator.~\cite{schindler2018b}

The presence of a crystalline symmetry is not a necessary requirement for the
boundary phenomenology associated with a higher-order phase. Indeed, early
examples of protected codimension-two boundary states include the superfluid
$^3$He-B phase\cite{volovik2010} and a three-dimensional topological insulator
with a suitable time-reversal breaking perturbation,\cite{sitte2012,zhang2013}
neither of which rely on the protection by a bulk crystalline symmetry.
Instead, in these cases the appearance of higher-order protected boundary
states can be solely attributed to a boundary termination that is itself
topologically nontrivial, whereas the underlying bulk is essentially trivial.
In Ref.\ \onlinecite{geier2018} we called these termination-dependent
higher-order topological phases {\em extrinsic}, to contrast them with the {\em
anomalous} (intrinsic), termination-independent higher-order boundary states of
topological crystalline phases. Although for anomalous higher-order topological
phases, too, the precise form of the $(d-2)$-dimensional boundary states may
still depend on details of the lattice termination, their very existence is a
consequence of a nontrivial bulk topology and is protected as long as the
crystal termination remains compatible with the crystalline symmetry.

While a complete classification of higher-order topological phases (HOTPs) is
still lacking, several authors have obtained partial classifications of
higher-order topological phases, restricted to certain crystalline symmetries
or for a certain ten-fold way
class.\cite{langbehn2017,geier2018,khalaf2018,khalaf2018b} (The
tenfold way or Altland-Zirnbauer classes are defined with respect to the presence or absence
of the fundamental non-spatial symmetry operations time-reversal ${\cal T}$,
particle-hole conjugation ${\cal P}$ and the chiral operation ${\cal C} = {\cal
P} {\cal T}$.\cite{altland1997}) Two approaches have been taken for the
classification of anomalous, termination-independent HOTPs: A bulk-based
approach, which starts from the classification of the bulk band structure and
then shows under which circumstances a nontrivial bulk topology implies a
higher-order topological phase,\cite{langbehn2017,geier2018} and a
boundary-based approach, in which all topologically nontrivial boundaries of
HOTPs are classified first, and a classification of anomalous,
termination-independent HOTPs is obtained upon identification of boundary
states that are related by a change of
termination.\cite{khalaf2018,khalaf2018b,geier2018} For crystalline phases with
an order-two crystalline symmetry, for which a complete classification of the
bulk topology exists,~\cite{shiozaki2014} the two approaches were found to be
in complete agreement for the second-order topological
phases.\cite{geier2018,khalaf2018b} The boundary-based approach not only
classifies the anomalous, termination-independent HOTPs, but also the extrinsic
higher-order topological phases, for which the higher-order boundary states are
a manifestation of a nontrivial boundary topology rather than a nontrivial bulk
topology.

In this work we provide a full classification of higher-order topological phases with an order-two crystalline symmetry or antisymmetry, for arbitrary order $n$ of the topological phase and in arbitrary spatial dimension $d$. A crystalline symmetry or antisymmetry $\mathcal{S}$ is called ``order-two'' if $\mathcal{S}^2=\pm1$. Its spatial type is determined by the number $d_\parallel$ of inverted dimensions, such that $d_\parallel=0$ corresponds to on-site (anti)symmetry, and $d_\parallel=1,2,3$ to mirror, twofold rotation, and inversion (anti)symmetry, respectively. 

\begin{figure}
	\includegraphics[width=\columnwidth]{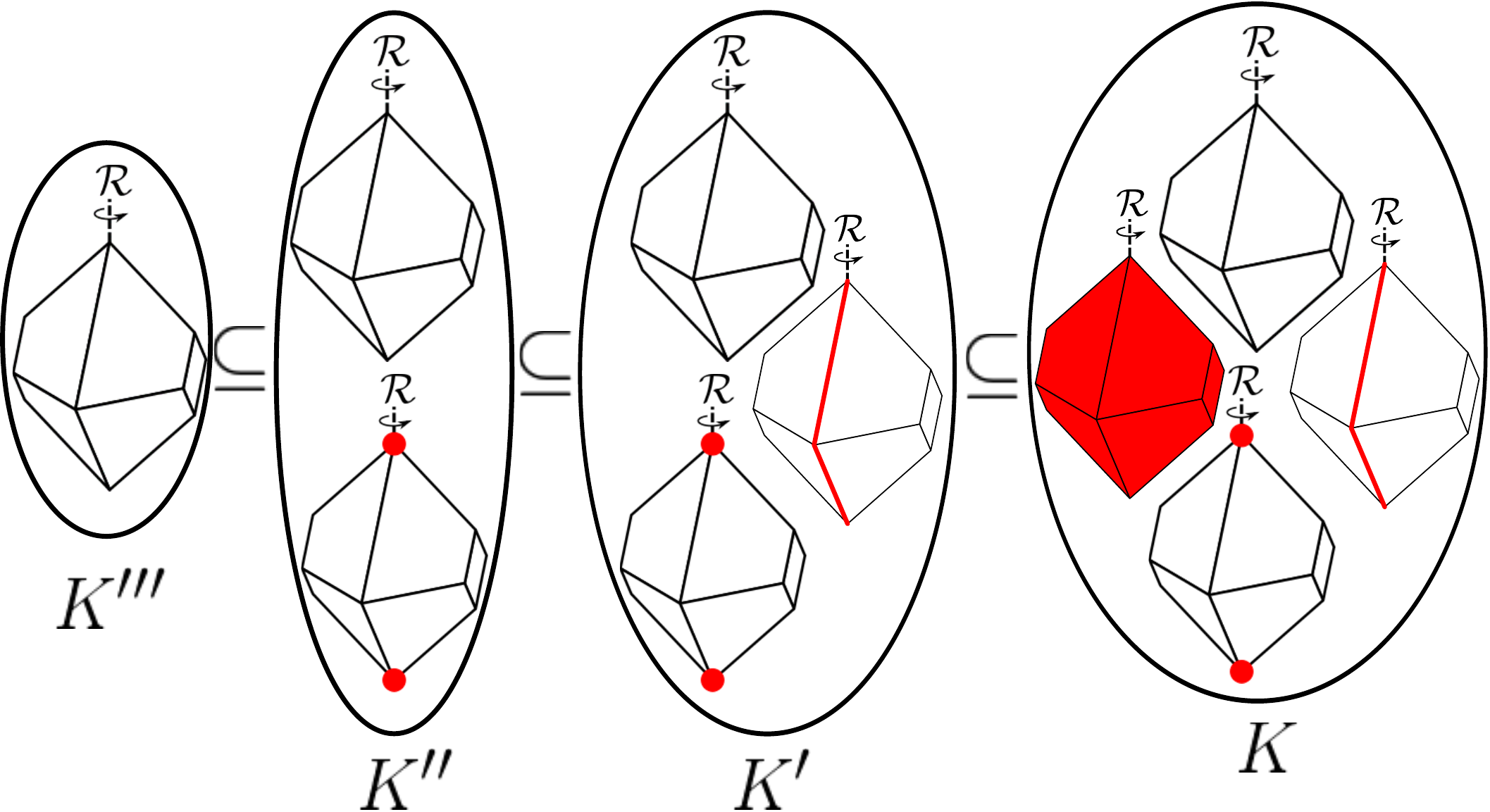}
	\caption{\label{fig:subgroups} The bulk classifying group $K$ classifies all bulk phases, regardless of the existence or type of anomalous boundary states. Refined classification groups $\K[n]$ are obtained by excluding topological phases with anomalous boundary states of codimensions $\le n$. The figure illustrates this procedure for a crystal with a twofold rotation symmetry $\mathcal{R}$. Anomalous boundary states are indicated in red.}
\end{figure}

We present classifications both from a bulk perspective and from a boundary
perspective. Our bulk classification of HOTPs with an order-two crystalline
symmetry refines the existing classification of Shiozaki and
Sato,\cite{shiozaki2014} who classified topological crystalline phases without
accounting for the type of the boundary signatures. Whereas Ref.\
\onlinecite{shiozaki2014} described the topological classification in terms of
a single classifying group $K$, our refined classification takes the form of a
{\em subgroup series}
\begin{equation}
	\K[d] \subseteq \ldots \subseteq \K[2] \subseteq \K[1] \subseteq \K[0],
	\label{eq:subgroup}
\end{equation}
which resolves the topological crystalline phases according to their associated
anomalous boundary signature. The last term in Eq.~(\ref{eq:subgroup})
$\K[0]\equiv K^{(0)}$ is the classifying group of Ref.\
\onlinecite{shiozaki2014}, which classifies the bulk band structure with an
order-two symmetry or antisymmetry. The other terms $\K[n] \subseteq \K[0]$ are
subgroups that exclude topological phases that are of order $n$ and lower for
any crystal shape consistent with the crystalline symmetry. An illustration of
the definitions of the groups $\K[n]$ is shown in Fig.~\ref{fig:subgroups} for
the case of a three-dimensional crystal with twofold rotation symmetry. The
subgroup $\K[1]$, which classifies topological crystalline phases that are not
first-order was previously studied in Ref.~\onlinecite{geier2018} in the
context of crystals with mirror, twofold rotation, or inversion symmetry, where
it was called the ``purely crystalline subgroup''. Note that the definition of
the groups $\K[n]$ excludes crystals with boundary states that can be removed
by a symmetry-respecting deformation of the crystal, such as the gapless
surface states on a mirror-symmetric surface of a mirror-symmetric crystal,
compare Fig.~\ref{fig:1}a and b. 

The bulk classification is complemented with a classification of anomalous
boundary states of codimension $n$, described by the boundary classification
group $\ki{n}$. We show that there exists a ``bulk-boundary correspondence'', a
general relation between the boundary classification group $\ki{n}$ and the
subgroup series (\ref{eq:subgroup}) of the bulk classification groups,
\begin{equation}
  \ki{n+1} = \K[n]/\K[n+1],\ n=0,1,2,\ldots,d.
  \label{eq:bb}
\end{equation}
%where the group $\ki{n}$ classifies anomalous boundary states of codimension
%$n$. 
In case the number of inverted dimensions $d_{\parallel} < d$ the subgroup
series (\ref{eq:subgroup}) starts with one or more trivial groups,
\begin{equation}
  \K[n] = 0 \ \mbox{for $n > d_{\parallel}$},
  \label{eq:bb1}
\end{equation}
so that Eq.~(\ref{eq:bb}) yields a complete bulk-boundary correspondence for
order-two crystalline symmetries: A topologically nontrivial bulk is
{\em uniquely} associated with a higher-order topological phase. On the other
hand, if $d_{\parallel} = d$ (inversion symmetry), the first group in the
subgroup series (\ref{eq:subgroup}) $\K[d]$ may be nontrivial. In that case
there is only a partial bulk-boundary correspondence and $\K[d]$ classifies the
topological crystalline phases without topologically protected boundary states,
see Fig.~\ref{fig:subgroups}. Topological phases contained in $\K[d]$ are
smoothly connected to atomic-limit insulators or superconductors. A nontrivial
classifying group $\K[d]$ indicates the existence of multiple topologically
distinct atomic-limit phases.~\cite{hughes2011,lau2016,shiozaki2016} 

All of the above results will be expanded and made more precise in
Sec.~\ref{sec:omap}. There, we present a fully algebraic proof of the central
relation (\ref{eq:bb}), making essential use of an ``order-raising''
homomorphism $\omap$, which maps the classifying group $\K[n]$ for a
$d$-dimensional crystal onto the group $\K[n+1]$ of a $(d+1)$-dimensional
crystal, while keeping the spatial and non-spatial symmetries the same, except
for an increase of the number of inverted dimensions $d_{\parallel}$ by one. We
construct the homomorphism $\omap$ by combination of two maps known from the
literature: The ``dimension-raising isomorphisms'', which relate bulk
classifying groups $\K[0] \equiv K^{(0)}$ in different dimensions and different
tenfold-way symmetry classes, see Refs.~\onlinecite{teo2010,shiozaki2014}, and
the dimension-lowering ``boundary map'' of Ref.~\onlinecite{turner2012}. The
map $\omap$ represents the same homomorphism between classifying groups as the
``stacking construction'' of Refs.~\onlinecite{isobe2015,fulga2016,huang2017},
which was used recently for the construction of higher-order topological phases
out of lower-dimensional structures.\cite{khalaf2018,khalaf2018b} It is the
algebraic formulation in terms of the dimension-raising isomorphism and the
boundary map that allows us to obtain the obtain the bulk-boundary
correspondence~(\ref{eq:bb}) by purely algebraic methods.

The remainder of this article is organized as follows: In
Sec.~\ref{sec:ShiozakiSato} we review the classification of the topological
crystalline phase stabilized by an order-two symmetry, and introduce the
dimension-raising isomorphisms, closely following previous work by Shiozaki and
Sato.\cite{shiozaki2014} In Sec.~\ref{sec:cform} we discuss Hamiltonians of
``canonical form'' and show how higher-order phases
naturally arise from the presence of crystalline-symmetry-breaking mass terms,
generalizing the conclusions of
Refs.~\onlinecite{langbehn2017,schindler2018,geier2018} for second-order
topological phases. In Secs.~\ref{sec:omap} and \ref{sec:omapexplicit}
we give the formal definitions of
the classifying groups $\K[n]$ and $\ki{n}$, construct the
order-raising homomorphism $\omap$, obtain explicit expressions, and
establish the bulk-boundary correspondence (\ref{eq:bb}) using algebraic
methods. Section~\ref{sec:examples} discusses a few representative examples 
of this general classification and shows how the homomorphism $\omap$ relates
classification results in different dimensions to each other. In Sec.\ 
\ref{sec:omegainverse} we construct a procedure for lowering the dimension $d$
of the crystal, while preserving the dimension of the anomalous boundary
states, providing a general realization of an idea put forward by Matsugatani and
Watanabe.~\cite{matsugatani2018} Finally, turning the arguments of our article
around, the bulk-boundary correspondence~(\ref{eq:bb}) can be used to obtain
the bulk classifying groups from the classification of anomalous boundaries
together with the classification of the topologically non-trivial atomic
limits, thus providing a boundary-based alternative to the
$K$-theory-based classification of Ref.~\onlinecite{shiozaki2014}. This
program is carried out in Sec.~\ref{sec:classification}.  We conclude
in Sec.~\ref{sec:conclusions}. The appendices contain derivations not presented in the main text as well as a few additional results.

\section{Shiozaki-Sato classes for topological phases with an order-two
symmetry}\label{sec:ShiozakiSato}
\begin{table}
  \begin{tabular*}{\columnwidth}{c @{\extracolsep{\fill}} cccc}
    \hline\hline
    Cartan & $s$ & ${\cal T}$      & ${\cal P}$      & ${\cal C}$ \\\hline
    A      & $0$ & -               & -               & - \\
    AIII   & $1$ & -               & -               & $\mathcal{C}$ \\ \hline
    AI     & $0$ & $\mathcal{T}^+$ & -               & - \\
    BDI    & $1$ & $\mathcal{T}^+$ & $\mathcal{P}^+$ & $\mathcal{C}$ \\
    D      & $2$ & -               & $\mathcal{P}^+$ & - \\
    DIII   & $3$ & $\mathcal{T}^-$ & $\mathcal{P}^+$ & $\mathcal{C}$ \\
    AII    & $4$ & $\mathcal{T}^-$ & -               & - \\
    CII    & $5$ & $\mathcal{T}^-$ & $\mathcal{P}^-$ & $\mathcal{C}$ \\
    C      & $6$ & -               & $\mathcal{P}^-$ & - \\
    CI     & $7$ & $\mathcal{T}^+$ & $\mathcal{P}^-$ & $\mathcal{C}$ \\ \hline \hline
  \end{tabular*}
\caption{\label{tab:0} The ten-fold way classes are defined according to the
presence or absence of time-reversal symmetry (${\cal T}$),
particle-hole antisymmetry (${\cal P}$), and chiral antisymmetry (${\cal C}$).
The entries $\mathcal{T}^\pm$ ($\mathcal{P}^\pm$) denote that
$\mathcal{T}^2=\pm1$ ($\mathcal{P}^2=\pm1$). The chiral antisymmetry is assumed to
square to one.}
\end{table}

The ten-fold way or Altland-Zirnbauer\cite{altland1997} classes are defined according to the presence or
absence of time-reversal symmetry ${\cal T}$, particle-hole antisymmetry
${\cal P}$, and chiral antisymmetry ${\cal C}$, see Table \ref{tab:0}. 
Shiozaki and Sato~\cite{shiozaki2014} extend the ten-fold way classes to
include an additional crystalline unitary
symmetry,~\cite{chiu2013,morimoto2013,trifunovic2017} unitary antisymmetry,
antiunitary symmetry or antiunitary antisymmetry $\mathcal{S}$. The crystalline
symmetry is an order-two symmetry, which means that its square is proportional
to the identity operation. 

It is sufficient to distinguish symmetry operations
that square to one (labeled by $\eta_{\cal S} = +$) and to minus one
($\eta_{\cal S} = -$). Further, the algebraic structure of the crystalline
symmetry is characterized by signs $\eta_{{\cal T},{\cal P},{\cal C}}$
indicating whether ${\cal S}$ commutes ($\eta = +$) or anticommutes ($\eta=-$)
with the time-reversal operation ${\cal T}$, particle-hole conjugation ${\cal
P}$, or the chiral symmetry operation ${\cal C}$. Following
Ref.~\onlinecite{shiozaki2014}, we denote the number of spatial degrees of
freedom that are inverted under the crystalline symmetry operation by
$d_{\parallel}$, so that on-site symmetries $\cO$ have $d_\parallel=0$,
reflections $\cR{}$ have $d_\parallel = 1$, twofold rotations $\mathcal{R}$ have
$d_\parallel =2$, and inversion ${\cal I}$ has $d_{\parallel} = 3$. Specifically, unitary symmetry ($\sigma_{\cal S}=1$) and
antisymmetry ($\sigma_{\cal S}=-1$) operations are represented by unitary
matrices $U_{\cal S}$,
\begin{align}
  H(\bm k,\mm)&= {\cal S} H(\vk,\mm) \equiv \sigma_{\cal S} U_{\cal S}
  H(\mathcal{S}\bm k,\mm) U_{\cal S}^{-1},
\end{align}
with $\mathcal{S}\bm k=(-\bm k_\parallel,\bm k_\perp)$, $\bm
k_\parallel=(k_1,\dots,k_{d_\parallel})$, $\bm
k_\perp=(k_{d_\parallel+1},\dots,k_{d})$ and
$U_\mathcal{S}^2=\eta_\mathcal{S}$,
$U_\mathcal{S}U_\mathcal{T}=\eta_\mathcal{T}U_\mathcal{T}U^*_\mathcal{S}$,
$U_\mathcal{S}U_\mathcal{P}=\eta_\mathcal{P}U_\mathcal{P}U^*_\mathcal{S}$ and
$U_\mathcal{S}U_\mathcal{C}=\eta_\mathcal{C}U_\mathcal{C}U_\mathcal{S}$.
Similarly, antiunitary symmetry and antisymmetry operations are
represented as
\begin{align}
  H(\bm k,\mm) &= {\cal S} H(\vk,\mm) \equiv \sigma_{\cal S} U_{\cal S}
  H^*(-\mathcal{S}\bm k,\mm) U_{\cal S}^{-1},
\end{align}
such that $U_{\cal S}U_{\cal S}^* = \eta_{\cal S}$, $U_{\cal S} U_{\cal T}^*
= \eta_{\cal T} U_{\cal T} U_{\cal S}^*$, $U_{\cal S} U_{\cal P}^* = \eta_{\cal
P} U_{\cal P} U_{\cal S}^*$, and $U_{\cal S} U_{\cal C}^* = \eta_{\cal C}
U_{\cal C} U_{\cal S}$.

\begin{table*}
  \begin{center}
    \begin{tabular}[t]{l @{\extracolsep{\fill}} cccccccc}
      \hline\hline 
                   & & & $d=0$ & $d=1$ & $d=1$ & $d=2$ & $d=2$ & $d=2$  \\
  class & $s$ & $t$ & ${\cal O}$ & ${\cal O}$        & ${\cal M}$          & ${\cal O}$                  & ${\cal M}$                    & ${\cal R}$ 
  \\ \hline
      A$^{\cal S}$      & $0$ & $0$ & $\ZZ^2$    & $0\subseteq0$     & $\ZZ\subseteq\ZZ$   & $0\subseteq0\subseteq\ZZ^2$ & $0\subseteq0\subseteq0$       & $\ZZ\subseteq\ZZ\subseteq\ZZ^2$  \\
      AIII$^{\cal S_+}$ & $1$ & $0$ & $0$        & $0\subseteq\ZZ^2$ & $0\subseteq0$       & $0\subseteq0\subseteq0$     & $0\subseteq\ZZ\subseteq\ZZ$   & $0\subseteq0\subseteq0$          \\ \hline
      A$^{\cal CS}$     & $0$ & $1$ & $0$        & $0\subseteq\ZZ$   & $0\subseteq0$       & $0\subseteq0\subseteq0$     & $0\subseteq\ZZ\subseteq\ZZ^2$ & $0\subseteq0\subseteq0$          \\
      AIII$^{\cal S_-}$ & $1$ & $1$ & $\ZZ$      & $0\subseteq0$     & $\ZZ\subseteq\ZZ^2$ & $0\subseteq0\subseteq\ZZ$   & $0\subseteq0\subseteq0$       & $2\ZZ\subseteq\ZZ\subseteq\ZZ$   \\ \hline\hline
    \end{tabular}
    \caption{Bulk classification sequence $\K[d] \subseteq \ldots \subseteq
      \K[1] \subseteq \K[0]$ for zero- ($d=0$), one- ($d=1$), and two-dimensional
      ($d=2$) typological crystalline insulators and superconductors
 with an order-two crystalline symmetry or antisymmetry for
      the complex ten-fold way classes.  The symbols ${\cal O}$, ${\cal M}$ and
      ${\cal R}$ refer to a local on-site ($d_{\parallel} = 0$), mirror
      ($d_{\parallel} = 1$) and twofold rotation symmetry ($d_{\parallel} =
      2$), respectively.
    \label{tab:subgroupC2d}}
  \end{center}
\end{table*}
\begin{table*}
\begin{center}
    \begin{tabular}[t]{l @{\extracolsep{\fill}} ccccccc}
      \hline\hline
   & & $d=0$ & $d=1$ & $d=1$ & $d=2$ & $d=2$ & $d=2$ 
  \\       class             & $s$ & ${\cal O}$ & ${\cal O}$        & ${\cal M}$          & ${\cal O}$                  & ${\cal M}$                    & ${\cal R}$  \\ \hline
      A$^{{\cal T}^+{\cal S}}$      & $0$ & $\ZZ$   & $0\subseteq0$     & $\ZZ_2\subseteq\ZZ_2$ & $0\subseteq0\subseteq0$     & $0\subseteq0\subseteq\ZZ$       & $\ZZ_2\subseteq\ZZ_2\subseteq\ZZ_2$ \\
      AIII$^{{\cal P}^+{\cal S}_+}$ & $1$ & $\ZZ_2$ & $0\subseteq\ZZ$   & $\ZZ_2\subseteq\ZZ_2$  & $0\subseteq0\subseteq0$     & $0\subseteq\ZZ_2\subseteq\ZZ_2$ & $0\subseteq0\subseteq0$             \\
      A$^{{\cal P}^+{\cal S}}$      & $2$ & $\ZZ_2$ & $0\subseteq\ZZ_2$ & $0\subseteq0$          & $0\subseteq0\subseteq\ZZ$   & $0\subseteq\ZZ_2\subseteq\ZZ_2$ & $0\subseteq0\subseteq2\ZZ$          \\
      AIII$^{{\cal T}^-{\cal S}_-}$ & $3$ & $0$     & $0\subseteq\ZZ_2$ & $0\subseteq2\ZZ$       & $0\subseteq0\subseteq\ZZ_2$ & $0\subseteq0\subseteq0$         & $0\subseteq0\subseteq0$             \\
      A$^{{\cal T}^-{\cal S}}$      & $4$ & $2\ZZ$  & $0\subseteq0$     & $0\subseteq0$          & $0\subseteq0\subseteq\ZZ_2$ & $0\subseteq0\subseteq2\ZZ$      & $0\subseteq0\subseteq0$             \\
      AIII$^{{\cal P}^-{\cal S}_+}$ & $5$ & $0$     & $0\subseteq2\ZZ$  & $0\subseteq0$          & $0\subseteq0\subseteq0$     & $0\subseteq0\subseteq0$         & $0\subseteq0\subseteq0$             \\
      A$^{{\cal P}^-{\cal S}}$      & $6$ & $0$     & $0\subseteq0$     & $0\subseteq0$          & $0\subseteq0\subseteq2\ZZ$  & $0\subseteq0\subseteq0$         & $0\subseteq0\subseteq\ZZ$           \\
      AIII$^{{\cal T}^+{\cal S}_-}$ & $7$ & $0$     & $0\subseteq0$     & $0\subseteq\ZZ$         & $0\subseteq0\subseteq0$     & $0\subseteq0\subseteq0$         & $0\subseteq\ZZ_2\subseteq\ZZ_2$     \\ \hline\hline
    \end{tabular}
    \caption{Same as table \ref{tab:subgroupC2d}, but for antiunitary symmetries and antisymmetries.\label{tab:subgroupA2d}}
  \end{center}
\end{table*}
\begin{table*}
  \begin{center}
    \begin{tabular}[t]{l @{\extracolsep{\fill}} cccccccc}
      \hline\hline 
                   & & & $d=0$ & $d=1$ & $d=1$ & $d=2$ & $d=2$ & $d=2$  \\
      class                     & $s$ & $t$ & ${\cal O}$ & ${\cal O}$          & ${\cal M}$              & ${\cal O}$                    & ${\cal M}$                        & ${\cal R}$  \\ \hline
      AI$^{\mathcal{S}_+}$      & $0$ & $0$ & $\ZZ^2$    & $0\subseteq0$       & $\ZZ\subseteq\ZZ$       & $0\subseteq0\subseteq0$       & $0\subseteq0\subseteq0$           & $2\ZZ\subseteq2\ZZ\subseteq2\ZZ$       \\
      BDI$^{\mathcal{S}_{++}}$  & $1$ & $0$ & $\ZZ_2^2$  & $0\subseteq\ZZ^2$   & $\ZZ_2\subseteq\ZZ_2$   & $0\subseteq0\subseteq0$       & $0\subseteq\ZZ\subseteq\ZZ$       & $0\subseteq0\subseteq0$               \\
      D$^{\mathcal{S}_{+}}$     & $2$ & $0$ & $\ZZ_2^2$  & $0\subseteq\ZZ_2^2$ & $\ZZ_2\subseteq\ZZ_2$   & $0\subseteq0\subseteq\ZZ^2$   & $0\subseteq\ZZ_2\subseteq\ZZ_2$   & $0\subseteq0\subseteq\ZZ$               \\
      DIII$^{\mathcal{S}_{++}}$ & $3$ & $0$ & $0$        & $0\subseteq\ZZ_2^2$ & $0\subseteq0$           & $0\subseteq0\subseteq\ZZ_2^2$ & $0\subseteq\ZZ_2\subseteq\ZZ_2$   & $0\subseteq0\subseteq0$               \\
      AII$^{\mathcal{S}_{+}}$   & $4$ & $0$ & $2\ZZ^2$   & $0\subseteq0$       & $2\ZZ\subseteq2\ZZ$     & $0\subseteq0\subseteq\ZZ_2^2$ & $0\subseteq0\subseteq0$           & $4\ZZ\subseteq4\ZZ\subseteq2\ZZ$       \\
      CII$^{\mathcal{S}_{++}}$  & $5$ & $0$ & $0$        & $0\subseteq2\ZZ^2$  & $0\subseteq0$           & $0\subseteq0\subseteq0$       & $0\subseteq2\ZZ\subseteq2\ZZ$     & $0\subseteq0\subseteq0$               \\
      C$^{\mathcal{S}_{+}}$     & $6$ & $0$ & $0$        & $0\subseteq0$       & $0\subseteq0$           & $0\subseteq0\subseteq2\ZZ^2$  & $0\subseteq0\subseteq0$           & $0\subseteq0\subseteq\ZZ$             \\
      CI$^{\mathcal{S}_{++}}$   & $7$ & $0$ & $0$        & $0\subseteq0$       & $0\subseteq0$           & $0\subseteq0\subseteq0$       & $0\subseteq0\subseteq0$           & $0\subseteq0\subseteq0$               \\
      \hline
      AI$^{\mathcal{CS}_-}$     & $0$ & $1$ & $0$        & $0\subseteq0$       & $0\subseteq0$           & $0\subseteq0\subseteq0$       & $0\subseteq0\subseteq0$           & $0\subseteq0\subseteq0$               \\
      BDI$^{\mathcal{S}_{+-}}$  & $1$ & $1$ & $\ZZ$      & $0\subseteq0$       & $\ZZ\subseteq\ZZ^2$     & $0\subseteq0\subseteq0$       & $0\subseteq0\subseteq0$           & $2\ZZ\subseteq\ZZ\subseteq\ZZ$       \\
      D$^{\mathcal{CS}_{+}}$    & $2$ & $1$ & $\ZZ_2$    & $0\subseteq\ZZ$     & $\ZZ_2\subseteq\ZZ_2^2$ & $0\subseteq0\subseteq0$       & $0\subseteq\ZZ\subseteq\ZZ^2$     & $0\subseteq\ZZ_2\subseteq\ZZ_2$               \\
      DIII$^{\mathcal{S}_{-+}}$ & $3$ & $1$ & $\ZZ_2$    & $0\subseteq\ZZ_2$   & $\ZZ_2\subseteq\ZZ_2^2$ & $0\subseteq0\subseteq\ZZ$     & $0\subseteq\ZZ_2\subseteq\ZZ_2^2$ & $0\subseteq\ZZ_2\subseteq\ZZ_2$               \\
      AII$^{\mathcal{CS}_{-}}$  & $4$ & $1$ & $0$        & $0\subseteq\ZZ_2$   & $0\subseteq0$           & $0\subseteq0\subseteq\ZZ_2$   & $0\subseteq\ZZ_2\subseteq\ZZ_2^2$ & $0\subseteq0\subseteq0$               \\
      CII$^{\mathcal{S}_{+-}}$  & $5$ & $1$ & $2\ZZ$     & $0\subseteq0$       & $2\ZZ\subseteq2\ZZ^2$   & $0\subseteq0\subseteq\ZZ_2$   & $0\subseteq0\subseteq0$           & $4\ZZ\subseteq2\ZZ\subseteq2\ZZ$       \\
      C$^{\mathcal{CS}_{+}}$    & $6$ & $1$ & $0$        & $0\subseteq2\ZZ$    & $0\subseteq0$           & $0\subseteq0\subseteq0$       & $0\subseteq2\ZZ\subseteq2\ZZ^2$   & $0\subseteq0\subseteq0$               \\
      CI$^{\mathcal{S}_{-+}}$   & $7$ & $1$ & $0$        & $0\subseteq0$       & $0\subseteq0$           & $0\subseteq0\subseteq2\ZZ$    & $0\subseteq0\subseteq0$           & $0\subseteq0\subseteq0$               \\
      \hline
      AI$^{\mathcal{S}_-}$      & $0$ & $2$ & $\ZZ$      & $0\subseteq0$       & $0\subseteq0$           & $0\subseteq0\subseteq2\ZZ$    & $0\subseteq0\subseteq0$           & $0\subseteq0\subseteq0$               \\
      BDI$^{\mathcal{S}_{--}}$  & $1$ & $2$ & $0$        & $0\subseteq\ZZ$     & $0\subseteq0$           & $0\subseteq0\subseteq0$       & $0\subseteq0\subseteq0$           & $0\subseteq0\subseteq0$               \\
      D$^{\mathcal{S}_{-}}$     & $2$ & $2$ & $2\ZZ$     & $0\subseteq0$       & $2\ZZ\subseteq\ZZ$      & $0\subseteq0\subseteq\ZZ$     & $0\subseteq0\subseteq0$           & $2\ZZ\subseteq\ZZ\subseteq\ZZ^2$      \\
      DIII$^{\mathcal{S}_{--}}$ & $3$ & $2$ & $0$        & $0\subseteq2\ZZ$    & $0\subseteq\ZZ_2$       & $0\subseteq0\subseteq0$       & $0\subseteq2\ZZ\subseteq\ZZ$      & $0\subseteq\ZZ_2\subseteq\ZZ_2^2$     \\
      AII$^{\mathcal{S}_{-}}$   & $4$ & $2$ & $\ZZ$      & $0\subseteq0$       & $\ZZ_2\subseteq\ZZ_2$   & $0\subseteq0\subseteq2\ZZ$    & $0\subseteq0\subseteq\ZZ_2$       & $\ZZ_2\subseteq\ZZ_2\subseteq\ZZ_2^2$ \\
      CII$^{\mathcal{S}_{--}}$  & $5$ & $2$ & $0$        & $0\subseteq\ZZ$     & $0\subseteq0$           & $0\subseteq0\subseteq0$       & $0\subseteq\ZZ_2\subseteq\ZZ_2$   & $0\subseteq0\subseteq0$               \\
      C$^{\mathcal{S}_{-}}$     & $6$ & $2$ & $2\ZZ$     & $0\subseteq0$       & $2\ZZ\subseteq2\ZZ$     & $0\subseteq0\subseteq\ZZ$     & $0\subseteq0\subseteq0$           & $2\ZZ\subseteq2\ZZ\subseteq2\ZZ^2$    \\
      CI$^{\mathcal{S}_{--}}$   & $7$ & $2$ & $0$        & $0\subseteq2\ZZ$    & $0\subseteq0$           & $0\subseteq0\subseteq0$       & $0\subseteq2\ZZ\subseteq2\ZZ$     & $0\subseteq0\subseteq0$               \\
      \hline
      AI$^{\mathcal{CS}_+}$     & $0$ & $3$ & $\ZZ_2$    & $0\subseteq\ZZ$     & $0\subseteq0$           & $0\subseteq0\subseteq0$       & $0\subseteq2\ZZ\subseteq2\ZZ$     & $0\subseteq0\subseteq0$               \\
      BDI$^{\mathcal{S}_{-+}}$  & $1$ & $3$ & $\ZZ_2$    & $0\subseteq\ZZ_2$   & $0\subseteq\ZZ$         & $0\subseteq0\subseteq\ZZ$     & $0\subseteq0\subseteq0$           & $0\subseteq0\subseteq0$               \\
      D$^{\mathcal{CS}_{-}}$    & $2$ & $3$ & $0$        & $0\subseteq\ZZ_2$   & $0\subseteq0$           & $0\subseteq0\subseteq\ZZ_2$   & $0\subseteq0\subseteq\ZZ$         & $0\subseteq0\subseteq0$               \\
      DIII$^{\mathcal{S}_{+-}}$ & $3$ & $3$ & $2\ZZ$     & $0\subseteq0$       & $4\ZZ\subseteq2\ZZ$     & $0\subseteq0\subseteq\ZZ_2$   & $0\subseteq0\subseteq0$           & $4\ZZ\subseteq2\ZZ\subseteq\ZZ$       \\
      AII$^{\mathcal{CS}_{+}}$  & $4$ & $3$ & $0$        & $0\subseteq2\ZZ$    & $0\subseteq0$           & $0\subseteq0\subseteq0$       & $0\subseteq4\ZZ\subseteq2\ZZ$     & $0\subseteq0\subseteq\ZZ_2$           \\
      CII$^{\mathcal{S}_{-+}}$  & $5$ & $3$ & $0$        & $0\subseteq0$       & $0\subseteq\ZZ$         & $0\subseteq0\subseteq2\ZZ$    & $0\subseteq0\subseteq0$           & $0\subseteq\ZZ_2\subseteq\ZZ_2$       \\
      C$^{\mathcal{CS}_{-}}$    & $6$ & $3$ & $0$        & $0\subseteq0$       & $0\subseteq0$           & $0\subseteq0\subseteq0$       & $0\subseteq0\subseteq\ZZ$         & $0\subseteq0\subseteq0$               \\
      CI$^{\mathcal{S}_{+-}}$   & $7$ & $3$ & $\ZZ$      & $0\subseteq0$       & $2\ZZ\subseteq2\ZZ$     & $0\subseteq0\subseteq0$       & $0\subseteq0\subseteq0$           & $2\ZZ\subseteq2\ZZ\subseteq2\ZZ$      \\
      \hline\hline
    \end{tabular}
    \caption{Bulk classification sequence $\K[d] \subseteq \ldots \subseteq
      \K[1] \subseteq \K[0]$ for zero- ($d=0$), one- ($d=1$), and two-dimensional
      ($d=2$) topological crystalline phases with an order-two crystalline symmetry or antisymmetry for
    the real ten-fold way classes.  The symbols ${\cal O}$,
    ${\cal M}$ and ${\cal R}$ refer to a local on-site ($d_{\parallel} =
    0$), mirror ($d_{\parallel} = 1$) and twofold rotation symmetry
    ($d_{\parallel} = 2$), respectively.
    \label{tab:subgroupR2d}}
  \end{center}
\end{table*}

The above characterization of unitary and antiunitary symmetry operations by
the signs $\eta_{{\cal S},{\cal T},{\cal P},{\cal C}}$ and $\sigma_{\cal S}$
may be redundant,~\cite{shiozaki2014} because symmetry operations that are
characterized differently may be mapped onto each other. For example, if $H$
satisfies a crystalline unitary symmetry operation ${\cal S}$ which squares to
one, then it also satisfies the unitary symmetry operation $i{\cal S}$, which
squares to minus one, or (provided ${\cal T}$-symmetry is present) it satisfies
the antiunitary symmetry ${\cal T}{\cal S}$. Using such equivalences, Shiozaki
and Sato group the symmetry operations ${\cal S}$ into ``equivalence classes'',
which, together with the ten-fold way class of Table~\ref{tab:0}, are
labeled by one integer $s$ or by two integers $s$ and $t$. In this work (as in
Ref.~\onlinecite{geier2018}) we label the equivalence classes by representative
(anti)symmetries that consist of a unitary crystalline symmetry $\mathcal{S}$
squaring to one or the product of such a crystalline symmetry and
$\mathcal{T}$, $\mathcal{P}$, or $\mathcal{C}$. These representatives are
summarized in the first column of
Tables~\ref{tab:subgroupC2d}-\ref{tab:subgroupR2d} for the complex
ten-fold way classes with unitary (anti)symmetries, the complex
ten-fold way classes with antiunitary (anti)symmetries, and the real 
ten-fold way classes with unitary (anti)symmetries, respectively. For
the complex  ten-fold way classes with antiunitary (anti)symmetries we
implicitly assume that ${\cal T}$, ${\cal P}$ commute with ${\cal S}$ 
when constructing these representatives, see Table \ref{tab:subgroupA}.

\begin{table*}[t]
	\begin{center}
		\begin{tabular}{l @{\extracolsep{\fill}} ccccccc}
			\hline\hline 
			class & $s$ &  $t$ & ${\cal O}$ & ${\cal M}$   & ${\cal R}$ &  ${\cal I}$  \\ \hline
			A$^{\cal S}$      & $0$ & $0$ & $0\subseteq0\subseteq0\subseteq0$         & $0\subseteq0\subseteq\ZZ\subseteq\ZZ$   & $0\subseteq0\subseteq0\subseteq0$         & $2\ZZ\subseteq2\ZZ\subseteq\ZZ\subseteq\ZZ$  \\
			AIII$^{\cal S_+}$ & $1$ & $0$ & $0\subseteq0\subseteq0\subseteq\ZZ^2$ & $0\subseteq0\subseteq0\subseteq0$       & $0\subseteq\ZZ\subseteq\ZZ\subseteq\ZZ^2$ & $0\subseteq0\subseteq0\subseteq0$            \\ \hline
			A$^{\cal CS}$     & $0$ & $1$ & $0\subseteq0\subseteq0\subseteq\ZZ$  & $0\subseteq0\subseteq0\subseteq0$       & $0\subseteq2\ZZ\subseteq\ZZ\subseteq\ZZ$  & $0\subseteq0\subseteq0\subseteq0$            \\
			AIII$^{\cal S_-}$ & $1$ & $1$ & $0\subseteq0\subseteq0\subseteq0$         & $0\subseteq0\subseteq\ZZ\subseteq\ZZ^2$ & $0\subseteq0\subseteq0\subseteq0$         & $2\ZZ\subseteq\ZZ\subseteq\ZZ\subseteq\ZZ^2$ \\ \hline\hline
		\end{tabular}
		  \caption{Bulk classification sequence~(\ref{eq:subgroup}) for
		  three-dimensional topological crystalline phases with an order-two unitary crystalline
		  (anti)symmetry for the complex ten-fold way classes. The
		  symbols ${\cal O}$, ${\cal M}$, ${\cal R}$ and ${\cal I}$
		  refer to local on-site ($d_{\parallel} = 0$), mirror
		  ($d_{\parallel} = 1$), twofold rotation ($d_{\parallel} =
		  2$), and inversion symmetry ($d_{\parallel} = 3$),
		  respectively.\label{tab:subgroupC}}
	\end{center}
\end{table*}
\begin{table*}[t]
	  \begin{center}
		  \begin{tabular}{l@{\extracolsep{\fill}} ccccccc}
			  \hline\hline 
			  class                         & $s$ & ${\cal O}$ & ${\cal M}$  & ${\cal R}$ & ${\cal I}$ \\ \hline
			  A$^{{\cal T}^+{\cal S}}$      & $0$ & $0\subseteq0\subseteq0\subseteq0$  & $0\subseteq0\subseteq0\subseteq0$         & $0\subseteq0\subseteq\ZZ_2\subseteq\ZZ_2$     & $0\subseteq0\subseteq0\subseteq0$             \\
			  AIII$^{{\cal P}^+{\cal S}_+}$ & $1$ & $0\subseteq0\subseteq0\subseteq0$  & $0\subseteq0\subseteq0\subseteq\ZZ$       & $0\subseteq\ZZ_2\subseteq\ZZ_2\subseteq\ZZ_2$ & $0\subseteq0\subseteq0\subseteq2\ZZ$          \\ 
			  A$^{{\cal P}^+{\cal S}}$      & $2$ & $0\subseteq0\subseteq0\subseteq0$         & $0\subseteq0\subseteq\ZZ_2\subseteq\ZZ_2$ & $0\subseteq0\subseteq0\subseteq0$             & $0\subseteq0\subseteq0\subseteq0$             \\
			  AIII$^{{\cal T}^-{\cal S}_-}$ & $3$ & $0\subseteq0\subseteq0\subseteq\ZZ$       & $0\subseteq0\subseteq\ZZ_2\subseteq\ZZ_2$ & $0\subseteq0\subseteq0\subseteq2\ZZ$          & $0\subseteq0\subseteq0\subseteq0$             \\ 
			  A$^{{\cal T}^-{\cal S}}$      & $4$ & $0\subseteq0\subseteq0\subseteq\ZZ_2$ & $0\subseteq0\subseteq0\subseteq0$         & $0\subseteq0\subseteq0\subseteq0$             & $0\subseteq0\subseteq0\subseteq0$             \\
			  AIII$^{{\cal P}^-{\cal S}_+}$ & $5$ & $0\subseteq0\subseteq0\subseteq\ZZ_2$ & $0\subseteq0\subseteq0\subseteq2\ZZ$      & $0\subseteq0\subseteq0\subseteq0$             & $0\subseteq0\subseteq0\subseteq\ZZ$           \\ 
			  A$^{{\cal P}^-{\cal S}}$      & $6$ & $0\subseteq0\subseteq0\subseteq0$         & $0\subseteq0\subseteq0\subseteq0$         & $0\subseteq0\subseteq0\subseteq0$             & $0\subseteq0\subseteq\ZZ_2\subseteq\ZZ_2$     \\
			  AIII$^{{\cal T}^+{\cal S}_-}$ & $7$ & $0\subseteq0\subseteq0\subseteq2\ZZ$      & $0\subseteq0\subseteq0\subseteq0$         & $0\subseteq0\subseteq0\subseteq\ZZ$           & $0\subseteq\ZZ_2\subseteq\ZZ_2\subseteq\ZZ_2$ \\ \hline\hline
		  \end{tabular}
		\caption{Same as table \ref{tab:subgroupC}, but for antiunitary (anti)symmetries.\label{tab:subgroupA}}
	  \end{center}
\end{table*}
\begin{table*}[!]
\begin{center}
\begin{tabular}[t]{l @{\extracolsep{\fill}} cccccc}
\hline\hline 
				class                     & $s$ & $t$ & ${\cal O}$                              & ${\cal M}$                                  & ${\cal R}$                                      & ${\cal I}$  \\ \hline
				AI$^{\mathcal{S}_+}$      & $0$ & $0$ & $0\subseteq0\subseteq0\subseteq0$       & $0\subseteq0\subseteq0\subseteq0$           & $0\subseteq0\subseteq0\subseteq0$               & $2\ZZ\subseteq2\ZZ\subseteq2\ZZ\subseteq2\ZZ$      \\
				BDI$^{\mathcal{S}_{++}}$  & $1$ & $0$ & $0\subseteq0\subseteq0\subseteq0$       & $0\subseteq0\subseteq0\subseteq0$           & $0\subseteq2\ZZ\subseteq2\ZZ\subseteq2\ZZ$      & $0\subseteq0\subseteq0\subseteq0$               \\
				D$^{\mathcal{S}_{+}}$     & $2$ & $0$ & $0\subseteq0\subseteq0\subseteq0$       & $0\subseteq0\subseteq\ZZ\subseteq\ZZ$       & $0\subseteq0\subseteq0\subseteq0$               & $0\subseteq0\subseteq0\subseteq0$               \\
				DIII$^{\mathcal{S}_{++}}$ & $3$ & $0$ & $0\subseteq0\subseteq0\subseteq\ZZ^2$   & $0\subseteq0\subseteq\ZZ_2\subseteq\ZZ_2$   & $0\subseteq0\subseteq0\subseteq\ZZ$             & $0\subseteq0\subseteq0\subseteq0$               \\
				AII$^{\mathcal{S}_{+}}$   & $4$ & $0$ & $0\subseteq0\subseteq0\subseteq\ZZ_2^2$ & $0\subseteq0\subseteq\ZZ_2\subseteq\ZZ_2$   & $0\subseteq0\subseteq0\subseteq0$               & $4\ZZ\subseteq4\ZZ\subseteq2\ZZ\subseteq\ZZ$    \\
				CII$^{\mathcal{S}_{++}}$  & $5$ & $0$ & $0\subseteq0\subseteq0\subseteq\ZZ_2^2$ & $0\subseteq0\subseteq0\subseteq0$           & $0\subseteq4\ZZ\subseteq4\ZZ\subseteq2\ZZ$      & $0\subseteq0\subseteq0\subseteq\ZZ_2$           \\
				C$^{\mathcal{S}_{+}}$     & $6$ & $0$ & $0\subseteq0\subseteq0\subseteq0$       & $0\subseteq0\subseteq2\ZZ\subseteq2\ZZ$     & $0\subseteq0\subseteq0\subseteq0$               & $0\subseteq0\subseteq\ZZ_2\subseteq\ZZ_2$       \\
				CI$^{\mathcal{S}_{++}}$   & $7$ & $0$ & $0\subseteq0\subseteq0\subseteq2\ZZ^2$  & $0\subseteq0\subseteq0\subseteq0$           & $0\subseteq0\subseteq0\subseteq\ZZ$             & $0\subseteq0\subseteq0\subseteq0$               \\
				\hline
				AI$^{\mathcal{CS}_-}$     & $0$ & $1$ & $0\subseteq0\subseteq0\subseteq2\ZZ$    & $0\subseteq0\subseteq0\subseteq0$           & $0\subseteq0\subseteq0\subseteq0$               & $0\subseteq0\subseteq0\subseteq0$               \\
				BDI$^{\mathcal{S}_{+-}}$  & $1$ & $1$ & $0\subseteq0\subseteq0\subseteq0$       & $0\subseteq0\subseteq0\subseteq0$           & $0\subseteq0\subseteq0\subseteq0$               & $4\ZZ\subseteq2\ZZ\subseteq2\ZZ\subseteq2\ZZ$    \\
				D$^{\mathcal{CS}_{+}}$    & $2$ & $1$ & $0\subseteq0\subseteq0\subseteq0$       & $0\subseteq0\subseteq0\subseteq0$           & $0\subseteq2\ZZ\subseteq\ZZ\subseteq\ZZ$        & $0\subseteq0\subseteq0\subseteq0$               \\
				DIII$^{\mathcal{S}_{-+}}$ & $3$ & $1$ & $0\subseteq0\subseteq0\subseteq0$       & $0\subseteq0\subseteq\ZZ\subseteq\ZZ^2$     & $0\subseteq0\subseteq\ZZ_2\subseteq\ZZ_2$       & $0\subseteq0\subseteq0\subseteq\ZZ$               \\
				AII$^{\mathcal{CS}_{-}}$  & $4$ & $1$ & $0\subseteq0\subseteq0\subseteq\ZZ$     & $0\subseteq0\subseteq\ZZ_2\subseteq\ZZ_2^2$ & $0\subseteq0\subseteq\ZZ_2\subseteq\ZZ_2$       & $0\subseteq0\subseteq0\subseteq0$               \\
				CII$^{\mathcal{S}_{+-}}$  & $5$ & $1$ & $0\subseteq0\subseteq0\subseteq\ZZ_2$   & $0\subseteq0\subseteq\ZZ_2\subseteq\ZZ_2^2$ & $0\subseteq0\subseteq0\subseteq0$               & $8\ZZ\subseteq4\ZZ\subseteq4\ZZ\subseteq2\ZZ$    \\
				C$^{\mathcal{CS}_{+}}$    & $6$ & $1$ & $0\subseteq0\subseteq0\subseteq\ZZ_2$   & $0\subseteq0\subseteq0\subseteq0$           & $0\subseteq4\ZZ\subseteq2\ZZ\subseteq2\ZZ$      & $0\subseteq0\subseteq0\subseteq0$               \\
				CI$^{\mathcal{S}_{-+}}$   & $7$ & $1$ & $0\subseteq0\subseteq0\subseteq0$       & $0\subseteq0\subseteq2\ZZ\subseteq2\ZZ^2$   & $0\subseteq0\subseteq0\subseteq0$               & $0\subseteq0\subseteq0\subseteq\ZZ$             \\
				\hline
				AI$^{\mathcal{S}_-}$      & $0$ & $2$ & $0\subseteq0\subseteq0\subseteq0$       & $0\subseteq0\subseteq2\ZZ\subseteq2\ZZ$     & $0\subseteq0\subseteq0\subseteq0$               & $0\subseteq0\subseteq0\subseteq0$               \\
				BDI$^{\mathcal{S}_{--}}$  & $1$ & $2$ & $0\subseteq0\subseteq0\subseteq2\ZZ$    & $0\subseteq0\subseteq0\subseteq0$           & $0\subseteq0\subseteq0\subseteq0$               & $0\subseteq0\subseteq0\subseteq0$               \\
				D$^{\mathcal{S}_{-}}$     & $2$ & $2$ & $0\subseteq0\subseteq0\subseteq0$       & $0\subseteq0\subseteq0\subseteq0$           & $0\subseteq0\subseteq0\subseteq0$               & $4\ZZ\subseteq2\ZZ\subseteq\ZZ\subseteq\ZZ$    \\
				DIII$^{\mathcal{S}_{--}}$ & $3$ & $2$ & $0\subseteq0\subseteq0\subseteq\ZZ$     & $0\subseteq0\subseteq0\subseteq0$           & $0\subseteq2\ZZ\subseteq\ZZ\subseteq\ZZ^2$      & $0\subseteq0\subseteq\ZZ_2\subseteq\ZZ_2$               \\
				AII$^{\mathcal{S}_{-}}$   & $4$ & $2$ & $0\subseteq0\subseteq0\subseteq0$       & $0\subseteq0\subseteq2\ZZ\subseteq\ZZ$      & $0\subseteq0\subseteq\ZZ_2\subseteq\ZZ_2^2$     & $0\subseteq0\subseteq\ZZ_2\subseteq\ZZ_2$               \\
				CII$^{\mathcal{S}_{--}}$  & $5$ & $2$ & $0\subseteq0\subseteq0\subseteq2\ZZ$    & $0\subseteq0\subseteq0\subseteq\ZZ_2$       & $0\subseteq\ZZ_2\subseteq\ZZ_2\subseteq\ZZ_2^2$ & $0\subseteq0\subseteq0\subseteq0$               \\
				C$^{\mathcal{S}_{-}}$     & $6$ & $2$ & $0\subseteq0\subseteq0\subseteq0$       & $0\subseteq0\subseteq\ZZ_2\subseteq\ZZ_2$   & $0\subseteq0\subseteq0\subseteq0$               & $4\ZZ\subseteq4\ZZ\subseteq2\ZZ\subseteq2\ZZ$    \\
				CI$^{\mathcal{S}_{--}}$   & $7$ & $2$ & $0\subseteq0\subseteq0\subseteq\ZZ$     & $0\subseteq0\subseteq0\subseteq0$           & $0\subseteq2\ZZ\subseteq2\ZZ\subseteq2\ZZ^2$    & $0\subseteq0\subseteq0\subseteq0$               \\
				\hline
				AI$^{\mathcal{CS}_+}$     & $0$ & $3$ & $0\subseteq0\subseteq0\subseteq0$       & $0\subseteq0\subseteq0\subseteq0$           & $0\subseteq2\ZZ\subseteq2\ZZ\subseteq2\ZZ$      & $0\subseteq0\subseteq0\subseteq0$               \\
				BDI$^{\mathcal{S}_{-+}}$  & $1$ & $3$ & $0\subseteq0\subseteq0\subseteq0$       & $0\subseteq0\subseteq2\ZZ\subseteq2\ZZ$     & $0\subseteq0\subseteq0\subseteq0$               & $0\subseteq0\subseteq0\subseteq0$               \\
				D$^{\mathcal{CS}_{-}}$    & $2$ & $3$ & $0\subseteq0\subseteq0\subseteq\ZZ$     & $0\subseteq0\subseteq0\subseteq0$           & $0\subseteq0\subseteq0\subseteq0$               & $0\subseteq0\subseteq0\subseteq0$               \\
				DIII$^{\mathcal{S}_{+-}}$ & $3$ & $3$ & $0\subseteq0\subseteq0\subseteq\ZZ_2$   & $0\subseteq0\subseteq0\subseteq\ZZ$         & $0\subseteq0\subseteq0\subseteq0$               & $4\ZZ\subseteq2\ZZ\subseteq\ZZ\subseteq\ZZ^2$   \\
				AII$^{\mathcal{CS}_{+}}$  & $4$ & $3$ & $0\subseteq0\subseteq0\subseteq\ZZ_2$   & $0\subseteq0\subseteq0\subseteq0$           & $0\subseteq4\ZZ\subseteq2\ZZ\subseteq\ZZ$       & $0\subseteq0\subseteq\ZZ_2\subseteq\ZZ_2^2$     \\
				CII$^{\mathcal{S}_{-+}}$  & $5$ & $3$ & $0\subseteq0\subseteq0\subseteq0$       & $0\subseteq0\subseteq4\ZZ\subseteq2\ZZ$     & $0\subseteq0\subseteq0\subseteq\ZZ_2$           & $0\subseteq\ZZ_2\subseteq\ZZ_2\subseteq\ZZ_2^2$ \\
				C$^{\mathcal{CS}_{-}}$    & $6$ & $3$ & $0\subseteq0\subseteq0\subseteq2\ZZ$    & $0\subseteq0\subseteq0\subseteq0$           & $0\subseteq0\subseteq\ZZ_2\subseteq\ZZ_2$       & $0\subseteq0\subseteq0\subseteq0$               \\
				CI$^{\mathcal{S}_{+-}}$   & $7$ & $3$ & $0\subseteq0\subseteq0\subseteq0$       & $0\subseteq0\subseteq0\subseteq\ZZ$         & $0\subseteq0\subseteq0\subseteq0$               & $2\ZZ\subseteq2\ZZ\subseteq2\ZZ\subseteq2\ZZ^2$\\
				\hline\hline
			\end{tabular}
			\caption{Bulk classification sequence~(\ref{eq:subgroup}) for
			three-dimensional topological crystalline phases with an order-two crystalline symmetry or
			antisymmetry for the real ten-fold way classes.  The symbols ${\cal
			O}$, ${\cal M}$, ${\cal R}$ and ${\cal I}$ refer to local on-site
			($d_{\parallel} = 0$), mirror ($d_{\parallel} = 1$), twofold rotation
			($d_{\parallel} = 2$), and inversion symmetry ($d_{\parallel} = 3$), respectively.
			\label{tab:subgroupR}}
		\end{center}
	\end{table*}

The classification of topological phases (with or without the additional
crystalline symmetry or antisymmetry) has a group structure, and the symbol $K$
(or ${\cal K}$) is used to denote the corresponding classifying group.
Formally, the group structure is obtained by the Grothendieck
construction,\cite{nakahara2003,turner2012}  where one considers equivalence
classes of ordered pairs $(H_1,H_2)$ of Hamiltonians represented by hermitian
matrix-valued functions $H(\vk)$ of equal dimension, the equivalence relation
being that two pairs $(H_1,H_2)$ and $(H_1^\prime,H_2^\prime)$ are
topologically equivalent if $H_1\oplus H_2^\prime$ is continuously deformable
to $H_1^\prime\oplus H_2$. Loosely speaking the ordered pair $(H_1,H_2)$
represents the ``difference'' of the two Hamiltonians $H_1$ and $H_2$. Without
loss of generality, one may take $H_1$ or $H_2$ to be a reference Hamiltonian
$H_{\rm ref}$. With this convention, the trivial element is represented by
$(H_{\rm ref},H_{\rm ref})$, whereas the inverse of the group element
$(H,H_{\rm ref})$ is $(H_{\rm ref},H)$.  Alternatively, instead of the ordered
pair $(H,H_{\rm ref})$ one may consider a one-parameter family of Hamiltonians
$H(\mm)$ that interpolates between $H$ and the reference Hamiltonian $H_{\rm
ref}$.\cite{shiozaki2015,shiozaki2016} In this work, we take the latter
approach and consider one-parameter family of Hamiltonians $H(\mm)$, such that
$H(\mm)$ is in the topological class of $H$ for $-2<\mm<0$ and in the
topological class of $H_{\rm ref}$ for $0<\mm<2$, with the transition between
topological classes (if any) taking place at $\mm=0$. When considering
Hamiltonian families $H(\mm)$, we will often omit the parameter $\mm$ and refer
to it simply as the ``Hamiltonian $H$''. The ``canonical-form'' Hamiltonians introduced in Sec.\ \ref{sec:cform} are examples of such $\mm$-dependent families of Hamiltonians.

The classification of topological crystalline phases of
Ref.~\onlinecite{shiozaki2014} is based on isomorphisms between the groups
$K(s,t|d,d_{\parallel})$ and $K(s|d,d_{\parallel})$
classifying $d$-dimensional Hamiltonians with the symmetries labeled by the
corresponding indices, where $d_{\parallel}$ is the number of inverted spatial
dimensions. The above mentioned isomorphisms are extensions of Teo and Kane's
dimension-raising isomorphism~\cite{teo2010} $\kappa$ increasing the spatial
dimension by one to the systems with an order-two crystalline symmetry or
antisymmetry.\cite{shiozaki2014} Shiozaki and Sato introduce two
isomorphisms $\kappa_\parallel$ and $\kappa_\perp$, where the isomorphism
$\kappa_\parallel$ increases both the spatial dimension $d$ and the number of
the inverted momenta $d_\parallel$, whereas the isomorphism $\kappa_\perp$
increases only the spatial dimension $d$ while keeping $d_\parallel$ unchanged.
%We review these isomorphisms in App.~\ref{app:dmaps}.

For the complex and real classes with unitary (anti)symmetry the classifying
groups are denoted $\K[0](s,t\vert d,d_\parallel)$ and these isomorphisms are
(with $d_{\parallel} < d$)
\begin{align}
  \K[0](s,t|d,d_{\parallel}) &\overset{\kappa_\parallel}= \K[0](s+1,t+1|d+1,d_{\parallel}+1) \nonumber \\
  &\overset{\kappa_\perp}= \K[0](s+1,t|d+1,d_{\parallel}),
  \label{eq:K1}
\end{align}
with the integers $s$ and $t$ taken mod $2$ for complex classes, and mod $8$
and mod $4$, respectively, for the real classes. We use the same notation for
the classifying groups for the real and complex classes. When discussing
specific examples we will always specify the ten-fold way class using its
Cartan symbol, so that no confusion is possible. For complex classes with
antiunitary (anti)symmetry these isomorphisms are
\begin{align}
  \K[0](s|d,d_{\parallel}) &\overset{\kappa_\parallel}= \K[0](s-1|d+1,d_{\parallel}+1) \nonumber \\
  &\overset{\kappa_\perp}= \K[0](s+1|d+1,d_{\parallel}).
  \label{eq:K2}
\end{align}
When applied repeatedly, these isomorphisms can be used to relate the
classification problem of $d$-dimensional Hamiltonians with an order-two
crystalline symmetry to a zero-dimensional classification problem with an on-site
symmetry,~\cite{shiozaki2014,thorngren2018} which can be solved with elementary
methods. 

Following Teo and Kane, Shiozaki and Sato also introduce an isomorphism
$\rho_{\parallel}$ relating a topological class of Hamiltonians $H(\vk)$ with
an additional crystalline (anti)symmetry ${\cal S}$ to the topological class of
one-parameter family of Hamiltonians $H(\vk,\varphi)$, $0 \le \varphi \le 2
\pi$, with the additional conditions $H(\vk,0) = H(\vk,2 \pi)$ and $H(\vk,\varphi) = {\cal S} H(\vk,2 \pi -
\varphi)$. This isomorphism and the dimension-raising isomorphism
$\kappa_{\parallel}$ introduced above play a central role in our algebraic
construction of a higher-order bulk-boundary correspondence for topological
crystalline phases, see Secs.~\ref{sec:omap} and \ref{sec:omapexplicit}.
Further details of these isomorphisms are given in App.~\ref{app:dmaps}

\begin{table}
  \begin{tabular*}{\columnwidth}{c @{\extracolsep{\fill}} ccc}
    \hline\hline
    spin-orbit & $d=1$                            & $d=2$                            & $d=3$ \\\hline
    mirror & A$^{\cal M}$, AII$^{{\cal M}_-}$ & A$^{\cal M}$, AII$^{{\cal M}_-}$ & A$^{\cal M}$, AII$^{{\cal M}_-}$    \\
    twofold rotation & A$^{\cal M}$, AII$^{{\cal M}_-}$ & A$^{\cal R}$, AII$^{{\cal R}_-}$ & A$^{\cal R}$, AII$^{{\cal R}_-}$    \\ 
    inversion & A$^{\cal M}$, AII$^{{\cal M}_+}$ & A$^{\cal R}$, AII$^{{\cal R}_+}$ & A$^{\cal I}$, AII$^{{\cal I}_+}$    \\ \hline
  \end{tabular*}
  \begin{tabular*}{\columnwidth}{c @{\extracolsep{\fill}} ccc}
    no spin-orbit & $d=1$                            & $d=2$                            & $d=3$ \\\hline
    mirror & A$^{\cal M}$, AI$^{{\cal M}_+}$ & A$^{\cal M}$, AI$^{{\cal M}_+}$ & A$^{\cal M}$, AI$^{{\cal M}_+}$    \\
    twofold rotation & A$^{\cal M}$, AI$^{{\cal M}_+}$ & A$^{\cal R}$, AI$^{{\cal R}_+}$ & A$^{\cal R}$, AI$^{{\cal R}_+}$    \\ 
    inversion & A$^{\cal M}$, AI$^{{\cal M}_+}$ & A$^{\cal R}$, AI$^{{\cal R}_+}$ & A$^{\cal I}$, AI$^{{\cal I}_+}$    \\ \hline \hline
  \end{tabular*}
\caption{\label{tab:insulators} Shiozaki-Sato classes that correspond to natural physical realizations of the order-two symmetries for insulators. The top and bottom panels are for crystals with and without strong spin-orbit coupling, respectively. Time-reversal symmetric insulators have ten-fold way class AII or AI, otherwise the class is A.}
\end{table}

The 44 Shiozaki-Sato classes represent all mathematically possible algebraic
relations between a twofold crystalline symmetry or antisymmetry and the
fundamental non-spatial symmetries ${\cal T}$, ${\cal P}$, and ${\cal C}$. Not
all of these classes are naturally realized in crystals, however. One important
reason why it is nevertheless important to classify all mathematically allowed
possibilities is the existence of the isomorphisms (\ref{eq:K1}) and
(\ref{eq:K2}), which connect different symmetry classes in different
dimensions. Another reason is that symmetry classes which at first sight may
appear ``unphysical'' may be realized in condensed matter systems as {\em
effective} symmetries, see, {\em e.g.}, the examples presented in
Refs.~\onlinecite{kells2012,tewari2012,benalcazar2017,dwivedi2018}. To
facilitate the translation between the Shiozaki-Sato classes used in this
article and the ``physical'' symmetries of crystals,
Tables~\ref{tab:insulators} and \ref{tab:SC} list the relevant Shiozaki-Sato
classes for crystals with mirror, twofold rotation, or inversion symmetry. Here
we note that, whereas the ``physical'' inversion symmetry does not affect the
spin degree of freedom, the ``physical'' mirror and twofold rotation operations
do. With our convention that (unitary) symmetries square to one, this implies
that inversion commutes with the time-reversal operation ${\cal T}$, whereas
mirror and twofold rotation anticommute with ${\cal T}$ in a crystal with
strong spin orbit coupling. For a superconducting system with an order-two
crystalline symmetry, a crystalline symmetry ${\cal S}$ must only leave the
normal-state Hamiltonian unchanged, whereas the superconducting order parameter
$\Delta$ may eventually change sign under ${\cal S}$. The parity of $\Delta$
under ${\cal S}$ determines whether ${\cal S}$ commutes or anticommutes with
particle-hole conjugation. (Please note that there are physical symmetries not
included in Tables \ref{tab:insulators} and \ref{tab:SC}, such as magnetic
symmetries.)

\begin{table}
  \begin{tabular*}{\columnwidth}{c @{\extracolsep{\fill}} ccc}
    \hline\hline
    spin-orbit & $d=1$                                               & $d=2$                                               & $d=3$ \\\hline
    mirror & D$^{{\cal M}_\alpha}$, DIII$^{{\cal M}_{-\alpha}}$ & D$^{{\cal M}_\alpha}$, DIII$^{{\cal M}_{-\alpha}}$ & D$^{{\cal M}_\alpha}$, DIII$^{{\cal M}_{-\alpha}}$    \\
    rotation & D$^{{\cal M}_\alpha}$, DIII$^{{\cal M}_{-\alpha}}$ & D$^{{\cal R}_\alpha}$, DIII$^{{\cal R}_{-\alpha}}$ & D$^{{\cal R}_\alpha}$, DIII$^{{\cal R}_{-\alpha}}$    \\
    inversion & D$^{{\cal M}_\alpha}$, DIII$^{{\cal M}_{+\alpha}}$ & D$^{{\cal R}_\alpha}$, DIII$^{{\cal R}_{+\alpha}}$ & D$^{{\cal I}_\alpha}$, DIII$^{{\cal I}_{+\alpha}}$    \\ \hline
  \end{tabular*}
  \begin{tabular*}{\columnwidth}{c @{\extracolsep{\fill}} ccc}
    no spin-orbit & $d=1$                                             & $d=2$                                             & $d=3$ \\\hline
    mirror & C$^{{\cal M}_\alpha}$,CI$^{{\cal M}_{+\alpha}}$ & C$^{{\cal M}_\alpha}$, CI$^{{\cal M}_{+\alpha}}$ & C$^{{\cal M}_\alpha}$, CI$^{{\cal M}_{+\alpha}}$    \\
    rotation & C$^{{\cal M}_\alpha}$,CI$^{{\cal M}_{+\alpha}}$ & C$^{{\cal R}_\alpha}$, CI$^{{\cal R}_{+\alpha}}$ & C$^{{\cal R}_\alpha}$, CI$^{{\cal R}_{+\alpha}}$    \\
    inversion & C$^{{\cal M}_\alpha}$,CI$^{{\cal M}_{+\alpha}}$ & C$^{{\cal R}_\alpha}$, CI$^{{\cal R}_{+\alpha}}$ & C$^{{\cal I}_\alpha}$, CI$^{{\cal I}_{+\alpha}}$    \\ \hline \hline
  \end{tabular*}
\caption{\label{tab:SC} Shiozaki-Sato classes that correspond to natural
physical realizations of the order-two symmetries for superconductors with
(top) and without (bottom) strong spin-orbit coupling. For the time-reversal
invariant superconductors the ten-fold way class is DIII or CI, otherwise it is
class D or C. The parity of the superconducting order parameter under the
order-two symmetry ${\cal S}$ is denoted $\alpha=\pm$. For classes C and CI we
assume an $s$-wave superconductor.}
\end{table}

The Shiozaki-Sato classifying groups $K$ are the largest groups in the
sequence~(\ref{eq:subgroup}), which for crystals of dimension $d=0$, $1$,
and $2$ are listed in
Tables~\ref{tab:subgroupC2d}-\ref{tab:subgroupR2d} for the complex
ten-fold way classes with unitary (anti)symmetries, the complex
ten-fold way classes with antiunitary (anti)symmetries, and the real
ten-fold way classes with unitary (anti) symmetries, respectively. The
corresponding classification of three-dimensional systems is given in
Tables~\ref{tab:subgroupC}-\ref{tab:subgroupR}.
When no confusion is possible, we will omit the arguments $s$
and $t$ in what follows, and write $\K[0](d,d_{\parallel})$ instead of
$\K[0](s,t\vert d,d_{\parallel})$ or $\K[0](s\vert d,d_{\parallel})$. 

The classifying groups for the ten-fold way classes (\textit{i.e.}, without additional crystalline symmetries) are denoted by $K_{\rm TF}(s|d)$. (See Table \ref{tab:0} for the symmetry label $s$; the symmetry label $t$ does not apply to the ten-fold way classes.) We further define the subgroup
\begin{equation}
  K_{{\rm TF},{\cal S}}(s,t|d) \subseteq K_{\rm TF}(s|d),
\end{equation}
which consists of those ten-fold way phases that are compatible with the crystalline (anti)symmetry ${\cal S}$. Since ten-fold way phases are first-order topological phases and since the (anti)symmetry ${\cal S}$ is a nonlocal symmetry at a generic boundary for $d_{\parallel} \ge 1$, for $d_{\parallel} \ge 1$ we have the isomorphism
\begin{equation}
  K_{{\rm TF},{\cal S}}(s,t|d) =
  \K[0](s,t|d,d_{\parallel})/\K[1](s,t|d,d_{\parallel}),
  \label{eq:KTFisomorphism}
\end{equation}
which identifies the quotient group
$\K[0](s,t|d,d_{\parallel})/\K[1](s,t|d,d_{\parallel})$ as a regular subgroup
of $K_{\rm TF}(s|d)$ for $d_{\parallel} \ge 1$. No such isomorphism exists if
$d_{\parallel} = 0$ because in that case ${\cal S}$ is a local symmetry at a
generic crystal boundary, allowing for a richer boundary classifying group than
the one obtained from the ten-fold way classification.

The ten-fold way classification and the 
Shiozaki-Sato classification of topological phases with a crystalline
order-two symmetry contains only ``strong'' topological crystalline invariants,
\textit{i.e.}, they address topological features that are unaffected by resizing of the
unit cell, thus allowing the addition of perturbations that break the
translation symmetry of the original (smaller) unit cell, while preserving the
crystalline symmetries. Throughout this work we only consider HOTPs originating
form such ``strong'' topology.

\section{Crystalline-symmetry-breaking mass terms}\label{sec:cform}

In this Section we consider model Hamiltonians of a simple, ``canonical'' form,
which are still sufficiently general that the model description can be applied
to all ten-fold way and Shiozaki-Sato classes. We count how many independent
``mass terms'' can be added to the Hamiltonian that satisfy the fundamental
non-spatial (anti)symmetries ${\cal T}$, ${\cal P}$, and ${\cal C}$ defining
the ten-fold way class, but break the crystalline (anti)symmetry ${\cal S}$
that determines the Shiozaki-Sato class and show that such mass terms can be
used to construct fully ${\cal S}$-(anti)symmetric models in which a ``boundary
mass term'' appears on boundaries that are not invariant under the crystalline
(anti)symmetry ${\cal S}$. This naturally explains the phenomenology of
higher-order topological phases in these models. This Section serves as the
summary of the approach of Refs.~\onlinecite{langbehn2017,geier2018} and as
an interlude to the subsequent, more formal Section.

Explicitly, the model Hamiltonians we consider have the form
\begin{align}
  H_0(\bm k,m)&= \sum_{j=0}^d d_j(\bm k) \Gamma_j,
  \label{eq:canonical}
\end{align}
with matrices $\Gamma_j$ that anticommute mutually and square to the identity.
For the functions $d_j$ we choose
\begin{align}
  d_0(\vk,m) =&\, \mm + \sum_{i=1}^{d} (1 - \cos k_{i}),\nonumber \\
  d_j(\vk) =&\, \sin k_j \ \ \mbox{for $j=1,\ldots,d$},
  \label{eq:d(k)}
\end{align}
although our considerations do not change if a different choice for the
functions $d_j$ is made, as long as the map $\bm d/|\bm d|:T^{d}\rightarrow
S^{d}$ has winding number equal to one for $-2<m<0$ and to zero for $0<m<2$, and
the vector $\bm d = (d_0,d_1,\ldots,d_{d})$ transforms the same as $(1,\bm k)$
under the crystalline (anti)symmetry ${\cal S}$ and the non-spatial
(anti)symmetries ${\cal T}$, ${\cal P}$, and ${\cal C}$. The non-spatial
(anti)symmetries ${\cal T}$, ${\cal P}$, and ${\cal C}$ and the crystalline
(anti)symmetry ${\cal S}$ impose restrictions on the possible choices for the
matrices $\Gamma_j$, $j=0,1,\ldots,d$, which we do not specify explicitly here.

We consider the regime $-2 < \mm < 0$, for which the Hamiltonian
(\ref{eq:canonical}) has a band inversion near $\vk = 0$ but not elsewhere in
the Brillouin zone. In this parameter range, a Hamiltonian of the form
(\ref{eq:canonical}) describes a nontrivial topological crystalline phase if
there exists no ``mass term'' $\rM$ --- a hermitian matrix $\rM$ squaring
to the identity and anticommuting with the Hamiltonian ---, that satisfies the
constraints imposed by ${\cal S}$ and by ${\cal T}$, ${\cal P}$, and/or ${\cal
C}$. The topological phase is a ``tenfold-way phase'' --- 
{\em i.e.}, it remains nontrivial if
the crystalline (anti)symmetry ${\cal S}$ is broken --- if there exists no mass
term $\rM$ which
satisfies the constraints imposed by the non-spatial (anti)symmetries ${\cal
T}$, ${\cal P}$, and/or ${\cal C}$ alone, {\em irrespective} of the crystalline
(anti)symmetry ${\cal S}$. On the other hand, if such an ${\cal S}$-breaking
mass terms exist, the Hamiltonian (\ref{eq:canonical}) describes a ``purely
crystalline'' topological phase, which relies on the crystalline
(anti)symmetry ${\cal S}$ for its protection. Whereas a nontrivial tenfold-way phase
is always a first-order phase, the purely crystalline phases can be higher-order
topological phases.

In principle, a Hamiltonian of the form (\ref{eq:canonical}) may allow for more than one ${\cal S}$-breaking mass term --- where we require that different ${\cal S}$-breaking mass terms $\rM_l$ not only anticommute with $H$, but also with each other. If a canonical-form Hamiltonian has the minimum possible dimension for a given topological class, the ${\cal S}$-breaking mass terms $\rM_l$ all change sign under the (anti)symmetry ${\cal S}$. In this case, as we argue below, there is a connection between the number of mutually anticommuting ${\cal S}$-breaking mass terms and the order $n$ of the topological phase: The presence of $n-1$ ${\cal S}$-breaking mass terms $\rM_l$, $l=1,\ldots,n-1$ gives rise to a topological phase of order $\min(n,d_\parallel+1)$ if $\min(n,d_\parallel+1)\le d$, and to a boundary without protected in-gap states if $d_\parallel=d$ and $n>d$.\footnote{In a non-minimal model (which can be obtained, {\em e.g.}, by adding trivial bands to a minimal model of a topological phase), the number of mutually anticommuting ${\cal S}$-breaking mass terms $M_l$ may be less than $n-1$ or the $M_l$ do not all change sign under ${\cal S}$. However, in that case a Hamiltonian with $n-1$ mutually anticommuting ${\cal S}$-breaking mass terms may be recovered by removing or adding trivial bands.}
%The number of crystalline-symmetry breaking mass terms For a minimal canonical-form Hamiltonian in a nontrivial topological crystalline phase ({\em i.e.}, for which there exist no ${\cal S}$-preserving mass terms), we denote the number of mutually anticommuting ${\cal S}$-breaking mass terms $\rM_l$ with $n-1$. The number of crystalline-symmetry-breaking mass terms $n-1$ is uniquely defined only for minimal canonical Hamiltonians. For an arbitrary Hamiltonian $H$, $n-1$ is defined as the maximum number, within the topological equivalence class, of mutually anticommuting $\mathcal{S}$-breaking mass terms that anticommute with the Hamiltonian $H$.

To establish this connection one constructs the low-energy boundary theory for a Hamiltonian of the form (\ref{eq:canonical}).\cite{khalaf2018,geier2018} This requires considering a Hamiltonian with a slowly position-dependent parameter $\mm(\bf r)$, such that the topological phase occupies the region for which $\mm < 0$, whereas the region $\mm > 0$ hosts a trivial gapped phase. Whereas the Hamiltonian (\ref{eq:canonical}) becomes gapless at the boundary at $\mm=0$, with the help of ${\cal S}$-breaking mass terms $\rM_l$ one may construct a perturbation $H_1$ that respects the (anti)symmetry ${\cal S}$ and that gaps out the boundary, up to a region of codimension $\max(n,d_{\parallel}+1)$,
%for an order-two crystalline (anti)symmetry $\mathcal{S}$ with $d_{\parallel}>0$ inverted dimensions, we can add a perturbation $H_1(\vk)$ to the canonical Hamiltonian $H_0(\vk)$ that respects all the (anti)symmetries
\begin{align}
  H_1&= i \sum_{l=1}^{n-1}\sum_{j=1}^{d_\parallel} b_j^{(l)}\rM_l\Gamma_0\Gamma_j,
  \label{eq:H1}
\end{align}
where, for technical convenience, we take the coefficients $b_j^{(l)}$ numerically small. The relation between the number of ${\cal S}$-breaking mass terms and the order of the topological phase then follows immediately.

We now verify this statement explicitly for $d=2$. 
% that the perturbation~(\ref{eq:H1}) gaps the boundaries not invariant under the (anti)symmetry ${\cal S}$ and that the boundary Hamiltonian has $n-1$ mass
%terms that can be derived from the crystalline-symmetry-breaking mass terms
%$\rM_l$. 
The construction is easily generalized to higher dimensions.
% and explains the relation between the number $n-1$ of the crystalline-symmetry-breaking mass terms and the order $n$ of the topological phase.
Starting from the low-energy limit of the Hamiltonian $H_0$ of Eq.\
(\ref{eq:canonical}) with $d=2$ in the vicinity of a boundary with normal $\bm
n=(n_1,n_2)=(\cos\phi,\sin\phi)$, we find that the projection operator onto low-energy
boundary states is \cite{geier2018}
\begin{align}
  P(\phi) &= \frac{1}{2}(i \Gamma_1 \Gamma_0 \cos \phi + i \Gamma_2 \Gamma_0 \sin \phi + 1)
  \nonumber \\
  &= e^{\phi \Gamma_2 \Gamma_1/2} P(0) e^{-\phi \Gamma_2 \Gamma_1/2}.
\end{align}
Projecting the bulk Hamiltonian $H_0 + H_1$ to the low-energy boundary states gives
\begin{align}
  P(\vn) H P(\vn) =&\,
  e^{\phi \Gamma_2 \Gamma_1/2} P(0) \nonumber \\ &\, \mbox{} \times
  [-i \hbar \Gamma_2 \partial_{x_{\rm b}} + \sum_{l=1}^{n-1} m_l(\phi) \rM_l] \nonumber \\ &\, \mbox{} \times
  P(0) e^{-\phi \Gamma_2 \Gamma_1/2},
  \label{eq:HP}
\end{align}
where $m_l(\phi) = \sum_{j=1}^{d_\parallel} b_j^{(l)} n_j$ and
$\partial_{x_{\rm b}} = \cos \phi\, \partial_{x_2} - \sin \phi\,
\partial_{x_1}$ is the derivative with respect to a coordinate along the edge.
We conclude that the effective boundary Hamiltonian reads
\begin{equation}
  H_{\rm boundary} = -i \hbar \Gamma_2' \partial_{x_{\rm b}} + \sum_{l=1}^{n-1} m_l(\phi) \rM_l^\prime,
  \label{eq:HboundaryGamma}
\end{equation}
where $\Gamma_2' = P(0) \Gamma_2 P(0)$ and $\rM_l' = P(0) \rM_l P(0)$.
Alternatively, one may arrive at the effective boundary
Hamiltonian~(\ref{eq:HboundaryGamma}) by starting from the canonical-form
Hamiltonian~(\ref{eq:canonical}) and adding the perturbation $\rM_l$ locally at
the boundary, provided the boundary is not itself invariant under $\mathcal{S}$
and the prefactor $m_l(\phi)$ obeys the restrictions imposed by $\mathcal{S}$
(as it does in Eq.~(\ref{eq:HboundaryGamma})).

The boundary Hamiltonian (\ref{eq:HboundaryGamma}) hosts zero-energy corner
states between crystal edges with opposite sign of $m_l(\phi)$, provided all
mass terms $m_l(\phi)$ go through zero at the same value of $\phi$. For an
on-site order-two symmetry $\mathcal{O}$ with $d_\parallel=0$, the mass terms
$\rM_l$ cannot be used to construct an $\mathcal{O}$-preserving perturbation,
which is consistent with the absence of $\mathcal{O}$-symmetry breaking
boundaries.
%Accordingly a local order-two symmetry does not allow for topological phases of order two or higher, consistent with the relation~(\ref{eq:bb1}). 
Mirror symmetry has
$d_\parallel=1$ flipped coordinates, which gives $m_l(\phi)=b_1^{(l)}\cos\phi$:
all mass terms $m_l(\phi)$ vanish simultaneously on the mirror line and 
one obtains a second-order phase whenever there is at least one mass term, {\em i.e.}, if $n\ge2$.
%We conclude that for mirror symmetry the order of the bulk phase must be $\le 2$, again consistent with the relation~(\ref{eq:bb1}). 
Finally, a
twofold rotation symmetry has $d_\parallel=2$, and zero-energy corner states
are obtained only if the number $n-1$ of crystalline-symmetry-breaking terms is
exactly one. 
%Therefore $n$ corresponds to the order of the phase. 
For $n>2$, the
coefficients $b_j^{(l)}$ can be chosen to yield a fully gapped boundary, which
describes the situation where the bulk is topologically nontrivial but the
boundary does not host any anomalous states --- in this case the group $\K[d]$ is
nontrivial. 
%Generalizing these arguments to higher dimensions, one verifies for
%Hamiltonians of the canonical form~(\ref{eq:canonical}), that the presence of
%$n-1$ crystalline-symmetry-breaking mass terms gives rise to a topological
%phase of order $\min(n,d_\parallel+1)$ if $\min(n,d_\parallel+1)\le d$, and to
%a boundary without protected in-gap states if $d_\parallel=d$ and $n>d$.

\section{Bulk and boundary classification of topological crystalline insulators} \label{sec:omap}

We now turn to a general topological classification of the electronic structure
of a $d$-dimensional crystal with an order-two crystalline symmetry or
antisymmetry ${\cal S}$ with $d_{\parallel}$ inverted dimensions. We assume
that the crystal shape, including the lattice termination, is compatible with
the crystalline symmetry. We recall that the system is in an $n$th order
topological phase if it has protected boundary states of codimension $n$,
whereas the bulk and all boundaries of codimension smaller than $n$ are gapped.
In this Section we establish the formal framework for a classification of such
$n$th-order topological phases, both from a bulk perspective and from a
boundary perspective, and show the extent to which they are related.

As announced in the Introduction of this article, the bulk-perspective
classification amounts to the construction of the subgroup series
(\ref{eq:subgroup}) of classifying groups $\K[n]$, where $\K[n]$ classifies the
topology of bulk band structures {\em excluding} topological phases of order
$\le n$. Since taking the direct sum of two topological phases can not reduce
the codimension of anomalous boundary states, the $\K[n]$ defined this way have
a well-defined group structure in the Grothendieck construction. Figure
\ref{fig:subgroups} illustrates the definitions of the subgroup sequence for
the case of a three-dimensional crystal with twofold rotation symmetry.

The definition of the boundary-perspective classification groups requires a
little more care, because for boundary states of order $n > 1$ their location,
number, and type may depend on the crystal shape and crystal termination. A
classifying group that is independent of such details is obtained by
considering {\em equivalence classes} of configurations of codimension-$n$
boundary states that differ by a change in lattice termination only. This is
the classifying group $\ki{n}$ of anomalous boundary states that appears in the
bulk-boundary correspondence (\ref{eq:bb}). In this Section we pursue a further
resolution of the boundary classification, by defining boundary classification
groups $\ki[k]{n}$ of equivalence classes of codimension-$n$ boundary states
that differ by the lattice termination along boundaries of codimension $\ge k$
only, $k=1,\ldots,n-1$. With that definition, the classifying group of anomalous
boundary states
\begin{equation}
  \ki{n} = \ki[1]{n}.
\end{equation}
Specifically, for a three-dimensional crystal, $\ki{2}=\ki[1]{2}$ classifies
configurations of protected gapless modes along hinges, where configurations
that differ by termination only are identified. Similarly, $\ki{3}=\ki[1]{3}$
classifies configurations of protected zero-energy states at crystal corners,
again identifying configurations of corner states that differ by a change of
lattice termination. The group $\ki[2]{3}$ classifies configurations of
protected zero-energy corner states, identifying configurations that differ by
changing the termination along crystal hinges, without affecting the crystal
faces.

\begin{figure}
	\includegraphics[width=0.8\columnwidth]{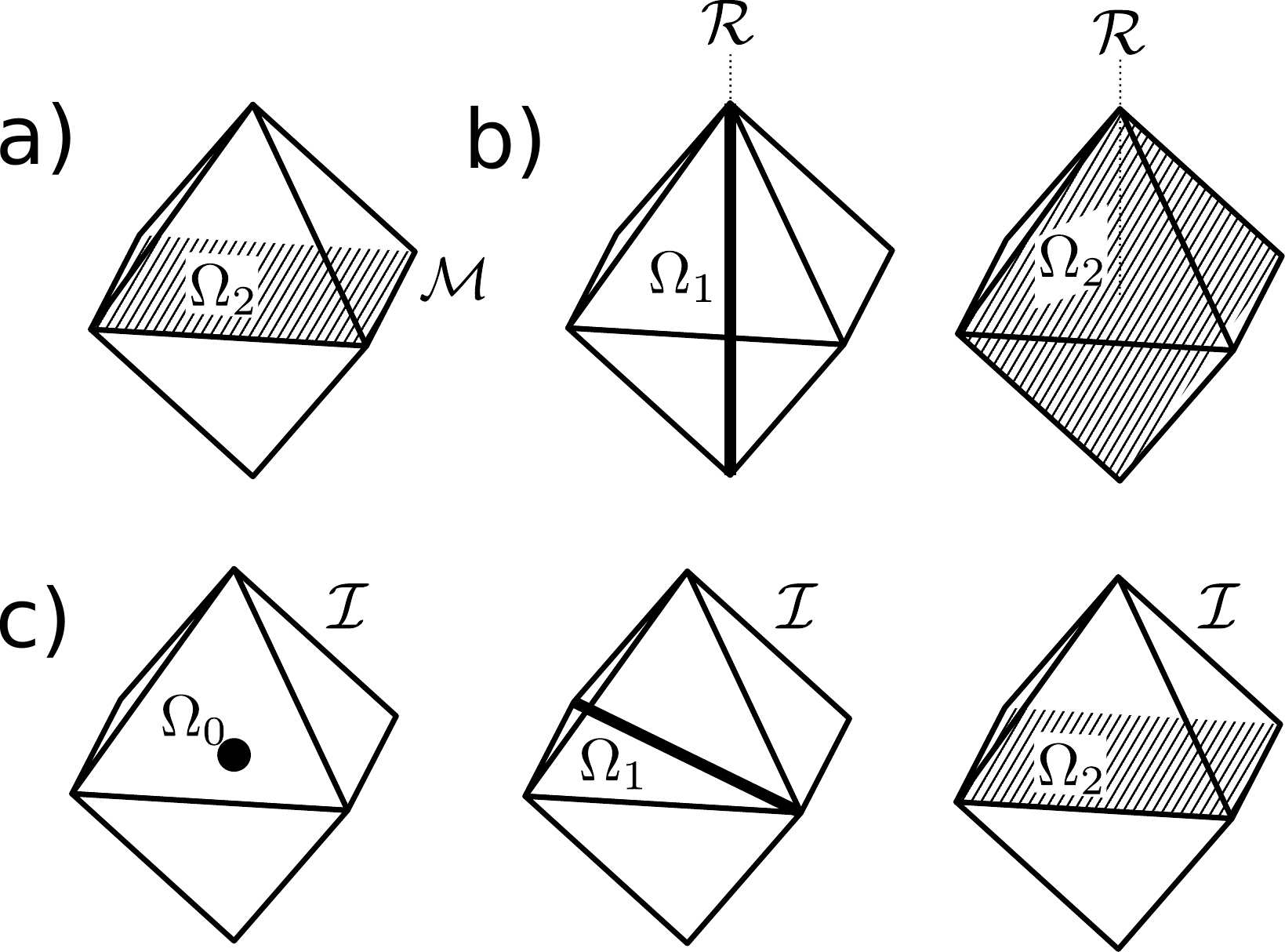}
	\caption{\label{fig:proper} Manifolds $\Omega_k$ with $d-d_{\parallel} \le k < d$ for three-dimensional
	crystals with mirror (a), two-fold rotation (b), and inversion (c)
	symmetry. The manifold $\Omega_d$ is equal to the entire
	crystal, see Eq.~(\ref{eq:invariantOmega}), and is not shown
	in the figure.}
\end{figure}

\begin{figure}
	\includegraphics[width=0.8\columnwidth]{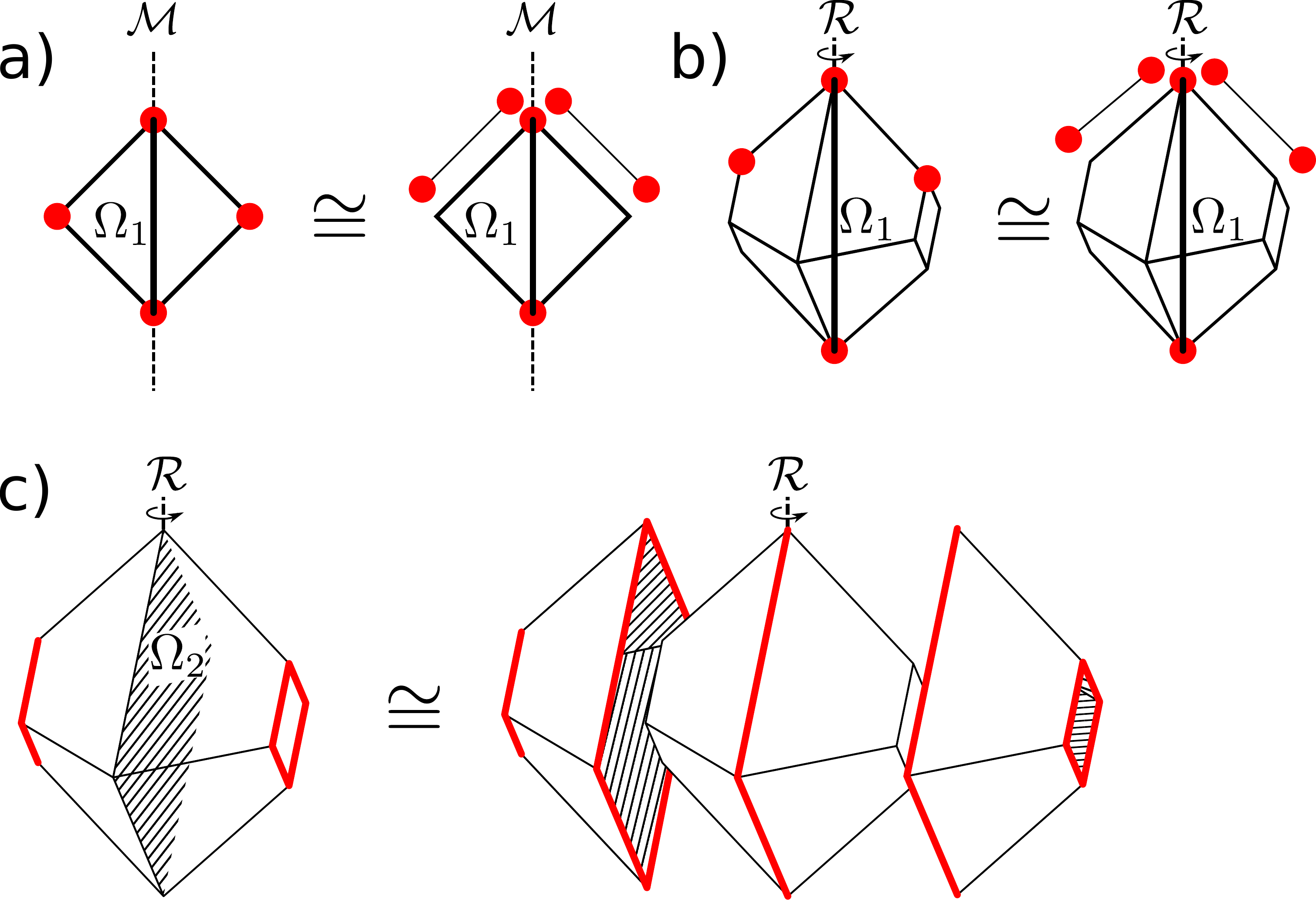}
	\caption{\label{fig:2proper} By attaching a ``decoration'' consisting of a first-order topological phase on a boundary of codimension $n-1$, an arbitrary configuration of boundary state of codimension $n$ can be moved to the subset $\partial\Omega_{d+1-n}$. The figure shows three examples: Corner states of a two-dimensional crystal with mirror symmetry ($n=2$), which can always be moved to the intersection of the mirror line and the crystal boundary upon changing the crystal termination (a), corner states of a three-dimensional crystal with twofold rotation symmetry ($n=3$), which can always be moved to the intersection points of the twofold rotation axis and the crystal boundary upon changing the termination along crystal hinges (b), and hinge states of a three-dimensional crystal with twofold rotation symmetry, which can always be moved to the intersection of the two-dimensional manifold $\Omega_2$ and the crystal boundary upon changing the termination at crystal faces (c).}
\end{figure}

{\em Location of boundary states for $n > d_{\parallel}$.---}
A crystalline symmetry ${\cal S}$ with $d_{\parallel}$ inverted dimensions
necessarily leaves a manifold $\Omega_{d-d_{\parallel}}$ invariant.  For
$d_{\parallel} = 1$ this is the mirror plane; for $d_{\parallel}=2$ it is the
twofold rotation axis, see Fig.\ \ref{fig:proper}. For boundary states of
codimension $n > d_{\parallel}$ it is always possible to change the crystal
termination along boundaries of codimension $n-1$ only, such that all boundary
states end up on the intersection $\partial \Omega_{d-d_{\parallel}}$ of the invariant manifold $\Omega_{d-d_{\parallel}}$ and the
crystal boundary. Examples of such a procedure are shown schematically in
Fig.\ \ref{fig:2proper} for a two-dimensional crystal with mirror symmetry and
for a three-dimensional crystal with twofold rotation symmetry. We conclude,
that for $n > d_{\parallel}$ it is sufficient to consider configurations of
codimension-$n$ boundary states with support on $\partial
\Omega_{d-d_{\parallel}}$ only. 

{\em Classifying groups for $n > d_{\parallel} + 1$.} We now combining this
conclusion with the observation that the crystalline symmetry ${\cal S}$ is a
{\em local} ({\em i.e.}, on-site) symmetry inside the invariant manifold $\Omega_{d-d_{\parallel}}$.
The ``conventional'' ten-fold way bulk-boundary correspondence, according to
which any anomalous states are the {\em first-order} boundary phase of a
topological phase, remains valid in the presence of a local crystalline
symmetry. Applying this bulk-boundary correspondence to protected gapless
boundary states of codimension $n > d_{\parallel}+1$ within $\partial
\Omega_{d-d_{\parallel}}$, such states can be interpreted as the first-order
boundary states of a codimension-$(n-1)$ topological phase, still located {\em
within} the invariant part $\partial \Omega_{d-d_{\parallel}}$ of the crystal
boundary. Obviously, such boundary states can be removed by changing crystal
termination along boundaries of codimension $n-1$. It follows that the boundary
classification groups $\ki[k]{n}$ are all trivial,
\begin{equation}
  \ki[k]{n} = 0,\ \ \mbox{for $n > d_{\parallel} + 1$ and $k=1,\ldots,n-1$}.
  \label{eq:Ki0}
\end{equation}
A similar argument can be made for the bulk classification groups $\K[n]$ for $n > d_{\parallel}+1$. Again, because ${\cal S}$ is a local symmetry on $\partial \Omega_{d-d_{\parallel}}$, a nontrivial bulk topology implies the presence of protected gapless boundary states of codimension $d_{\parallel}+1$ or less (see, {\em e.g.}, Sec.~\ref{sec:cform} and Refs.~\onlinecite{geier2018,khalaf2018b}). Equation (\ref{eq:bb1}) follows immediately from this
observation, which, combined with the relation~(\ref{eq:Ki0}), yields the
bulk-boundary correspondence~(\ref{eq:bb}) advertised in the introduction for
$n>d_\parallel$.

{\em Boundary classification for $n = d_{\parallel} + 1$.---} The
calculation the groups $\ki[k]{n}$ for $n=d_{\parallel}+1$ 
proceeds via a series of auxiliary groups
$\ks[k]{n}$. The first of these, $\ks[0]{n}$, is defined as the classifying
group of codimension-$n$ boundary states with support entirely within $\partial
\Omega_{d-d_{\parallel}}$. We refer to this group
%, which may in principle depend on the crystal shape, 
as the ``extrinsic boundary
classification group''. To relate $\ks[0]{n}$ to the known classification groups of first-order topological phases, we argue that (i) the $(d-d_{\parallel}-1)$-dimensional boundary states on $\partial \Omega_{d-d_{\parallel}}$ may be interpreted as first-order boundary states of $\Omega_{d-d_{\parallel}}$ and (ii) ${\cal S}$ is a local symmetry on $\Omega_{d-d_{\parallel}}$. For (i) it is essential that the crystal boundary is fully gapped away from $\partial \Omega_{d-d_{\parallel}}$, so that the crystal away from
$\Omega_{d-d_{\parallel}}$ may be considered effectively topologically trivial and one may consider the manifold $\Omega_{d-d_{\parallel}}$ in isolation. This immediately gives the identification
\begin{align}
  \ks[0]{n} = K(d-d_{\parallel},0)
  \label{eq:Kb1a}
\end{align}
where $K(d-d_{\parallel},0)$ is the Shiozaki-Sato classifying group for a $(d-d_{\parallel})$-dimensional crystal
with an on-site crystalline symmetry. Before we proceed with the definition of the
remaining groups $\ks[k]{n}$ and the construction of the boundary
classification groups $\ki[k]{n}$, we first discuss how the above construction
is generalized to boundary states of codimension $n \le d_{\parallel}$.

{\em Boundary classification for $n \le d_{\parallel}$.---}
To calculate the boundary classifying groups $\ki[k]{n}$ for $n \le
d_{\parallel}$, we define a sequence of manifolds
\begin{align}
  \Omega_{d-d_\parallel}\subset\Omega_{d-d_\parallel+1}\subset\dots\subset\Omega_d,
  \label{eq:invariantOmega}
\end{align}
where $\Omega_d$ equals the entire crystal, whereas
the $k$-dimensional manifolds $\Omega_k$ for $d-d_\parallel\le k<d$ are mapped
into themselves under the crystalline symmetry $\cal S$. We additionally
require that 
%$\Omega_k$ cannot be continuously deformed to a manifold
%$\Omega_k^\prime\subset\partial\Omega_d$ while preserving the symmetry $\cal
%S$. Assuming that 
the intersection $\partial \Omega_{d+1-n}$ with the crystal
boundary is along crystal boundaries of codimension $n$. Examples of such a sequence of manifolds $\Omega_k$ are shown in Fig.\ \ref{fig:proper}b and c.
With this construction, one easily
verifies that by changing crystal termination along boundaries of codimension
$n-1$ only, any configuration of codimension-$n$ boundary states can be made to
lie entirely within $\partial \Omega_{d+1-n}$.

Generalizing the above discussion for the case $n = d_{\parallel} + 1$, we define the extrinsic classifying group $\ks[0]{n}$ as the classifying group of codimension-$n$ boundary states with support entirely within $\partial \Omega_{d+1-n}$. 
%We note that $\ks[0]{n}$ not only depends on the crystal shape, but also on  (in general non-unique) choice of the manifold $\Omega_{d+1-n}$. 
To relate $\ks[0]{n}$ to the known classification groups of first-order topological phases, we again interpret boundary states on $\partial \Omega_{d+1-n}$ as first-order boundary states of $\Omega_{d+1-n}$. A difference with the case $n=d_{\parallel}+1$ is that now the order-two crystalline (anti)symmetry ${\cal S}$ is a non-local symmetry with $d_{\parallel} + 1 - n$ inverted dimensions. We thus find
\begin{align}
  \ks[0]{n} = K_{{\rm TF},\cal S}(d+1-n,d_{\parallel}+1 - n),
  \label{eq:Kb1}
\end{align}
where $K_{{\rm TF},\cal S}\subseteq K_{\rm TF}$ classifies the ten-fold way phases compatible with the non-local crystalline symmetry ${\cal S}$, see Sec.~\ref{sec:ShiozakiSato}. 

\begin{figure}
	\includegraphics[width=0.8\columnwidth]{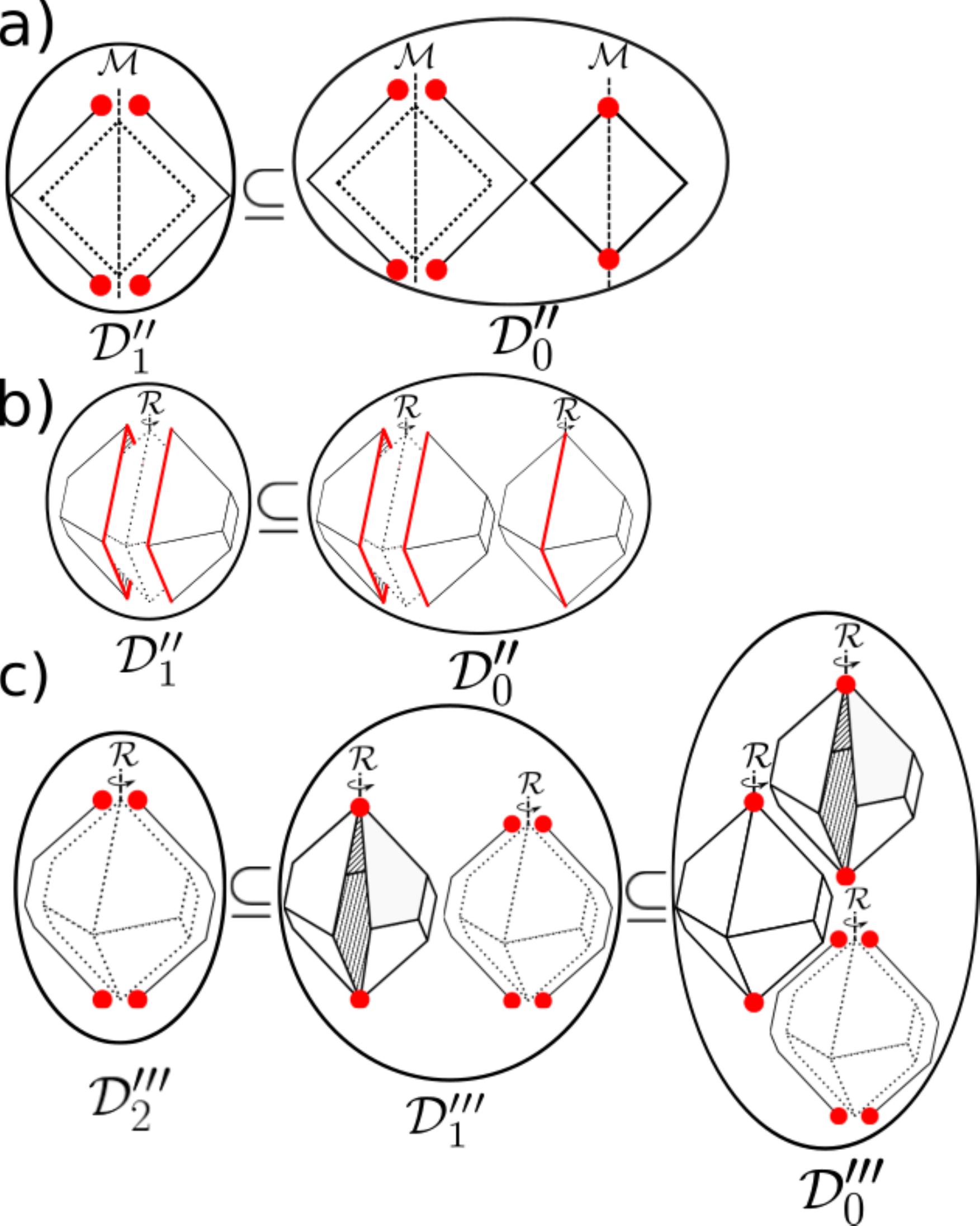}
	\caption{\label{fig:Dk} Examples of the subgroup sequence
	(\ref{eq:SubgroupD}) for the decoration subgroups $\ks[k]{n}$ and the
	extrinsic boundary classifying group $\ks[0]{n}$. (a) For a
	two-dimensional crystal with mirror symmetry, the group $\ks[0]{2}$
	classifies all possible configurations of corner states on the mirror
	axis, whereas the decoration subgroup $\ks[1]{2}$ classifies corner
	states on the mirror axis that can be obtained by ``glueing'' two
	mirror-related first-order topological phases to the crystal boundary.
	(b) For a three-dimensional crystal with twofold rotation symmetry,
	$\ks[0]{2}$ classifies all configurations of protected hinge modes
	along the intersection $\partial \Omega_2$ of the reference manifold
	$\Omega_2$ and the crystal boundary, whereas the decoration subgroup
	$\ks[1]{2}$ classifies all protected hinge modes along $\partial
	\Omega_2$ that are obtained by ``glueing'' two rotation-related
	first-order topological phases to the crystal boundary.  (c) For a
	three-dimensional crystal with twofold rotation symmetry, $\ks[0]{3}$
	classifies all configurations of protected zero-energy states at the
	two corners on the rotation axis. The subgroups $\ks[2]{3}$ and
	$\ks[1]{3}$ contain corner states on the rotation axis that are
	obtained from decorating rotation-related hinges or surfaces with
	first-order and second-order topological phases, respectively. }
\end{figure}

The boundary classification groups $\ki[k]{n}$ describe equivalence classes of
configurations of codimension-$n$ boundary states with respect to changes of
the lattice termination for boundaries of codimension $\ge k$. In other
words, when seen as an element of $\ki[k]{n}$, a codimension-$n$ state on
$\partial \Omega_{d+1-n}$ is considered trivial if it can be obtained as the
boundary state of a $(d-k)$-dimensional $(n-k)$th order topological phase
entirely contained within the crystal boundary and respecting the global
(anti)symmetry ${\cal S}$. Denoting the classifying group for such ``trivial''
boundary states as $\ks[k]{n}$, the boundary classification group $\ki[k]{n}$
can thus be obtained as the quotient
\begin{equation}
   \ki[k]{n} = \ks[0]{n}/\ks[k]{n},\ \ k=2,\ldots,n,
  \label{eq:anomalous}
\end{equation} 
where $\ks[0]{n}$ is the extrinsic boundary classifying group introduced above.
The groups $\ks[k]{n}$, which we call ``decoration groups'', form a subgroup
sequence that also includes the extrinsic boundary classification group,
\begin{equation}
  \ks[n-1]{n} \subseteq \ldots \subseteq \ks[1]{n} \subseteq \ks[0]{n}.
  \label{eq:SubgroupD}
\end{equation}
Figure \ref{fig:Dk} shows examples of this subgroup sequence for a
two-dimensional crystal with mirror symmetry and a three-dimensional crystal
with twofold rotation symmetry. Since the crystalline symmetry ${\cal S}$ acts
nonlocally for a generic position in a $(d-k)$-dimensional boundary state,
the Hamiltonian of such a decoration state is ``separable'', {\em i.e.},
it may be written as 
$$
  \begin{pmatrix} h_{d-k}(\vk) & 0 \\ 0 & \tilde {\cal S} h_{d-k}(\vk)
  \end{pmatrix}
$$%
where $h_{d-k}$ is a $(d-k)$-dimensional Hamiltonian without crystalline
symmetries and
\begin{equation}
  {\cal S} = \begin{pmatrix} 0 & 1 \\ 1 & 0 \end{pmatrix} \tilde {\cal S},
\end{equation}
where $\tilde {\cal S}$ is an (anti)symmetry operator with $d_{\parallel}-k$
inverted dimensions acting on $(d-k)$-dimensional Hamiltonians. (Note that the
boundary of a decoration need not be a separable in this sense. This is
illustrated schematically in Fig.~\ref{fig:Dk}b. Further
examples are given in Sec.~\ref{sec:examples}.) 

\begin{table}[t]
\begin{tabular*}{\columnwidth}{l @{\extracolsep{\fill}} ccccccc}
\hline\hline 
TF class            & $s$ & $t$ & ${\cal O}$  & ${\cal M}$  & ${\cal R}$      & ${\cal I}$ \\ \hline
A$^{\cal S}$        & $0$ & $0$ & $0,0,0$     & $0,0,0$     & $0,0,0$         & $0,0,0$         \\
AIII$^{{\cal S}_+}$ & $1$ & $0$ & $\ZZ^2,0,0$ & $\ZZ^2,0,0$ & $\ZZ^2,\ZZ,\ZZ$ & $0,0,0$ \\ \hline
A$^{\cal CS}$       & $0$ & $1$ & $\ZZ,0,0$   & $\ZZ,0,0$  & $\ZZ,\ZZ,\ZZ$   & $0,0,0$   \\
AIII$^{{\cal S}_-}$ & $1$ & $1$ & $0,0,0$     & $0,0,0$     & $0,0,0$         & $\ZZ,\ZZ_2,\ZZ_2$         \\
\hline \hline
\end{tabular*}
\caption{Boundary classification of third-order phases in three-dimensional
	systems with an order-two symmetry (antisymmetry) for complex
	ten-fold way classes. The symbols ${\cal O}$, ${\cal M}$, ${\cal
	R}$ and ${\cal I}$ refer to a local on-site ($d_{\parallel} =
	0$), mirror ($d_{\parallel} = 1$), twofold rotation
	($d_{\parallel} = 2$), and inversion symmetry ($d_{\parallel} = 3$),
	respectively. The boundary classification groups are given in the
	order $\ks[0]{3}$, $\ki[2]{3}$, $\ki[1]{3}=\ki{3}$.\label{tab:boundaryC}}
\end{table}
\begin{table}[t]
\begin{tabular*}{\columnwidth}{l @{\extracolsep{\fill}} ccccccc}
\hline\hline 
Shiozaki-Sato class                      & $s$  & ${\cal O}$  & ${\cal M}$  & ${\cal R}$        & ${\cal I}$ \\ \hline
A$^{{\cal T}^+{\cal S}}$      & $0$  & $0,0,0$     & $0,0,0$     & $0,0,0$           & $0,0,0$ \\
AIII$^{{\cal P}^+{\cal S}_+}$ & $1$  & $\ZZ,0,0$   & $\ZZ,0,0$   & $\ZZ,\ZZ_2,\ZZ_2$ & $0,0,0$ \\ 
A$^{{\cal P}^+{\cal S}}$      & $2$  & $\ZZ_2,0,0$ & $\ZZ_2,0,0$ & $\ZZ_2,\ZZ_2,0$   & $0,0,0$ \\
AIII$^{{\cal T}^-{\cal S}_-}$ & $3$  & $\ZZ_2,0,0$ & $\ZZ_2,0,0$ & $\ZZ_2,0,0$       & $2\ZZ,0,0$ \\ 
A$^{{\cal T}^-{\cal S}}$      & $4$  & $0,0,0$     & $0,0,0$     & $0,0,0$           & $0,0,0$ \\
AIII$^{{\cal P}^-{\cal S}_+}$ & $5$  & $2 \ZZ,0,0$ & $2 \ZZ,0,0$ & $2 \ZZ,0,0$       & $0,0,0$ \\ 
A$^{{\cal P}^-{\cal S}}$      & $6$  & $0,0,0$     & $0,0,0$     & $0,0,0$           & $0,0,0$ \\
AIII$^{{\cal T}^+{\cal S}_-}$ & $7$  & $0,0,0$     & $0,0,0$     & $0,0,0$           & $\ZZ,\ZZ_2,\ZZ_2$           \\
\hline \hline
\end{tabular*}
\caption{Same as table~\ref{tab:boundaryC}, but for antiunitary symmetries and antisymmetries.\label{tab:boundaryA}}
\end{table}

\begin{table}[t]
\begin{tabular*}{\columnwidth}{l @{\extracolsep{\fill}} ccccccc}
\hline\hline 
class                         & $s$ & $t$ & ${\cal O}$    & ${\cal M}$    & ${\cal R}$             & ${\cal I}$ \\ \hline
    AI$^{\mathcal{S}_+}$      & $0$ & $0$ & $0,0,0$       & $0,0,0$       & $0,0,0$                & $0,0,0$     \\
    BDI$^{\mathcal{S}_{++}}$  & $1$ & $0$ & $\ZZ^2,0,0$   & $\ZZ^2,0,0$   & $\ZZ^2,\ZZ,\ZZ$        & $0,0,0$     \\
    D$^{\mathcal{S}_{+}}$     & $2$ & $0$ & $\ZZ_2^2,0,0$ & $\ZZ_2^2,0,0$ & $\ZZ_2^2,\ZZ_2,0$      & $0,0,0$     \\
    DIII$^{\mathcal{S}_{++}}$ & $3$ & $0$ & $\ZZ_2^2,0,0$ & $\ZZ_2^2,0,0$ & $\ZZ_2^2, \ZZ_2,0$ 	   & $0,0,0$     \\
    AII$^{\mathcal{S}_{+}}$   & $4$ & $0$ & $0,0,0$       & $0,0,0$       & $0,0,0$                & $0,0,0$     \\
    CII$^{\mathcal{S}_{++}}$  & $5$ & $0$ & $2 \ZZ^2,0$   & $2 \ZZ^2,0$   & $2 \ZZ^2, 2\ZZ,2\ZZ$   & $0,0,0$     \\
    C$^{\mathcal{S}_{+}}$     & $6$ & $0$ & $0,0,0$       & $0,0,0$       & $0,0,0$                & $0,0,0$     \\
    CI$^{\mathcal{S}_{++}}$   & $7$ & $0$ & $0,0,0$       & $0,0,0$       & $0,0,0$                & $0,0,0$     \\
    \hline
    AI$^{\mathcal{CS}_-}$     & $0$ & $1$ & $0,0,0$       & $0,0,0$       & $0,0,0$                & $0,0,0$     \\
    BDI$^{\mathcal{S}_{+-}}$  & $1$ & $1$ & $0,0,0$       & $0,0,0$       & $0,0,0$                & $\ZZ,\ZZ_2,\ZZ_2$ \\
    D$^{\mathcal{CS}_{+}}$    & $2$ & $1$ & $\ZZ,0,0$     & $\ZZ,0,0$     & $\ZZ,\ZZ,\ZZ$          & $\ZZ_2,\ZZ_2,0$ \\
    DIII$^{\mathcal{S}_{-+}}$ & $3$ & $1$ & $\ZZ_2,0,0$   & $\ZZ_2,0,0$   & $\ZZ_2, \ZZ_2,0$       & $\ZZ_2,\ZZ_2,0$ \\
    AII$^{\mathcal{CS}_{-}}$  & $4$ & $1$ & $\ZZ_2,0,0$   & $\ZZ_2,0,0$   & $\ZZ_2,\ZZ_2,0$        & $0,0,0$     \\
    CII$^{\mathcal{S}_{+-}}$  & $5$ & $1$ & $0,0,0$       & $0,0,0$       & $0,0,0$                & $2\ZZ,\ZZ_2,\ZZ_2$ \\
    C$^{\mathcal{CS}_{+}}$    & $6$ & $1$ & $2\ZZ,0,0$    & $2\ZZ,0,0$    & $2\ZZ,2\ZZ,2\ZZ$       & $0,0,0$     \\
    CI$^{\mathcal{S}_{-+}}$   & $7$ & $1$ & $0,0,0$       & $0,0,0$       & $0,0,0$                & $0,0,0$     \\
    \hline
    AI$^{\mathcal{S}_-}$      & $0$ & $2$ & $0,0,0$       & $0,0,0$       & $0,0,0$                & $0,0,0$     \\
    BDI$^{\mathcal{S}_{--}}$  & $1$ & $2$ & $2 \ZZ, 0,0$  & $2 \ZZ, 0,0$  & $2 \ZZ, 0,0$           & $0,0,0$     \\
    D$^{\mathcal{S}_{-}}$     & $2$ & $2$ & $0,0,0$       & $0,0,0$       & $0,0,0$                & $\ZZ_2,\ZZ_2,\ZZ_2$ \\
    DIII$^{\mathcal{S}_{--}}$ & $3$ & $2$ & $2\ZZ,0,0$    & $2\ZZ,0,0$    & $2\ZZ, 2\ZZ,2\ZZ$      & $\ZZ_2,\ZZ_2,0$ \\
    AII$^{\mathcal{S}_{-}}$   & $4$ & $2$ & $0,0,0$       & $0,0,0$       & $0,0,0$                & $0,0,0$     \\
    CII$^{\mathcal{S}_{--}}$  & $5$ & $2$ & $2 \ZZ,0,0$   & $2 \ZZ,0,0$   & $2 \ZZ, \ZZ_2,\ZZ_2$   & $0,0,0$     \\
    C$^{\mathcal{S}_{-}}$     & $6$ & $2$ & $0,0,0$       & $0,0,0$       & $0,0,0$                & $0,0,0$     \\
    CI$^{\mathcal{S}_{--}}$   & $7$ & $2$ & $2 \ZZ,0,0$   & $2 \ZZ,0,0$   & $2\ZZ,2\ZZ,2\ZZ $       & $0,0,0$     \\
    \hline
    AI$^{\mathcal{CS}_+}$     & $0$ & $3$ & $\ZZ,0,0$     & $\ZZ,0,0$     & $\ZZ,\ZZ,\ZZ$          & $0,0,0$     \\
    BDI$^{\mathcal{S}_{-+}}$  & $1$ & $3$ & $\ZZ_2, 0,0$  & $\ZZ_2, 0,0$  & $\ZZ_2, 0,0$           & $2\ZZ,0,0$     \\
    D$^{\mathcal{CS}_{-}}$    & $2$ & $3$ & $\ZZ_2,0,0$   & $\ZZ_2,0,0$   & $\ZZ_2,0,0$            & $0,0,0$     \\
    DIII$^{\mathcal{S}_{+-}}$ & $3$ & $3$ & $0,0,0$       & $0,0,0$       & $0,0,0$                & $\ZZ_2,\ZZ_2,\ZZ_2$ \\
    AII$^{\mathcal{CS}_{+}}$  & $4$ & $3$ & $2\ZZ,0,0$    & $2\ZZ,0,0$    & $2\ZZ,2\ZZ,2\ZZ$       & $0,0,0$     \\
    CII$^{\mathcal{S}_{-+}}$  & $5$ & $3$ & $0,0,0$       & $0,0,0$       & $0,0,0$                & $2\ZZ,\ZZ_2,\ZZ_2$ \\
    C$^{\mathcal{CS}_{-}}$    & $6$ & $3$ & $0,0,0$       & $0,0,0$       & $0,0,0$                & $0,0,0$     \\
    CI$^{\mathcal{S}_{+-}}$   & $7$ & $3$ & $0,0,0$       & $0,0,0$       & $0,0,0$                & $0,0,0$     \\
\hline \hline
\end{tabular*}
\caption{Boundary classification of third-order phases in three-dimensional
	systems with an order-two symmetry (antisymmetry) for real
	ten-fold way classes. The symbols ${\cal O}$, ${\cal M}$, ${\cal
	R}$ and ${\cal I}$ refer to a local on-site ($d_{\parallel} =
	0$), mirror ($d_{\parallel} = 1$), twofold rotation
	($d_{\parallel} = 2$), and inversion symmetry ($d_{\parallel} = 3$),
	respectively. The boundary classification groups are given in the
	order $\ks[0]{3}$, $\ki[2]{3}$, $\ki[1]{3}=\ki{3}$.\label{tab:boundaryR}}
\end{table}

{\em Bulk-boundary correspondence for $n \le d_{\parallel}$.---} To establish a relation between the bulk classifying groups $\K[n]$ and the decoration subgroups $\ks[k]{n}$ for $n \le d_{\parallel}$ we make use of a homomorphism
\begin{equation}
  K(d,d_{\parallel}) \overset{\omap}{\rightarrow} K(d+1,d_{\parallel}+1),
\end{equation}
which maps an equivalence class of $d$-dimensional Hamiltonians $H$ in
Shiozaki-Sato class $(s,t,d_{\parallel})$ to a $(d+1)$-dimensional Hamiltonian
in Shiozaki-Sato class $(s,t,d_{\parallel}+1)$. 
%The repeated application of
%$\omap$ allows us to connect the boundary classification groups $\ki[k]{n+1}$
%to the Shiozaki-Sato group $K(d,d_{\parallel})$, from which the bulk classifying
%groups $\K[n]$ are derived. 
The precise definition of the homomorphism will be
given in Sec.~\ref{sec:examples}. For the derivation of the bulk-boundary
correspondence~(\ref{eq:bb}), it will be sufficient to use three key properties
of $\omap$:
\begin{itemize}
\item $\omap(H)$ is in the trivial class if and only if $H$ is separable if $H$ can be deformed to a separable Hamiltonian,
\item the homomorphism $\omap$ commutes with the dimension-raising isomorphisms
  $\kappa_\parallel$ and $\kappa_\perp$, up to a possible sign change of the
  topological invariants, and
\item   If $H$ is a non-separable Hamiltonian with $n-1$
	crystalline-symmetry-breaking mass terms, then
	$\omap(H)$ is a Hamiltonian with $n$ crystalline-symmetry-breaking
	mass terms. The inverse is also true: $H \in \img \omap$ if $H$ has at least one ${\cal S}$-breaking mass term.
\end{itemize}
The third property ensures that $\omap$ does not change the dimension of the
protected boundary states (if any). For that reason, we refer to $\omap$ as the
``order-raising homomorphism'', as it increases the order of the topological
phase by one. A proof of these three properties will be given in
App.~\ref{app:omap} for the homomorphism that we will introduce in Sec.\
\ref{sec:examples}. The stacking construction previously considered in the
literature~\cite{isobe2015,fulga2016,song2017b} is another realization of the
order-raising homomorphism --- this is explicitly demonstrated in
Sec.~\ref{sec:smap}.

The proof of the bulk-boundary correspondence (\ref{eq:bb}) makes use of the three properties of $\omap$, without requiring knowledge of the specific form of the homomorphism. Hereto, we first note that the last of these properties can be used to calculate the bulk classifying groups $\K[n]$ in the subgroup series (\ref{eq:subgroup}), since the number $n-1$ of crystalline-symmetry-breaking mass terms is related to the order $n$ of the
topological phase (provided $n\le d-1$), see Sec.~\ref{sec:cform} and
Refs.~\onlinecite{langbehn2017,geier2018}. We conclude that Hamiltonians in
$\K[n]$ must have at least $n$ mass terms on a boundary if $n \le
d_{\parallel}$, so that
\begin{equation}
  \K[n](d,d_{\parallel}) = \omap^n [K(d-n,d_{\parallel}-n)]. \label{eq:Knexpr}
\end{equation}
In particular, the ``purely crystalline subgroup'' $\K[1](d,d_{\parallel})$
consists of the (classes of) Hamiltonians with at least one mass term on the
boundary,
\begin{equation}
  \K[1](d,d_{\parallel}) = \omap [K(d-1,d_{\parallel}-1)]. \label{eq:K1expr}
\end{equation}

Similarly, the first property of the order-raising homomorphism $\omap$ leads to
an expression for the decoration subgroups. We first consider the case $n =
d_{\parallel}+1$, for which one has $\ks[0]{n} = K(d+1-n,0)$, see Eq.\
(\ref{eq:Kb1}). In this case, we find that the decoration subgroups $\ks[k]{n}
\subseteq \ks[0]{n}$ are given by

\begin{equation}
  \ks[k]{n} = \ker \omap^{n-k},
\end{equation}
since $\ks[k]{n}$ classifies codimension-$n$ boundary states from separable
$(k-1)$th order Hamiltonians. For the classifying group $\ki[k]{n}$ this gives
\begin{equation}
  \ki[k]{n} = K(d+1-n,d_{\parallel}+1-n)/\ker \omap^{n-k} \label{eq:Kikspecial}
\end{equation}
if $n = d_{\parallel}+1$. For $n < d_{\parallel}+1$ one finds similarly, using
the isomorphism (\ref{eq:KTFisomorphism}),
\begin{align}
  \ks[k]{n} =&\, K'(d+1-n,d_{\parallel}+1-n) \ker \omap^{n-k}
  \nonumber \\ &\, \mbox{} /K'(d+1-n,d_{\parallel}+1-n),
\end{align}
where the subgroup $K'(d+1-n,d_{\parallel}+1-n) \ker \omap^{n-k} \subseteq
K(d+1-n,d_{\parallel}+1-n)$ consists of direct sums $g \oplus h$, with $g \in
K'(d+1-n,d_{\parallel}+1-n)$ and $h \in \ker \omap^{n-k}$.
%\footnote{Following common practice in the literature, we use ``multiplication'' when we refer to the group operation in abstract terms, whereas we use ``addition'' when specifically referring to the classifying groups as groups of integers. The group operation for the Grothendieck group is the direct sum $\oplus$.} 
(Note that all classifying
groups considered here are abelian.) This gives the compact expression
\begin{align}
  \ki[k]{n} =&\, K(d+1-n,d_{\parallel}+1-n) \nonumber \\ &\, 
  \mbox{} /K'(d+1-n,d_{\parallel}+1-n)\ker \omap^{n-k}. \label{eq:Kik}
\end{align}
Note that Eq.~(\ref{eq:Kikspecial}) can be considered a special case of
Eq.~(\ref{eq:Kik}) since $K'(d+1-n,d_{\parallel}+1-n)$ is trivial if $n =
d_{\parallel}+1$. The bulk-boundary correspondence (\ref{eq:bb}) now follows
from Eqs.~(\ref{eq:Kik}) with $k=1$ and Eqs.~(\ref{eq:K1expr}) and
(\ref{eq:Knexpr}) upon applying the general group isomorphism $K/G \ker \alpha
= \alpha[K]/\alpha[G]$ for any subgroup $G \subseteq K$ and homomorphism
$\alpha$ to the case $K = K(d+1-n,d_{\parallel}+1-n)$, $G =
\omap[K(d-n,d_{\parallel}-n)]$, and $\alpha = \omap^n$.\footnote{This property
  follows directly from the observation that the kernel of the natural
quotient map $\tilde \alpha: K \to \alpha[K]/\alpha[G]$ is $G \ker \alpha$.} In
App.~\ref{app:bb} we discuss a possible way to extend the above proof to an
arbitrary crystalline symmetry.

{\em Calculation of the subgroup sequence.---} The bulk classifying groups
$K(d,d_{\parallel})$ were calculated by Shiozaki and Sato in
Ref.~\onlinecite{shiozaki2014}. The purely crystalline subgroups
$K'(d,d_{\parallel})$ were calculated in
Refs.~\onlinecite{langbehn2017,geier2018} by explicit calculation for each
Shiozaki-Sato symmetry class separately.  (Although Ref.~\onlinecite{geier2018}
considered $d_{\parallel} \ge 1$ for $d=2$ and $d=3$ only, the results can be
transferred to all other Shiozaki-Sato classes using the dimension-raising and
lowering isomorphisms $\kappa_{\parallel}$ and $\kappa_{\perp}$.) As shown
in App.~\ref{app:kerk}, the kernels $\ker\, \omap \subseteq K(d,d_{\parallel})$
can be obtained from the known results for $K'$ and $K$.

The remainder of the calculation of the classifying groups can be done without
further explicit calculations. This relies on the key observation that the
nontrivial groups in the sequence
\begin{align}
  \K[1](d+1,d_\parallel+1) &\rightarrow \K[1](d+2,d_\parallel+2) \nonumber \\ &  \rightarrow \K[1](d+3,d_\parallel+3) \rightarrow \ldots
  \label{eq:seq}
\end{align} 
are isomorphic to $\ZZ$ or to $\ZZ_2$ and that the succession $\ZZ \to \ZZ_2$
does not occur. Since both $\ZZ$ and $\ZZ_2$ have a single generator, it
follows that {\em any} homomorphism $\K[1](d+l,d_\parallel+l) \to
\K[1](d+l+1,d_\parallel+l+1)$ is either injective, or it maps
$\K[1](d+l,d_\parallel+l)$ to the trivial element. Applying this observation to
the order-raising homomorphism $\omap$ and denoting the first instance in
which $\omap$ maps $\K[1](d+l,d_\parallel+l)$ to the trivial element by
$\K[1](d+q,d_{\parallel}+q)$, we obtain the sequence
\begin{align}
  \K[1](d+1,d_\parallel+1) &\overset{\omap}{\hookrightarrow} \K[1](d+2,d_\parallel+2) \overset{\omap}{\hookrightarrow} \ldots \nonumber \\ & \overset{\omap}{\hookrightarrow} \K[1](d+q,d_\parallel+q) \overset{\omap}{\rightarrow} 0,
  \label{eq:seq2}
\end{align}
where the symbol ``$\hookrightarrow$'' denotes an injection. 
Since $\K[1](d+1,d_{\parallel}+1) = \omap[\K[0](d,d_{\parallel})]$, it follows that
\begin{align}
  \ker\omap^k=
  \begin{cases}
    \ker\omap & \text{for } 0<k\le q,\\
    \K[0](d,d_\parallel) & \text{for } k>q,
  \end{cases}
  \label{eq:kerq}
\end{align}
where $\ker\omap\subseteq \K[0](d,d_\parallel)$. The cut-off $q$ can be
obtained from the calculation of $\ker \omap$, see App.~\ref{app:kerk}.

Once $\ker \omap^k$ and $\K[1]$ are known, the boundary classification groups
$\ki[k]{n+1}$ follow from Eq.~(\ref{eq:Kik}), whereas the subgroup sequence of
bulk classification groups follows from the bulk-boundary
correspondence~(\ref{eq:bb}). The results of this calculation are summarized in Tables \ref{tab:subgroupC2d}--\ref{tab:subgroupR} for the bulk classifying sequence for $d=0$, $1$, $2$, and $3$, and in Tables \ref{tab:boundaryC}--\ref{tab:boundaryR} for the boundary classifying groups for $d=3$. Tables with boundary classifying groups for $d=2$ can be found in Ref.\ \onlinecite{geier2018}.

\section{Order-raising homomorphism $\omap$} \label{sec:omapexplicit}

In this Section we give an explicit expression for the order-raising
homomorphism $\omap$ in terms of the dimension-raising maps
$\kappa_{\parallel}$ and $\rho_{\parallel}$ introduced by Shiozaki and
Sato\cite{shiozaki2014} and the boundary map $\delta$ of Turner {\em et al.},
give explicit expressions for the action of $\omap$ on the Hamiltonians
$H(\vk,m)$ introduced in Sec.\ \ref{sec:ShiozakiSato}, and discuss the relation between $\omap$ and the ``layer stacking
construction'',\cite{isobe2015,fulga2016,khalaf2018,khalaf2018b} which was
previously used to construct topological crystalline phases and classify the
boundary states. The expression of $\omap$ in terms of the maps
$\kappa_{\parallel}$, $\rho_{\parallel}$, and $\delta$ relates it to the
$K$-theory approaches to the classification of topological crystalline phases.
It plays a key role for establishing the properties of the homomorpishm used in
our demonstration of the bulk-boundary correspondence. On the other hand, the
explicit realization of the order-raising homomorphism and its relation to the
layer stacking construction are of more use for concrete examples.

\subsection{Construction using dimension-raising isomorphisms} \label{sec:omapraising}

The order-raising homomorphism $\omap$ is obtained by sequential application of
the dimension-raising maps $\kappa_{\parallel}$ and $\rho_{\parallel}$ of
Shiozaki and Sato\cite{shiozaki2014} (see also Sec.\ \ref{sec:ShiozakiSato} and App.\ \ref{app:dmaps}) and the boundary map $\delta$ of Turner
{\em et al.},\cite{turner2012}
\begin{align}
  \omap=\kappa_\parallel\circ\delta\circ\rho_\parallel.
  \label{eq:omap}
\end{align}
Here, the dimension-raising isomorphism $\rho_{\parallel}$ maps an equivalence
class of Hamiltonians $H$ to a one-parameter family $H(\varphi)$, $0 \le
\varphi \le 2 \pi$, with $H(0) = H(2 \pi)$, on which the crystalline symmetry
${\cal S}$ acts nonlocally,
$\sigma_\mathcal{S}U_\mathcal{S}H(\varphi)U_\mathcal{S}^\dagger=H(2\pi-\varphi)$;
the boundary map $\delta$ then maps the equivalence class of one-parameter
families $H(\varphi)$ to 
\begin{align}
	\delta[H(\varphi)]=H(\pi)\ominus H(0),
	\label{eq:delta}
\end{align}
which gives a Hamiltonian with the topological numbers equal to the difference
between the topological numbers of $H(\varphi)$ at $\varphi=0,\pi$. (The
operation ``$\ominus$'' formally requires the use of the Grothendieck
construction, see, {\em e.g.}, Ref.\ \onlinecite{nakahara2003,turner2012}.)  Lastly,
the dimension-raising isomorphism $\kappa_\parallel$ maps the equivalence class
of $d$-dimensional Hamiltonians thus obtained to an equivalence class of
$(d-1)$-dimensional Hamiltonians, thus defining an element in the group
$\K[0](d+1,d_\parallel+1)$. Although the dimension-raising isomorphisms
$\rho_{\parallel}$ and $\kappa_{\parallel}$ also change the Shiozaki-Sato
symmetry class, the symmetry class is not changed by combination of the two
maps in Eq.\ (\ref{eq:omap}).

The maps $\kappa_\parallel$, $\rho_\parallel$, and $\delta$ all respect the
group structure of the classifying groups and they commute with the
dimensional-raising maps $\kappa_{\parallel}$ and $\kappa_{\perp}$, immediately
proving the second property of the order-raising homomorphism advertised in the
previous Section. A proof of the remaining two properties is given in
App.~\ref{app:omap}.

\subsection{Explicit realization of the order-raising homomorphism $\omap$} \label{sec:omapcanonical}

\begin{table}
\begin{tabular*}{\columnwidth}{c @{\extracolsep{\fill}} cccccc}
\hline\hline 
TF class & $\mathcal{S}$     & $(H_{\omap},\Gamma_{\omap})$         & $\omap(U_\mathcal{C})$ & $\omap(U_\mathcal{S})$              & $\rM_n$\\
\hline
A        & $\mathcal{S}$     & $(\tau_3H,\tau_1)$ & -                      & $\tau_3U_\mathcal{S}$               & $\tau_2$     \\
AIII     & ${\mathcal{S}_+}$ & $(\tau_3H,\tau_2)$ & $\tau_1$               & $\tau_1U_\mathcal{C}U_\mathcal{S}$  & $\tau_3U_\mathcal{C}$\\
A        & $\mathcal{CS}$    & $(\tau_3H,\tau_1)$ & -                      & $\tau_0U_\mathcal{S}$               & $\tau_2$       \\
AIII     & ${\mathcal{S}_-}$ & $(\tau_3H,\tau_2)$ & $\tau_1$               & $\tau_3U_\mathcal{S}$               & $\tau_3U_\mathcal{C}$\\
\hline\hline
\end{tabular*}
\caption{The action~(\ref{eq:omapH}) of the order-raising homomorphism $\omap$
on a Hamiltonian $H$ in the complex ten-fold way classes with a unitary
order-two symmetry or antisymmetry.
$\rM_n$ is the crystalline-symmetry-breaking mass term generated by the
homomorphism $\omap$.}
\label{tab:omapC}
\end{table}

We recall that the group structure of the bulk and boundary classifying groups is given by the Grothendieck construction. As discussed in Sec.\ \ref{sec:ShiozakiSato} this motivates us to consider $m$-dependent Hamiltonians $H(\vk,m)$, such that $H(\vk,m)$ is in well-defined topological phases for $-2 < m < 0$ and for $0 < m < 2$, with the transition between topological classes (if any) taking place at $m=0$. The canonical-form Hamiltonians of Sec.\ \ref{sec:cform} are examples of such $m$-dependent Hamiltonians. The action of the order-raising homomorphism $\omap$ on Hamiltonians $H(\vk,m)$ follows from the known action of the maps $\kappa_{\parallel}$, $\rho_{\parallel}$, and $\delta$ on such $m$-dependent Hamiltonians $H(\vk,m)$.~\cite{teo2010,shiozaki2014} Specifically, for an equivalence class containing the $d$-dimensional Hamiltonian $H(\vk,m)$, the mapped class is represented by the Hamiltonian\cite{shiozaki2016}
\begin{align}
	\label{eq:omapH}
	\omap(H(\bm k,\mm))&= H_\omap(\bm k,\mm+1-\cos
	k^\prime)+\Gamma_\omap\sin k^\prime,\nonumber
\end{align}
where the pair $(H_\omap,\Gamma_\omap)$ is given in
Tables~\ref{tab:omapC}-\ref{tab:omap} and the $(d+1)$-dimensional momentum is
defined as $(k^\prime,\bm k)$. The maps $\kappa_{\parallel}$, $\delta$, and
$\rho_{\parallel}$ featuring in the definition~(\ref{eq:omap}) can be
represented in a similar way, see
App.~\ref{app:omap}.

\begin{table}
\begin{tabular*}{\columnwidth}{c @{\extracolsep{\fill}} cccccc}
\hline\hline
TF class     & $\mathcal{S}$                                                   & $(H_{\omap},\Gamma_{\omap})$         & $\omap(U_\mathcal{C})$ & $\omap(U_\mathcal{S})$ & $\rM_n$\\
\hline
A            & ${\mathcal{T}^+\mathcal{S}}$, ${\mathcal{T}^-\mathcal{S}}$      & $(\tau_3H,\tau_1)$ & -                      & $\tau_0U_\mathcal{S}$  & $\tau_2$\\
AIII         & ${\mathcal{P}^+\mathcal{S}_+}$, ${\mathcal{P}^-\mathcal{S}_+}$  & $(\tau_3H,\tau_2)$ & $\tau_1$               & $\tau_0U_\mathcal{S}$  & $\tau_3U_\mathcal{C}$\\
A            & ${\mathcal{P}^+\mathcal{S}}$, ${\mathcal{P}^-\mathcal{S}}$      & $(\tau_3H,\tau_1)$ & -                      & $\tau_3U_\mathcal{S}$  & $\tau_2$\\
AIII         & ${\mathcal{T}^+\mathcal{S}_-}$, ${\mathcal{T}^-\mathcal{S}_-}$  & $(\tau_3H,\tau_2)$ & $\tau_1$               & $\tau_3U_\mathcal{S}$  & $\tau_3U_\mathcal{C}$\\
\hline\hline
\end{tabular*}
\caption{The action~(\ref{eq:omapH}) of the order-raising homomorphism $\omap$
on a Hamiltonian in the complex ten-fold way classes with an
  antiunitary order-two symmetry or antisymmetry.
$\rM_n$ is the crystalline-symmetry-breaking mass term generated by
the homomorphism $\omap$.}
\label{tab:omapA}
\end{table}
\begin{table}
\begin{tabular*}{\columnwidth}{c @{\extracolsep{\fill}} cccccc}
\hline\hline 
TF classes   & $\mathcal{S}$ & $(H_{\omap},\Gamma_{\omap})$         & $\omap(U_\mathcal{T})$ & $\omap(U_\mathcal{P})$ & $\omap(U_\mathcal{S})$             & $\rM_n$\\
\hline
AI, AII      & ${\mathcal{S}_+,\mathcal{S}_-}$       & $(\tau_3H,\tau_1)$ & $\tau_3U_\mathcal{T}$  & -                      & $\tau_3U_\mathcal{S}$              & $\tau_2$  \\
AI, AII      & ${\mathcal{CS}_+,\mathcal{CS}_-}$     & $(\tau_3H,\tau_1)$ & $\tau_3U_\mathcal{T}$  & -                      & $\tau_0U_\mathcal{S}$              & $\tau_2$ \\
BDI, CII     & ${\mathcal{S}_{++},\mathcal{S}_{--}}$ & $(\tau_3H,\tau_2)$ & $\tau_0U_\mathcal{T}$  & $\tau_1U_\mathcal{T}$  & $\tau_1U_\mathcal{C}U_\mathcal{S}$ & $\tau_3U_\mathcal{C}$          \\
BDI, CII     & ${\mathcal{S}_{+-},\mathcal{S}_{-+}}$ & $(\tau_3H,\tau_2)$ & $\tau_0U_\mathcal{T}$  & $\tau_1U_\mathcal{T}$  & $\tau_3U_\mathcal{S}$              & $\tau_3U_\mathcal{C}$   \\
D, C         & ${\mathcal{S}_+,\mathcal{S}_-}$       & $(\tau_3H,\tau_1)$ & -                      & $\tau_0U_\mathcal{P}$  & $\tau_3U_\mathcal{S}$              & $\tau_2$    \\
D, C         & ${\mathcal{CS}_+,\mathcal{CS}_-}$     & $(\tau_3H,\tau_1)$ & -                      & $\tau_0U_\mathcal{P}$  & $\tau_0U_\mathcal{S}$              & $\tau_2$  \\
DIII, CI     & ${\mathcal{S}_{++},\mathcal{S}_{--}}$ & $(\tau_3H,\tau_2)$ & $\tau_2U_\mathcal{P}$  & $\tau_3U_\mathcal{P}$  & $\tau_1U_\mathcal{C}U_\mathcal{S}$ & $\tau_3U_\mathcal{C}$          \\
DIII, CI     & ${\mathcal{S}_{+-},\mathcal{S}_{-+}}$ & $(\tau_3H,\tau_2)$ & $\tau_2U_\mathcal{P}$  & $\tau_3U_\mathcal{P}$  & $\tau_3U_\mathcal{S}$              & $\tau_3U_\mathcal{C}$    \\
\hline\hline
\end{tabular*}
\caption{The action~(\ref{eq:omapH}) of the order-raising homomorphism $\omap$
on a Hamiltonian in the real ten-fold way classes with a unitary order-two
symmetry or antisymmetry.  $\rM_n$ is the crystalline-symmetry-breaking mass
term generated by the homomorphism $\omap$.}
\label{tab:omap}
\end{table}

\subsection{Stacking Construction} \label{sec:omapstacking}

References~\onlinecite{isobe2015,fulga2016,huang2017} construct higher-order
topological phases by stacking layers of lower-dimensional ones. Like the
order-raising homomorphism $\omap$ considered here, the stacking construction
also involves simultaneously increasing the spatial dimension $d$ and the
number of inverted dimensions $d_{\parallel}$ by one, so that it, too, provides
a homomorphism $\sigma$
\begin{align}
  \sigma:\,K(d,d_{\parallel}) \to K(d+1,d_{\parallel}+1).
\end{align}
Further, in Ref.~\onlinecite{huang2017} it is argued, from the boundary
perspective, that the stacking of $d$-dimensional ``layers'' that differ by a
separable phase yields topologically equivalent $(d+1)$-dimensional crystals.
This, too, is a property that is shared by the order-raising homomorphism
$\omap$. Indeed, below we show that the stacking homomorphism $\sigma$ has all
three defining properties of the order-raising homomorphism specified in
Sec.~\ref{sec:omap}.  The order-raising homomorphism $\omap$ of
App.~\ref{sec:omapraising} and the stacking construction are two realizations of the
same homomorphism. 

\begin{figure}
	\includegraphics[width=\columnwidth]{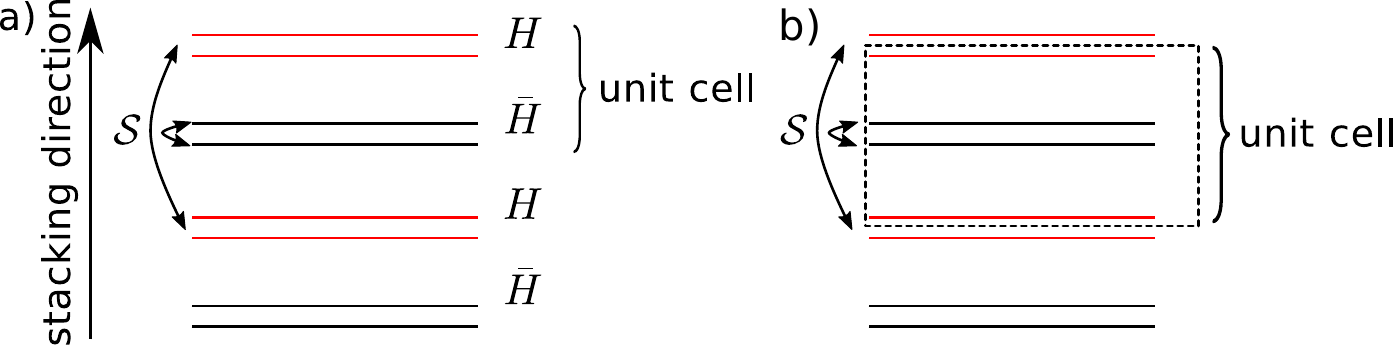}
	\caption{\label{fig:layers} Layer-stacking construction of higher-order
        topological phases: (a) A $(d+1)$-dimensional crystal is constructed
        out of alternating $d$-dimensional ``layers'' with opposite topological
	numbers. The unit cell consisting of two such layers is not mapped to
	itself under the (anti)symmetry operation ${\cal S}$, which inverts the
	coordinate $x_{d+1}$ in the stacking direction. (b) The unit cell may
	be redefined, so that it is mapped to itself under ${\cal S}$. This
	redefinition of the unit cell involves splitting the odd layers into
	two parts that are mapped onto each other under ${\cal S}$, eventually
      after adding topological trivial bands.}
\end{figure}

Specifically, the stacking procedure constructs a $(d+1)$-dimensional crystal
by alternating $d$-dimensional ``layers'' with opposite topological numbers as
shown schematically in Fig.~\ref{fig:layers}a. Denoting the Hamiltonians of the
alternating $d$-dimensional layers as $H_d(\bm k)$ and $\bar H_d(\bm k)$,
respectively, the Hamiltonian of the $(d+1)$-dimensional stack is
\begin{equation}
	H_{d+1}(\bm k,k_{d+1})=
	\begin{pmatrix}
		H_d(\bm k) & 0\\
		0 & \bar H_d(\bm k)
	\end{pmatrix}.
	\label{eq:Hstack}
\end{equation}
If the $d$-dimensional Hamiltonians $H_d$ and $\bar H_d$ have a crystalline
(anti)symmetry with $d_{\parallel}$ inverted dimensions encoded by the unitary
matrix $U_{\cal S}$, the $(d+1)$-dimensional Hamiltonian $H_{d+1}$ has {\em
two} crystalline (anti)symmetries, encoded by $\mbox{diag}\, (U_{\cal
S},U_{\cal S})$ and $\mbox{diag}\, (e^{i k_{d+1}} U_{\cal S},U_{\cal S})$, with
$d_{\parallel}$ and $d_{\parallel} + 1$ inverted dimensions, respectively. The
former (anti)symmetry yields a weak topological crystalline phase and will not
be considered here. The latter (anti)symmetry has a $k_{d+1}$-dependent
transformation matrix, which reflects the fact that it does not map the unit
cell defined by the representation (\ref{eq:Hstack}) of $H_{d+1}$ to itself,
see Fig.~\ref{fig:layers}a. To remedy this situation we replace
Eq.~(\ref{eq:Hstack}) by
\begin{align}
  \sigma(H_{d}) \equiv&\,
   	\begin{pmatrix}
  H'_{d+1}(\bm k,k_{d+1}) & 0 \\ 0 & \bar H_d(\bm k)
  \end{pmatrix} \nonumber \\ =&\,
	\begin{pmatrix}
		e^{i \cM k_{d+1}/2} H_d(\bm k) e^{-i \cM k_{d+1}/2} & 0\\
		0 & \bar H_d(\bm k)
	\end{pmatrix},
	\label{eq:smap}
\end{align}  
where $\cM$ is a matrix that commutes with the non-spatial (anti)symmetries
${\cal T}$, ${\cal P}$, and ${\cal C}$, and anticommutes with $U_{\cal S}$, and
the crystalline (anti)symmetry is represented by $\mbox{diag}\, (U_{\cal S},U_{\cal
S})$. (Being able to find a matrix $\cM$ with these properties may require the
addition of additional, topological trivial bands.) Loosely speaking, the
transformation described by Eq.~(\ref{eq:smap}) involves the redefinition
of the unit cell as in Fig.~\ref{fig:layers}b, so that the additional
crystalline symmetry ${\cal S}$ maps the $(d+1)$-dimensional unit cell to
itself for the new choice of the unit cell.

The form of the bulk Hamiltonian~(\ref{eq:smap}) immediately allows us to
conclude that for $H_d(\bm k)$ separable, the Hamiltonian $H_d(\bm k)$ can be
deformed to manifestly separable form and the matrix $\cM$ can be chosen to
commute with it, resulting in a $k_{d+1}$-independent, and therefore topologically
trivial Hamiltonian $\sigma(H_{d})$ (aside from possible weak invariants). The reverse is
also true: $\sigma(H_d)$ topologically trivial implies that the upper-right
block $H^\prime_{d+1}$ of Eq.~(\ref{eq:smap}) has only weak topological
invariants. Thus $H^\prime_{d+1}$ can be continuously deformed to a
$k_{d+1}$-independent Hamiltonian. The only possible way to remove
$k_{d+1}$-dependence from $e^{i \cM k_{d+1}/2} H_d(\bm k) e^{-i \cM k_{d+1}/2}$
is to continuously deform the Hamiltonian $H_d$ and/or the matrix $\rho$ to
mutually commute. We have therefore shown
\begin{itemize}
	\item $\sigma(H)$ is in the trivial class if and only if $H$ is separable. 
\end{itemize}
The above statement is obtained from the bulk perspective, accordingly, it also
holds for $d$-dimensional topological phases from $\K[d]$ that do not support
topologically protected boundary states.

The $d$-dimensional Hamiltonian $H_d(\bm k)$ in Eq.~(\ref{eq:smap}) is to be
understood as an $\mm$-dependent family $H_d(\bm k,\mm)$ that represents a
topologically trivial Hamiltonian for $\mm>0$. A topologically trivial
Hamiltonian is separable, and we choose a parameterization where $H_d(\bm
k,\mm)$ is manifestly separable for $\mm>0$. With this choice, the term $e^{i
\cM k_{d+1}/2} H_d(\bm k,\mm) e^{-i \cM k_{d+1}/2}$ is $k_{d+1}$-independent
for $\mm>0$, thus trivial without any additional weak invariants. 

Using the
definition~(\ref{eq:smap}) and Eq.\ (\ref{eq:omapH}) applied to the dimension-raising maps $\kappa_{\parallel}$ and $\kappa_{\perp}$, {\em i.e.}, with $\omap$ replaced by $\kappa_{\parallel}$ or $\kappa_{\perp}$, we obtain 
\begin{itemize}
	\item the stacking homomorphism $\sigma$ commutes with the dimension-raising
		isomorphisms $\kappa_\parallel$ and $\kappa_\perp$.
\end{itemize}

The stacking construction has the property that if a non-separable Hamiltonian
$H_d$ supports topologically protected states on its $(d-1)$-dimensional
boundary, $\sigma(H)$ also supports topologically protected states of the same
dimensionality on its $d$-dimensional boundary, see
Refs.~\onlinecite{isobe2015,fulga2016,huang2017}, --- combined with the above
property this gives
\begin{itemize}
	\item   If $H$ is a non-separable Hamiltonian with $n-1$
		crystalline-symmetry-breaking mass terms, then
		$\sigma(H)$ is a Hamiltonian with $n$ crystalline-symmetry-breaking
		mass terms.
\end{itemize}
To see this consider a $d$-dimensional Hamiltonian $H_d$ with $n$
crystalline-symmetry-breaking mass terms. By repeatedly applying
the dimension-raising isomorphism $\kappa_\perp$ and $\kappa_\parallel$ or their
inverse, we can change both the values of $d$ and $d_\parallel$ to $n+1$. The
resulting inversion-symmetric $(n+1)$-dimensional Hamiltonian $H_{n+1}$ is
guaranteed to have zero-dimensional protected boundary states, see
Sec.~\ref{sec:cform}. Thus $\sigma(H_{n+1})$ has also zero-dimensional
topologically protected boundary states,~\cite{isobe2015,fulga2016,huang2017}
and accordingly $\sigma(H_{n+1})$ has $n+1$ crystalline-symmetry-breaking mass
terms (boundary mass terms). Since the homomorphism $\sigma$ commutes with the
dimension-raising isomorphism $\kappa_\perp$ and $\kappa_\parallel$, the same
is true for $\sigma(H_d)$. We additionally checked that
$\omap(H_0)\cong\sigma(H_0)$ for zero-dimensional Hamiltonians $H_0$.

Although the realizations
$\sigma$ and $\omap$ are indistinguishable as homomorphisms between classifying groups, their action on a specific Hamiltonian is
rather different. When acting on a nearest-neighbor hopping Hamiltonian, the
homomorphism $\omap$ gives a Hamiltonian of the same form. In particular, if
$H$ is a minimal canonical-form Hamiltonian, $\omap(H)$ is also a minimal
canonical-form Hamiltonian. On the other hand, as evident from the
definition~(\ref{eq:smap}), the stacking homomorphism $\sigma$ generates
hopping elements beyond the nearest-neighbors. Section \ref{sec:examples} 
illustrate these differences for three examples.

\section{Examples}\label{sec:examples}
In this Section we 
illustrate the full classification using the subgroup sequence (\ref{eq:subgroup}) for a few representative examples and show how the order-raising homomorphism $\omap$ relates topological crystalline phases in different dimensions. We further
%give various tight-binding model realizations for the
%minimal generators of the higher-order crystalline phases obtained by applying
%the order-raising homomorphism $\omap$. Explicit examples for higher-order
%decoration subgroups are provided and their connection to the boundary
%classifying groups is clarified. We 
compare the realization of the
order-raising homomorphism $\omap$, see Secs.~\ref{sec:omapraising} and
App.~\ref{app:omap}, to that of the layer stacking procedure of Refs.\
\onlinecite{isobe2015,fulga2016,khalaf2018,khalaf2018b}, see Sec.\
\ref{sec:omapstacking}.  Additionally, we discuss the connection to recently
studied embedded topological phases.~\cite{tuegel2018} As in the previous
Section we reserve the symbol $\omap$ for the concrete realization of the
order-raising homomorphism given in Secs.~\ref{sec:omapraising} and
\ref{sec:omapcanonical}.

The models we consider can all be expressed in the canonical form of
Eq.~(\ref{eq:canonical}), where we add the perturbation (\ref{eq:H1}) or a
crystalline-symmetry-breaking mass term localized at the sample boundaries to
gap out the boundaries (if applicable). The action of the order-raising
homomorphism $\omap$ on $H_0$ is defined by Eq.\ (\ref{eq:omapH}) and
Tables~\ref{tab:omapC}-\ref{tab:omap}.

\subsection{Higher-order phases originating from the Quantum Hall phase}
In two dimensions, systems with broken time-reversal symmetry but without
crystalline symmetries admit a quantum Hall phase, which has chiral propagating
modes along crystal edges. This first-order topological phase in ten-fold way
class A is compatible with an on-site crystalline symmetry ${\cal O}$, with a
mirror antisymmetry ${\cal CM}$, and with a twofold rotation symmetry ${\cal
R}$. Further, in the presence of ${\cal CM}$ a two-dimensional
second-order topological phase
with protected zero-energy states at mirror-symmetric corners is possible, too.
The order-raising homomorphism links these two-dimensional topological phases
to three-dimensional topological phases with an additional mirror symmetry
${\cal M}$, rotation antisymmetry ${\cal CR}$, or inversion symmetry ${\cal
I}$, respectively. For each of these cases we describe the action of the
order-raising homomorphism in detail and show how it connects the subgroup sequences
classifying the bulk crystalline topology for the two-dimensional and
three-dimensional phases. 

{\em Higher-order phases originating from class A$^{\cal O}$ in two
dimensions.---} Without loss of generality we may represent the on-site
crystalline symmetry ${\cal O}$ using $U_{\cal O} = \tau_3$. The on-site
crystalline symmetry forces the Hamiltonian $H$ to have a block-diagonal
structure, $H = \mbox{diag}\, (h_+,h_-)$, with separate blocks $h_+$ and $h_-$
for even and odd-parity states, respectively. In two dimensions the bulk band
structure in class A$^{\cal O}$ is classified by the subgroup sequence $0
\subseteq 0 \subseteq \ZZ^2$, see Table \ref{tab:subgroupC2d}. The existence of
first-order topological phases with a $\ZZ^2$ classification follows directly
from the well-known $\ZZ$ classification for the ten-fold way class A, because
the individual blocks $h_{\pm}$ are not subject to any constraints from the crystalline
symmetry. The minimal canonical-form generators for the $\ZZ^2$ classifying
group are quantum-Hall phases for each parity block separately, which have
$\Gamma_0 = \sigma_1$, $\bm \Gamma = (\sigma_2,\sigma_3)$, {\em i.e.},
Hamiltonians
\begin{align}
  h_{\pm}(\vk,m) =&\, \sigma_2 (2+m-\cos k_x - \cos k_y)
  \nonumber \\ &\, \mbox{} + \sigma_3 \sin k_x + \sigma_1 \sin k_y.
  \label{eq:qhe}
\end{align}
(The Pauli matrices $\sigma_j$ act on a different degree of freedom than the
Pauli matrix $\tau_3$ used to represent the on-site crystalline symmetry.)
Applying the order-raising homomorphism $\omap$ to such a generator gives a
three-dimensional canonical-form Hamiltonian with (see Table \ref{tab:omapC})
\begin{equation}
  \Gamma_0 = \tau_3 \sigma_2,\ \ \bm\Gamma=(\tau_1,\tau_3\sigma_3,\tau_3\sigma_1),
	\label{eq:3dqhe}
\end{equation}
which satisfies an additional mirror symmetry with the representation $U_{\cal M} = \tau_3$. 
%(Recall that the order-raising homomorphism increases the number of inverted dimensions $d_{\parallel}$ by one, so that an on-site crystalline symmetry ${\cal O}$ is mapped to a mirror symmetry ${\cal M}$.) 
This Hamiltonian has one crystalline-symmetry-breaking mass term $M_1 =
\tau_2$, so that it represents a second-order topological phase. 

One verifies
that two-dimensional Hamiltonians for class A$^{\cal O}$ are separable if and
only if the even and odd parity blocks have equal $\ZZ$ topological indices.
This implies $\mbox{ker}\, \omap = \ZZ$, so that the image of the classifying
group $\omap(\K[0](2,0))=\K[1](3,1)=\ZZ$. [We recall our notation according to
which the groups $K^{(n)}(d,d_{\parallel})$ classify the bulk topology for
$d$-dimensional phases with a crystalline symmetry with $d_{\parallel}$
inverted dimensions.] Since the ten-fold way class A is trivial for $d=3$ there
are no first-order phases, {\em i.e.}, $\K[0](3,1) = \K[1](3,1)$, consistent
the subgroup sequence $0 \subseteq 0 \subseteq \ZZ \subseteq \ZZ$ for class
A$^{\cal M}$ in three dimensions, see Table \ref{tab:subgroupC}.

{}From the boundary perspective, for class A$^{\cal M}$ in three dimensions one
finds $\ks[0]{2}(3,1) = \ZZ^2$, where the two $\ZZ$ indices counts the number
of chiral hinge modes for each mirror parity. Equal numbers of of chiral hinge
modes for the two mirror parities correspond to a separable boundary phase, so
that $\ks[1]{2}(3,1) = \ZZ$. The anomalous boundary classifying group is,
hence, $\ki{2}(3,1) = \ks[0]{2}(3,1)/\ks[1]{2}(3,1) = \ZZ$.\cite{schindler2018}

\begin{figure}
	\includegraphics[width=0.5\columnwidth]{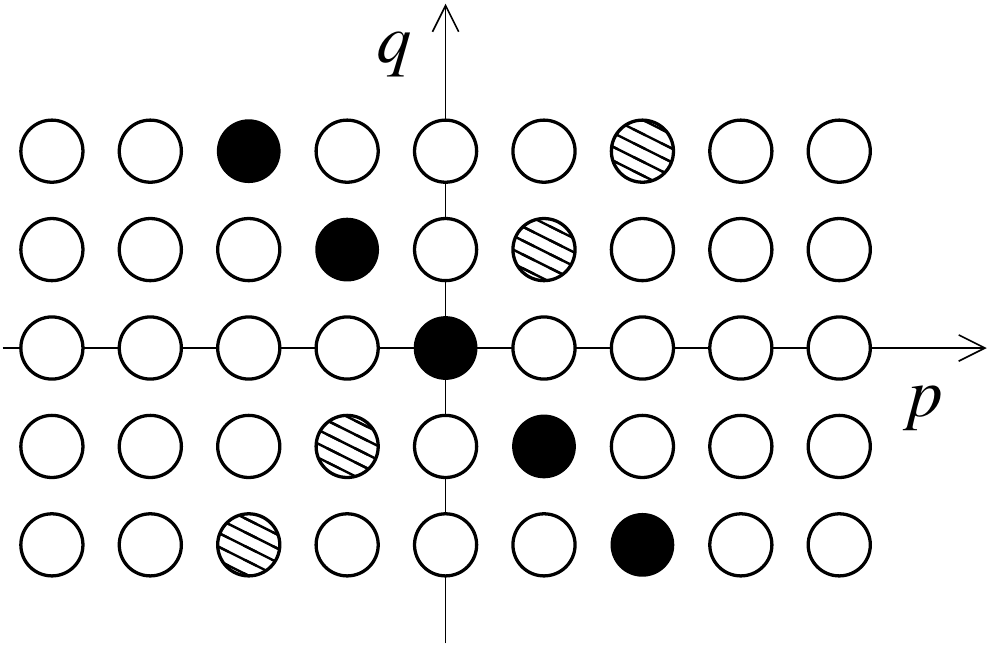}
	\caption{\label{fig:Klattice} The full bulk classifying group $\K[0](2,1)$ for classes A$^{\cal CM}$ is $\ZZ^2$. The purely crystalline subgroup $\K[1](2,1)$ and the separable subgroup $\ker \omega$ are denoted by solid and hatched circles, respectively.}
\end{figure}

{\em Higher-order phases originating from class A$^{\cal CM}$ in two
dimensions.---} This class has a $0 \subseteq \ZZ \subseteq \ZZ^2$ subgroup
sequence for its bulk topological classification in two dimensions, see Table
\ref{tab:subgroupC2d}. For definiteness we choose to represent the mirror
antisymmetry by $U_{\cal CM} = \sigma_3$, so that 
\begin{equation}
  H(k_x,k_y) = -\sigma_3 H(-k_x,k_y) \sigma_3.
\end{equation}
At the high-symmetry lines $k_x=0$ or $\pi$ the mirror antisymmetry ${\cal CM}$
effectively simplifies to a chiral antisymmetry ${\cal C}$ represented by
$U_{\cal C} = \sigma_3$, allowing one to define the difference $W$ of winding
numbers for $0 \le k_y \le 2 \pi$ at $k_x = 0$ and $k_x = \pi$ as a suitable
topological index. The second topological invariant of the Hamiltonian $H$ is
the Chern number $C$, which counts the number of chiral boundary modes. Since
$C$ and $W$ have the same parity, the $\ZZ^2$ bulk topological index $(p,q)$
can be defined setting $p = (C+W)/2$, $q=(C-W)/2$. A common set of generators
for the classifying groups $\K[0](2,1) = \ZZ^2$ and $\K[1](2,1) = \ZZ$ is given
by the canonical-form Hamiltonians $H_{(1,0)}$ and $H_{(1,-1)}$, with
\begin{align}
  & \Gamma_0 = \sigma_2,\ \ \bm \Gamma = (\sigma_3,\sigma_1),\ \
  && \mbox{for $H_{(1,0)}$}, \nonumber \\
  & \Gamma_0 = \sigma_2 \tau_0,\ \ \bm \Gamma = (\sigma_3 \tau_3, \sigma_1 \tau_0),\ \ && \mbox{for $H_{(1,-1)}$}.
  \label{eq:HgeneratorsHCM}
\end{align}
The Hamiltonian $H_{(1,0)}$ represents a first-order topological phase with a
single anomalous chiral boundary mode; it is a generator of $\K[0](2,1)$, but
not of $\K[1](2,1)$. The Hamiltonian $H_{(1,-1)}$, which has the ${\cal CM}$-breaking mass terms $\rM_1 = \sigma_3 \tau_1$ and $\rM_2 = \sigma_3 \tau_2$,
represents a second-order
topological phase, with anomalous zero-energy corner states at mirror-symmetric
corners; it is a generator of both $\K[0](2,1)$ and $\K[1](2,1)$, see Fig.\
\ref{fig:Klattice}.

The order-raising homomorphism $\omap$ maps a two-dimensional Hamiltonian with
a mirror antisymmetry ${\cal CM}$ to a three-dimensional Hamiltonian with a
rotation antisymmetry ${\cal CR}$. To see how the order-raising map $\omap$
maps between the subgroup sequence of the two classes, we first consider its
action on the Hamiltonian $H_{(1,0)}$ considered above. Application of the
order-raising homomorphism $\omap$ to $H_{(1,0)}$ gives the three-dimensional
canonical-form Hamiltonian with
\begin{equation}
  \Gamma_0 = \sigma_2 \tau_3,\ \ \bm\Gamma=(\sigma_0\tau_1,\sigma_3\tau_3,\sigma_1\tau_3),
  \label{eq:omapHgeneratorCM}
\end{equation}
which satisfies a twofold-rotation antisymmetry ${\cal CR}$ with the
representation $U_{\cal CR} = \sigma_2$. This Hamiltonian has a single
crystalline-symmetry-breaking mass term $M_1 = \tau_2$, corresponding to a
second-order topological phase with a single chiral mode along a hinge. To
further specify the action of the order-raising homomorphism $\omap$, we search
for separable two-dimensional Hamiltonians, since these are mapped to the
trivial class under $\omap$. With a little algebra one verifies that the
canonical-form Hamiltonian with
\begin{align}
  & \Gamma_0 = \sigma_2 \tau_0,\ \ \bm \Gamma = (\sigma_3 \tau_1, \sigma_1 \tau_1)
\end{align}
has topological indices $(p,q)=(1,1)$ and is separable. We thus identify
$H_{(1,1)}$ as the generator of the separable subgroup $\ker \omap$, see Fig.\
\ref{fig:Klattice}. Since $H_{(2,0)}$ differs from $H_{(1,-1)}$ by a separable
Hamiltonian, $H_{(2,0)}$ and $H_{(1,-1)}$ must be mapped to the same
topological class under $\omap$, {\em i.e.} $\omap(H_{(2,0)})$ must be a
representative of
a {\em third-order} topological phase. It follows that $\K[1](3,2) =
\omega(\K[0](2,1)) = \ZZ$ and $\K[2](3,2) = \omega(\K[1](2,1)) = 2 \ZZ$.
Combined with the observation that there are no first-order topological phases
for the ten-fold-way class A, we arrive at the subgroup sequence $0 \subseteq 2
\ZZ \subseteq \ZZ \subseteq \ZZ$ for three-dimensional Hamiltonians with a
twofold-rotation antisymmetry, consistent with Table \ref{tab:subgroupC}.

{}From the boundary perspective, we note that chiral hinge modes have a $\ZZ$
classification: The $\ZZ$ topological index simply counts the number of chiral
hinge modes. Hence $\ks[0]{2}(3,2) = \ZZ$. The presence of the ${\cal CR}$
antisymmetry plays no role here, as it does not leave any hinges invariant. An
even number of hinge modes represents a separable phase, so that
$\ks[1]{2}(3,2) = 2 \ZZ$. It follows that $\ki{2}(3,2) =
\ks[0]{2}(3,2)/\ks[1]{2}(3,2) = \ZZ_2$. To describe the third-order phases from
the boundary perspective, we note that the rotation antisymmetry is a local
symmetry for a corner on the rotation axis. For this situation one finds a
$\ZZ$ topological index, counting the difference of the number of zero-energy
corner states for even and odd ${\cal CR}$ parity. Since none of these boundary
classes is separable, one has $\ks[0]{3}(3,2) = \ZZ$, $\ks[1]{3}(3,2) =
\ks[2]{3}(3,2) = 0$, so that and $\ki[2]{3}(3,2) = \ki{3}(3,2) = \ZZ$, see
Table \ref{tab:boundaryC}.

{\em  Higher-order phases originating from class A$^{\cal R}$ in two
dimensions.---} The bulk topological classification of a two-dimensional
topological insulator with an additional twofold rotation symmetry ${\cal R}$
is given by the subgroup sequence $\ZZ \subseteq \ZZ \subseteq \ZZ^2$, implying
a $\ZZ$ topological index for first-order phases and a $\ZZ$ topological index
classifying topological phases without boundary states. (Such phases are
essentially atomic-limit insulators.) The two topological invariants are the
Chern number $C$ and the number $N = n_o(0,0) - n_o(\pi,0) - n_o(0,\pi) +
n_o(\pi,\pi)$, where $n(k_x,k_y)$ is the number of occupied odd-parity bands at
the high-symmetry momentum
$(k_x,k_y)$.~\cite{shiozaki2014,trifunovic2017,po2017} Since $C$ and $N$ have
the same parity, the $\ZZ^2$ bulk topological index $(p,q)$ is defined setting
$p = (C-N)/2$, $q=(C+N)/2$. A common set of generators for the classifying
groups $\K[n](2,2)$, $n=0,1,2$, is given by the canonical-form Hamiltonians
$H_{(1,0)}$ and $H_{(1,-1)}$ of Eq.~(\ref{eq:HgeneratorsHCM}), where we have
chosen the representation $U_{\cal R} = \sigma_2$. The Hamiltonian $H_{(1,0)}$, which has the ${\cal R}$-breaking mass terms $\sigma_3 \tau_1$ and $\sigma_3 \tau_2$,
represents a first-order topological phase with a single anomalous chiral
boundary mode; it is a generator of $\K[0](2,2)$, but not of $\K[1](2,2)$ and
$\K[2](2,2)$. The Hamiltonian $H_{(1,-1)}$ represents an atomic insulator with
no boundary states; it is a generator of $\K[0](2,2)$, $\K[1](2,2)$, and
$\K[2](2,2)$.

The order-raising homomorphism $\omap$ maps a two-dimensional Hamiltonian with
a twofold rotation symmetry ${\cal R}$ to a three-dimensional Hamiltonian with
inversion symmetry ${\cal I}$. To see how the order-raising map $\omap$ maps
between the subgroup sequence of the two classes, we first consider its action
on the Hamiltonian $H_{(1,0)}$ considered above. Application of the
order-raising homomorphism $\omap$ to $H_{(1,0)}$ gives the three-dimensional
canonical-form Hamiltonian specified by Eq.~(\ref{eq:omapHgeneratorCM}), which
satisfies an inversion symmetry represented by $U_{\cal I} = \tau_3 \sigma_2$.
As in the previous example, one verifies that the canonical-form Hamiltonian
with
\begin{equation}
  \Gamma_0 = \sigma_2 \tau_1,\ \ \bm\Gamma=(\sigma_3 \tau_1,\sigma_1\tau_0)
\end{equation}
has topological indices $(p,q) = (1,1)$ and is separable. Accordingly,
$H_{(1,1)}$ is the generator of the subgroup $\ker \omap \subseteq K(2,2)$.
Since $H_{(1,-1)}$ differs from $H_{(2,0)}$ by a separable Hamiltonian, we
conclude that $H_{(2,0)}$ and $H_{(1,1)}$ must be mapped to the same
topological class under $\omap$. Since $H_{(1,1)}$ represents an atomic
insulator without boundary states, its image $\omap(H_{(2,0)})$ must also
represent an atomic insulator without boundary states. It follows that
$\K[1](3,3) = \omap(\K[0](2,2)) = \ZZ$ and $\K[2](3,3) = \omap(\K[2](2,2)) =
\K[3](3,3) = \omap(\K[3](3,3)) = 2 \ZZ$. As in the previous example, since
there are no first-order topological phases for the ten-fold way class A, we
thus arrive at the subgroup sequence $2 \ZZ \subseteq 2 \ZZ \subseteq \ZZ
\subseteq \ZZ$ for class A$^{\cal I}$ in three dimensions, consistent with
Table \ref{tab:subgroupC}.

We conclude this example with a discussion of the classification from the
boundary perspective. We first note that chiral hinge modes have a $\ZZ$
classification, whereby the $\ZZ$ topological index simply counts the number of
chiral hinge modes. Hence $\ks[0]{2}(3,2) = \ZZ$. The presence of the inversion
symmetry plays no role here, as ${\cal I}$ does not leave any hinges invariant.
An even number of hinge modes represents a separable phase, so that
$\ks[1]{2}(3,2) = 2 \ZZ$. It follows that $\ki{2}(3,2) =
\ks[0]{2}(3,2)/\ks[1]{2}(3,2) = \ZZ_2$. Finally, since no protected zero-energy
corner states are possible in the absence of an antisymmetry, the boundary
classification of third-order phases is entirely trivial, $\ks[0]{3}(3,2) =
\ks[1]{3}(3,2) = \ks[2]{3}(3,2) = \ki[2]{3}(3,2) = \ki{3}(3,2) = 0$, see Table
\ref{tab:boundaryC}.

\subsection{Separable higher-order topological phases}

As discussed in Sec.\ \ref{sec:omap}, the boundary classification considers
classifying groups $\ki[k]{n}$ for protected states at boundaries of dimension
$d-n$, whereby such boundary states are considered equivalent if they differ by
a lattice termination along a boundary of dimension $\le d-k$. The boundary
classifying groups $\ki[k]{n} = \ks[0]{n}/\ks[k]{n}$ are the quotient of the
group $\ks[0]{n}$ classifying all possible $(d-n)$-dimensional boundary states
and the ``decoration subgroup'' $\ks[k]{n}$, which classifies
$(d-n)$-dimensional boundary states that can be attributed to the combination
of a topological nontrivial boundary of dimension $\le d-k$ and a topologically
trivial bulk. 
%The anomalous boundary states are obtained by dividing out the group $\ks{n}\equiv\ks[n]{n}$, which classifies all $(d-n)$-dimensional boundary states that can be obtained by changing the lattice termination, without affecting the bulk band structure.

In the examples of the previous Subsection, all decoration groups $\ks[k]{n}$
with $k=2,\ldots,n$ are equal, so that effectively it is sufficient to consider
the (a priori) smallest decoration subgroup $\ks[n-1]{n}$, which classifies the
$(d-n)$-dimensional boundary states of a symmetry-compatible
$(d-n+1)$-dimensional topological phase located on the crystal boundary. This
$(d-n+1)$-dimensional topological phase is a separable phase, {\em i.e.}, it
consists of two halves, which are mapped onto each other by the crystalline
symmetry ${\cal S}$, see Fig.\ \ref{fig:Dk}. Although the group $\ks[n-1]{n}$
describes codimension-$n$ boundary states of the crystal as a whole, it
describes first-order boundary states of the separable $(d-n+1)$-dimensional
topological phase located on the crystal boundary. 

There are seven Shiozaki-Sato symmetry classes, for which the decoration
subgroups $\ks[k]{n}$ are {\em not} the same for all $k$. For those classes,
the integer $q$ in Eq.~(\ref{eq:kerq}) is \textit{finite} and one must consider
higher-order separable phases to obtain the boundary classification. Five of
these classes are relevant for the boundary classification of third-order
phases in three dimensions. These classes originate from two-dimensional
separable second-order phases in classes DIII$^{\mathcal{M}_{++}}$,
DIII$^{\mathcal{M}_{-+}}$, D$^{\mathcal{M}_+}$, AII$^{\mathcal{CM}_-}$ and
A$^{\mathcal{P}^+\mathcal{M}}$.  Two of these classes are relevant for the
boundary classification of fourth-order phases in four dimensions. These
classes can be traced to separable three-dimensional third-order phases in
classes CII$^{\mathcal{R}_{--}}$ and AIII$^{\mathcal{T}^+\mathcal{R}_+}$. We
now discuss two of these classes in detail.

{\em Class DIII$^{\mathcal{M}_{++}}$ in two dimensions.---} We choose the
representation $U_\mathcal{T}=\sigma_2$ and $U_\mathcal{P}=\tau_1$ for
time-reversal and particle-hole conjugation, respectively. Starting from the
ten-fold way canonical-form Hamiltonian specified by 
\begin{equation}
  	\Gamma_0= \sigma_0\tau_3,\,\bm \Gamma=(\sigma_1\tau_1,\sigma_0\tau_2),
	\label{eq:AZ_DIII}
\end{equation}
which has a single helical Majorana boundary mode, we construct the manifestly
separable Hamiltonian $\mbox{diag}\, [H(k_1,k_2), \sigma_1 H(-k_1,k_2)
\sigma_1]$, which has the canonical-form representation
\begin{align}
	\Gamma_0=\mu_0\sigma_0\tau_3,\,\bm \Gamma=(\mu_3\sigma_1\tau_1,\mu_0\sigma_0\tau_2), 
	\label{eq:DIIIMpp}
\end{align}
where the $\mu_j$ are Pauli matrices acting on a different degree of freedom
than the Pauli matrices $\sigma_j$ and $\tau_j$. The Hamiltonian
(\ref{eq:DIIIMpp}) satisfies the mirror symmetry $U_{\cal M} = \sigma_1
\mu_1$, which commutes with ${\cal T}$ and ${\cal P}$. It has a single ${\cal
M}$-breaking mass term $\rM = \mu_2 \sigma_1 \tau_1$, 
which makes it a second-order topological
superconductor with protected zero-energy states at the two mirror-symmetric
corners.

\begin{figure}
	\includegraphics[width=\columnwidth]{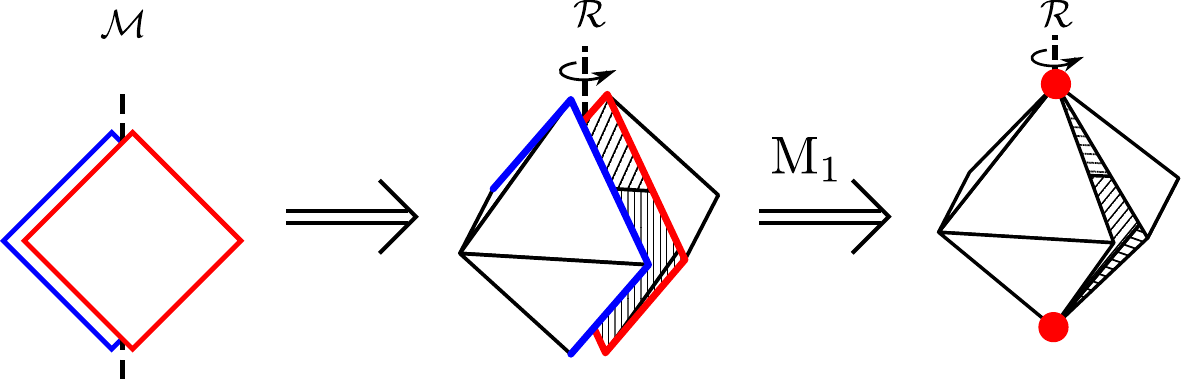}
	\caption{\label{fig:separable2nd} Two copies of a topologically
	nontrivial superconductor in class DIII that are related by mirror
	symmetry ${\cal M}$ form a separable topological phase (left). The
	separable phase can be deformed into a hollow shell, such that the
	mirror symmetry ${\cal M}$ becomes a twofold rotation symmetry ${\cal
	R}$ (center). Upon addition of a mass term along the ``seams'' of the
      shell, while preserving the global twofold rotation symmetry, any pairs
    of counterpropagating Majorana hinge modes can be gapped out, leaving
  behind Kramers pairs of zero-energy states at the two corners on the rotation
axis (right).}
\end{figure}

Two-dimensional separable Hamiltonians can be used to decorate a
three-dimensional bulk, as shown schematically in Fig.\ \ref{fig:separable2nd}.
Specifically, the separable two-dimensional system is deformed into a
two-dimensional ``shell'' embedded in three-dimensional space, where the mirror
symmetry ${\cal M}_{++}$ of the two-dimensional model becomes a twofold
rotation symmetry ${\cal R}_{++}$ in three dimensions. When localized near the
sample boundaries, the ${\cal M}$-breaking mass term $\mu_2\sigma_1\tau_1$ does
not obstruct this deformation procedure, while ensuring that any helical
boundary modes running along the two ``seams'' of the shell are gapped out
apart from the two corners on the rotation axis, see Fig.\
\ref{fig:separable2nd}. The corners on the rotation axis each host a Kramers
pair of zero-energy states.

We now discuss the consequences for the boundary classification of third-order
topological phases in class D$^{{\cal R}_{++}}$ in three dimensions. Since the
twofold rotation symmetry is a local symmetry at the two corners on the
rotation axis, zero-energy corner states have an extrinsic $\ks[0]{3} =
\ZZ_2^2$ classification, the two $\ZZ_2$ topological indices counting the
parities of the numbers of Kramers pairs of such corner states that are even or
odd under ${\cal R}$, respectively. Kramers pairs of zero-energy corner states
can be obtained by decoration with one-dimensional topological superconductors
along crystal hinges, but such a procedure always gives equal number of
even-parity and odd-parity states. Hence, corner states obtained from
decorations along hinges have classifying group $\ks[2]{3} = \ZZ_2$,
corresponding to the ``diagonal'' elements of $\ks[0]{3} = \ZZ_2^2$. To obtain
a {\em single} Kramers pair of zero-energy corner states one must decorate the
crystal boundary with a two-dimensional ``shell'' as constructed above. As a
result one finds $\ks[1]{3}(3,2) = \ks[0]{3} = \ZZ_2^2$. The resulting boundary
classifying groups then follow by taking quotients, $\ki{3}(3,2) = \ki[1]{3}(3,2) = 0$, $\ki[2]{3}(3,2) = \ZZ_2$, see Table \ref{tab:boundaryR}.

{\em Class A$^{{\cal P}^+{\cal M}}$ in two dimensions.---} A separable
second-order phase in class A$^{{\cal P}^+{\cal M}}$ can be constructed from two
copies of a quantum Hall system related to each other by particle-hole
conjugation. The minimal canonical-form Hamiltonian describing such a separable
phase is specified by
\begin{align}
 \Gamma_0=\tau_3\sigma_1,\,\bm\Gamma=(\tau_0\sigma_2,\tau_0\sigma_3).
\end{align}
This Hamiltonian has an antiunitary mirror antisymmetry represented by
$U_{{\cal PM}} = \tau_1$. It has two ${\cal PM}$-breaking mass terms
$\rM_1=\tau_1\sigma_1$ and $\rM_2=\tau_2\sigma_1$, which render it a
second-order topological phase with a single zero-energy state at
mirror-symmetric corners. Proceeding as before, we can use this Hamiltonian to
decorate a three-dimensional crystal, whereby the ${\cal P}^+{\cal M}$ mirror
antisymmetry of the two-dimensional Hamiltonian turns into a ${\cal P}^+{\cal
R}$ rotation antisymmetry in three dimensions. Again, the mass terms $\rM_1$
and $\rM_2$ do not obstruct the deformation procedure if they are localized
near the seams of the shell only, while the presence of such mass terms ensures
that the entire shell is gapped out, except for the corners at the twofold
rotation axis, which host zero-energy states. From the boundary perspective,
one has the extrinsic classifying group $\ks[0]{3} = \ZZ_2$, which counts the
parity of the number of such zero-energy corner states. One-dimensional
decorations along hinges can not result in any protected zero-energy corner states in
this symmetry class. However, zero-energy corner states can be obtained from a
two-dimensional shell-like decoration as described above. We conclude that
$\ks[1]{3}(3,2) = \ZZ_2$ and $\ks[2]{3}(3,2) = 0$, giving the boundary
classifying groups $\ki{3}(3,2) = \ki[1]{3}(3,2) = 0$ and 
$\ki[2]{3}(3,2) = \ZZ_2$, see Table \ref{tab:boundaryA}.

\subsection{Stacking construction}\label{sec:smap}

We now discuss three examples that compare the action of the order-raising homomorphism $\omap$ and the stacking homomorphism $\sigma$. 
The first example is the canonical-form Hamiltonian $H_d$ from Shiozaki-Sato class D$^{{\cal O}_-}$ with $d=0$
\begin{align}
  H_0(\mm)=\mm\sigma_1,
\end{align}
with $U_\mathcal{O}=\sigma_1$ and $U_\mathcal{P}=\sigma_3$. (Keeping the dependence on the parameter $\mm$ is necessary to allow for a meaningful distinction between topological phases in zero dimensions; For zero-dimensional Hamiltonians, one can uniquely assign topological invariants only to one-parameter family of Hamiltonians, but not to the Hamiltonian itself, see the discussion in Sec.\ \ref{sec:ShiozakiSato}.) The stacking procedure gives a one-dimensional Hamiltonian
$\sigma(H_0(m))$ in class D$^{{\cal M}_-}$. The upper-left block $H'_{1}$ of
Eq.~(\ref{eq:smap}) takes the form 
\begin{align}
  H'_{1}(k)=\mm(\sigma_1 \cos k+\sigma_2\sin k),
  \label{eq:DstrongF}
\end{align}
where we take $\hat\rho=\sigma_3$ in Eq.~(\ref{eq:smap}). The lower-right block
of the Hamiltonian $H_{1}$ of Eq.~(\ref{eq:smap}) is $k$-independent
and it does not carry any strong topological invariants. Since the above Hamiltonian is not in canonical form, we calculate
the topological invariant $N=n_o(\pi)-n_o(0)$ for the Hamiltonian
$H'_{1}(k_1)$, where $n_o(k)$ is the number of the odd-parity
negative-energy eigenvalues at the inversion symmetric momentum $k=0$, $\pi$. 
We find that $H'_{1}(k_1)$ has $N=1$ for $\mm<0$ and $N=-1$
for $\mm>0$, therefore the one-parameter family~(\ref{eq:DstrongF}) has
topological invariant $N=2$. The same is true for $\omap(H_0)$, as one 
verifies using the explicit representation of $\omap$ given in Table
\ref{tab:omap}.

For the second example, we consider a canonical-form Hamiltonian $H_d$ from
Shiozaki-Sato class D$^{{\cal M}_-}$ with $d=1$, specified by
\begin{align}
  \Gamma_0=\sigma_1,\,\bm\Gamma=(\sigma_2),
  \label{eq:Dstrong}
\end{align}
with $U_\mathcal{M}=\sigma_1$ and $U_\mathcal{P}=\sigma_3$. The above
Hamiltonian describes a one-dimensional $p$-wave superconductor with a
single Majorana mode localized at each end. The application of the stacking
construction to the one-dimensional superconductor with Hamiltonian $H_d$
specified by matrices~(\ref{eq:Dstrong}) gives the Hamiltonian $\sigma(H_d)$ with
$d=1$, and the upper-right block $H_{d+1}^\prime$ 

\begin{align}
	H_{2}^\prime=&\,(\mm+1-\cos k_1)(\sigma_1\cos k_2-\sigma_3\sin k_2)
  \nonumber \\ &\, +\sin k_1\sigma_2,
\end{align}
where we used $\cM=\sigma_2$, compare with Eq.\ (\ref{eq:smap}). Since this Hamiltonian is not of minimal
canonical form, its topological invariant cannot simply be determined by
counting the number of bands. The topological invariant $N$ in this class
takes integer values~\cite{turner2012,shiozaki2014}
\begin{align}
 N=&\,  n_o(\pi,\pi)-n_o(\pi,0)-n_o(0,\pi)+n_o(0,0),
\end{align}
where $n_o(\bm k)$ counts the number of odd-parity negative eigenvalues at the high-symmetry momentum $\bm k = (k_1,k_2)$.
Direct calculation gives that both $\sigma(H_d)$ and $\omap(H_d)$ have $N=2$
for $d=2$. 

Finally, we apply the stacking homomorphism $\sigma$ to a first-order 
non-separable
superconductor in class D$^{\mathcal{R}_-}$, with two-dimensional Hamiltonian
specified by
\begin{align}
  \Gamma_0=\sigma_1,\,\bm\Gamma=(\sigma_2,\sigma_3),
\end{align}
with $U_\mathcal{R}=\sigma_1$ and $U_\mathcal{P}=\sigma_3$. We choose
$\cM=\sigma_2$ and obtain the upper-left block $H'_{d+1}$ of
Eq.~(\ref{eq:smap}) as
\begin{align}
  H'_3(k_1,k_2,k_3) =&\, (\mm + 2 - \cos k_1)(\sigma_1 \cos k_3 - \sigma_3 \sin k_3) \nonumber \\ &\,
  +  \sigma_2 \sin k_1 - \sigma_1 \cos(k_1+k_3)
  \nonumber \\ &\,   + \sigma_3 \sin(k_1+k_3),
  \label{eq:Hstackl}
\end{align}
which has inversion symmetry with $U_{\cal I} = \sigma_1$, and particle-hole
antisymmetry $U_\mathcal{P}=\sigma_3$. For class D$^{\mathcal{I}_-}$ in
three-dimensions, similar to the previously considered classes, the topological
invariant $N$ can be evaluated via the inversion eigenvalues of the occupied
bands~\cite{turner2012,shiozaki2014}
\begin{align}
 N=&\, [ n_o(\pi,\pi,\pi)-n_o(\pi,\pi,0)-n_o(\pi,0,\pi)\nonumber\\
 &\, -n_o(0,\pi,\pi)+n_o(\pi,0,0)+n_o(0,\pi,0)\nonumber\\
 &\, +n_o(0,0,\pi)-n_o(0,0,0) ]/2,
\end{align}
We find that both $\omap(H_d)$ and $\sigma(H_d)$ have $N=1$ for $d=2$,
accordingly they are deformable into each other.

\subsection{Embedded topological phases}\label{sec:embedding}

It was pointed out recently~\cite{tuegel2018} that in the presence of
crystalline symmetries a lower-dimensional topological phase embedded in a 
higher-dimensional topologically trivial bulk --- a so-called ``embedded 
topological phases'' --- has the same boundary phenomenology as the 
higher-order topological phases considered in this work.
Can an embedded topological phase with Hamiltonian $H$ be deformed into
a higher-order topological crystalline phase with Hamitonian $\omap(H)$? 
The same question was
recently addressed by Matsugatani and Watanabe using a slightly
different approach.\cite{matsugatani2018}

\begin{figure}
	\includegraphics[width=0.9\columnwidth]{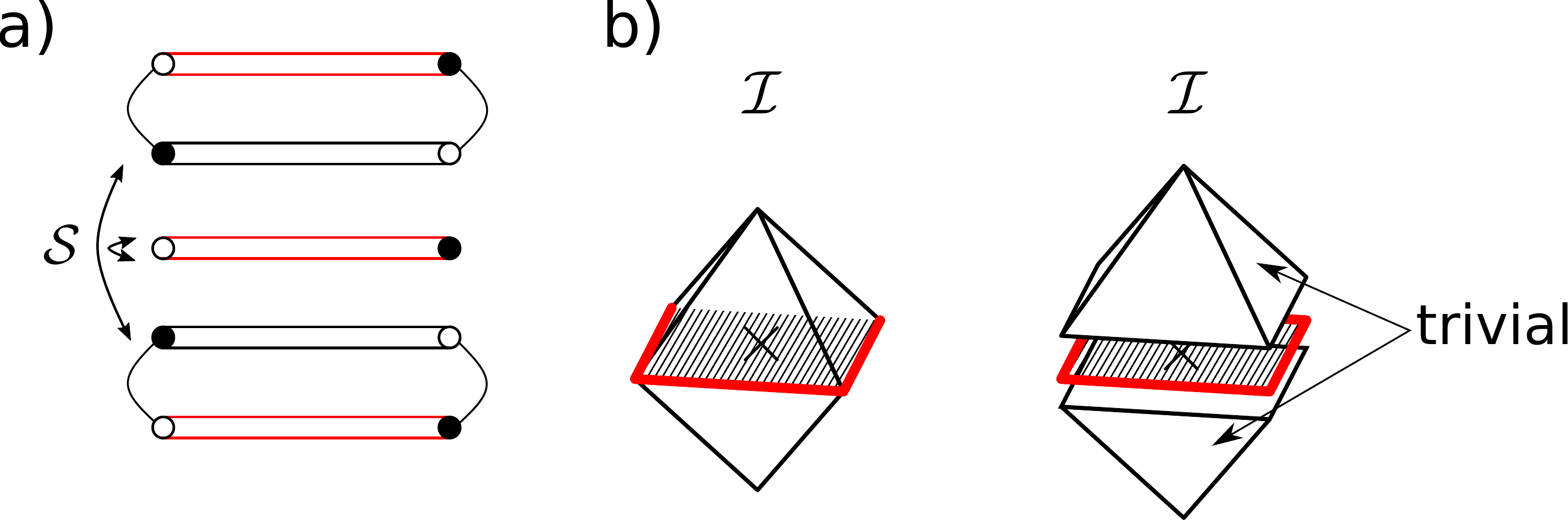}
	\caption{\label{fig:embedded} (a) Dimerization of a stacked-layer
	system that locally breaks $\mathcal{S}$-(anti)symmetry, while
	preserving $\mathcal{S}$ globally. (b) After breaking the local
	inversion symmetry, the upper and the lower halves of three-dimensional
	second-order Chern insulator can be trivialized, resulting in a
	embedded topological insulator.}
\end{figure}

Figure~\ref{fig:embedded}a shows that the stacked-layer system $\sigma(H)$ can
be deformed to the corresponding embedded topological system by breaking the
crystalline symmetry $\mathcal{S}$ locally by dimerizing the layers, while
globally preserving $\mathcal{S}$ symmetry. Using the conclusions of the
previous section we obtain that $\omap(H)\cong\sigma(H)$ is deformable to the
corresponding embedded system using a deformation that breaks $\mathcal{S}$
locally, while preserving it globally --- below we arrive at the same
conclusion using a different argument.

Assuming for concreteness that the Hamiltonian $\omap(H)$ is a
three-dimensional inversion-symmetric, second-order Chern insulator with a
single hinge mode at its boundary, Fig.~\ref{fig:embedded}b shows that its
halves above and below the hinge mode can be trivialized as the local symmetry
is broken, because $\omap(H)$ has only purely crystalline topological invariants.
This construction immediately enables us to conclude that $\omap(H)$ is
deformable to an embedded topological insulator.

\section{Order-lowering map $\bar\omap$}
\label{sec:omegainverse}

In this section we introduce an order-lowering map $\bar\omega$ that acts on a canonical-form Hamiltonian $H$ with anomalous boundary states of order $n>1$ and gives a Hamiltonian $\bar\omega(H)$ such that $\omap(\bar\omap(H))$ is continuously deformable to the original Hamiltonian $H$. Although the map $\bar\omap$ can be defined entirely algebraically, there is a simple geometric picture underlying the construction of $\bar \omap$, which we discuss first. 

To explain the geometrical picture underlying the construction of the order-lowering map $\bar\omap$, we recall the introduction of the sequence of manifolds (\ref{eq:invariantOmega}) in Sec.\ \ref{sec:omap} and the subsequent observation that an $n$th order topological phase with $n > 1$ is essentially trivial away from $\Omega_{d-n+1} \subseteq \Omega_{d-1}$. For a Hamiltonian in canonical form we choose $\Omega_{d-1}$ to be the intersection of the crystal with the hyperplane $x_1=0$. Since $x_1 \to -x_1$ under the crystalline (anti)symmetry ${\cal S}$, the hyperplane $x_1=0$ divides the crystal into two symmetry-related ``halves''. We smoothly deform the Hamiltonian by adding the term $m_1 {\rm M}_i \mbox{sign}\,(x_1)$, where ${\rm M}_i$ is a crystalline-symmetry breaking mass term and $m_1 > 0$. Although this extra term locally breaks the crystalline (anti)symmetry ${\cal S}$, ${\cal S}$ is preserved globally. Taking the limit $m_1 \to \infty$ amounts to a projection onto the hyperplane $x_1=0$, which gives the $(d-1)$-dimensional Hamiltonian
\begin{align}
  \bar\omap(H,{\rm M}_i)&=P_i H P_i,
  \label{eq:baromap}
\end{align}
where $P_i=(i{\rm M}_i{\rm \Gamma}_{1}+1)/2$ is a projection operator. The
Hamiltonian $\bar\omap(H,{\rm M}_i)$ obeys a crystalline (anti)symmetry with
$d_{\parallel}-1$ inverted dimensions, which is obtained by restricting ${\cal
S}$ to the plane $x_1=0$. If $H$ has anomalous boundary states of codimension
$n$, so has $\bar\omap(H,{\rm M}_i)$ anomalous boundary states of codimension
$n-1$. (The inverse is not true, see the discussion below!) A variant of this
construction was suggested by Matsugatani and Watanabe, who instead of adding a
crystalline-symmetry-breaking mass term proposed to symmetrically remove the
crystal on both sides of $\Omega_{d-1}$.\cite{matsugatani2018} 

We illustrate this procedure using the example of a three-dimensional
second-order Chern insulator with inversion symmetry ${\cal I}$. The
canonical-form Hamiltonian for this topological phase is specified by 
\begin{align}
	\Gamma_0&= \sigma_2\tau_3,\ \
        \bm \Gamma = (\sigma_0\tau_1,\sigma_3\tau_3,\sigma_1\tau_3),
  \label{eq:3dqhe_2}
\end{align}
where we use the representation $U_{\cal I} = \tau_3 \sigma_2$. There is one
crystalline-symmetry-breaking mass term ${\rm M}_1 = \tau_2$. Adding a term
$m_1 {\rm M}_1 \mbox{sign}\, (x_1)$ preserves the global inversion symmetry,
while manifestly opening up a spectral gap away from the plane $x_1=0$. If the
spectral gap from this additional mass term is much larger than the other
spectral gaps, we may perform a partial low-energy expansion along the $x_1$
direction, which yields the three-dimensional Hamiltonian
\begin{align}
  H=&\, (m+2-\cos k_1-\cos k_2)\Gamma_0+\sin k_2\Gamma_2+\sin k_3\Gamma_3\nonumber\\
  &\, \mbox{} +m_1 \rM_1 \mbox{sign}\,(x_1) -i\partial_{x_1}\Gamma_1.
\end{align}
Taking $m_1\gg |m|\sim1$, we obtain the two-dimensional effective Hamiltonian
\begin{align}
  H_2=P_1&\left[(m+2-\cos k_1-\cos k_2)\Gamma_0+\right.\nonumber\\
&\left.\sin k_2\Gamma_2+\sin k_3\Gamma_3\right]P_1,
\end{align}
which describes the Quantum Hall phase (compare with Eq.~(\ref{eq:qhe})).

\begin{figure}
	\includegraphics[width=0.8\columnwidth]{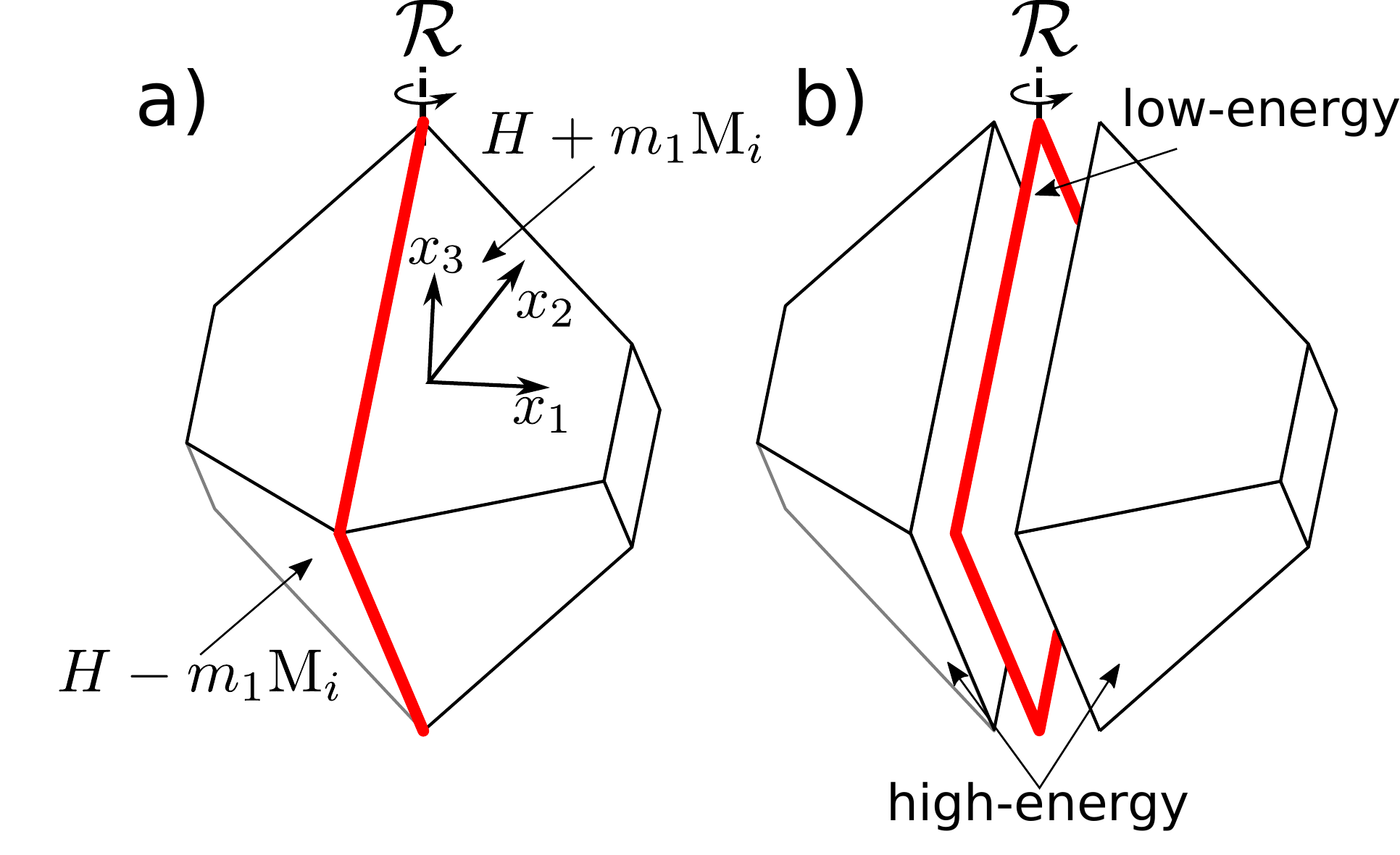}
	\caption{\label{fig:bar_omap} A three-dimensional, inversion
	symmetric, second-order phase can be mapped onto a two-dimensional
	first-order phase. The map is achieved by adding a 
        crystalline-symmetry-breaking mass term $\rM_i$ with opposite
        prefactors on both sides of a plane that symmetrically divides
        the crystal into two halves and that contains the anomalous
        boundary states (a). Such an additional mass term respects 
        the inversion rotation symmetry and manifestly trivializes the 
        three-dimensional bulk away from the boundary states.
        The low-energy theory consists of a first-order two-dimensional
        Hamiltonian (b).}
\end{figure}

The above construction of the inverse map $\bar\omap$ depends explicitly on the
form of the crystalline-symmetry-breaking mass term ${\rm M}_i$ and on the
choice of the hyperplane $\Omega_{d-1}$. Indeed, in general there is no unique
inverse map for the order-raising map $\omap$, because $\omap$ is a
homomorphism, not an isomorphism. However, the order-raising homomorphism
$\omap$ becomes an isomorphism when seen as a map between the quotient groups
$\K[n](d,d_\parallel)/\ker\omap$ and its image $\K[n+1](d+1,d_{\parallel}+1)$.
Regarded as a map between these two groups, the above-defined map $\bar\omap$
is a true inverse map that does not depend on the choice of ${\rm M}_i$. 

To illustrate the ``non-uniqueness'' of the order-lowering map, we consider the
trivial inversion-symmetric Hamiltonian $H$ in three dimensions specified by
\begin{align}
	\Gamma_0&=\mu_3\tau_0\sigma_1,\,\bm\Gamma=(\mu_1,\mu_3\tau_0\sigma_2,\mu_3\tau_0\sigma_3).
%	\label{eq:3dqhe}
\end{align}
Inversion symmetry is represented using $U_{\cal I}=\mu_3\tau_1\sigma_1$. The
above Hamiltonian has the crystalline-symmetry-breaking mass term ${\rm M}_1 =
\mu_2$, as well as a symmetry-preserving mass term ${\rm M} = \mu_2 \tau_3$. A
direct application of Eq.~(\ref{eq:baromap}) gives that $\bar\omap(H,{\rm
M}_1)$ is a canonical-form Hamiltonian specified by
\begin{align}
	\Gamma_0&=\tau_0\sigma_1,\,\bm\Gamma=(\tau_0\sigma_2,\tau_0\sigma_3),
	\label{eq:qhe2}
\end{align}
and with the twofold rotation symmetry $U_{\cal R}=\tau_1\sigma_1$. This
Hamiltonian has two chiral boundary modes, implying that it corresponds to a
nontrivial element of $K(2,2)$, despite the three-dimensional Hamiltonian $H$
being trivial (because of the existence of the mass term ${\rm M}$). However,
$\bar\omap(H,{\rm M}_1)$ is manifestly separable, so that it corresponds to the
trivial element of the group $K(2,2)/\ker\omap$. On the other hand, a different
choice of the crystalline-symmetry breaking mass term, ${\rm M}_1 = \mu_2
\tau_1$, gives a different inverse, $\bar\omap(H,\mu_2\tau_1)$, which is a
topologically trivial member of $K(2,2)$ since there is a nonzero
crystalline-symmetry-preserving mass term $P_1\rM P_1$.

\section{Bulk classifying groups from boundary and atomic-limits classification}\label{sec:classification}

The bulk-boundary correspondence (\ref{eq:bb}), together with the observation
that the classifying groups $\K[d]$ represent atomic-limit phases, can be used
to calculate the subgroup series (\ref{eq:subgroup}) of bulk classifying groups
(see Tables \ref{tab:subgroupC2d}--\ref{tab:subgroupR}) without using the
$K$-theory-based classification of Shiozaki and Sato.\cite{shiozaki2014} In
this section we explain how such a calculation proceeds. The advantage of such
an approach is that in principle it is not restricted for the order-two crystalline
(anti)symmetries, for which the $K$-theory approach of
Ref.~\onlinecite{shiozaki2014} was derived. 
%{\tt Is this correct? What about
%the use of $\bar\omap$? How to define $\bar\omap$ for other spatial
%symmetries than the order-two symmetries?}\color{red} Yes, at least for the
%situations in which one needs to use $\bar\omap$---let's discuss
%tomorrow.\color{black}

The input for the calculation described below are the ten-fold way classifying
groups in $d$ dimensions $K_{\rm TF}(d)$, their subgroups $K_{{\rm TF},{\cal
S}}(d,d_{\parallel})$ describing ten-fold way phases compatible with the
crystalline symmetry ${\cal S}$, as well as the groups $K(d,0)$ classifying
ten-fold way phases with an additional on-site symmetry, all of which can be
obtained by elementary means from the known ten-fold way classification. The
construction below requires an explicit realization of the order-raising
homomorphism $\omap$ and its inverse map $\bar\omap$, which can be done using
the geometric construction of the previous Section in the case of an order-two
symmetry.

\subsection{Calculation of $\K[d]$}

The group $\K[d]$ is obtained from the Shiozaki-Sato group $K(0,0)$ that
classifies topological zero-dimensional phases protected with an on-site ({\em
i.e.}, internal) symmetry. Following the discussion of Sec.\ \ref{sec:omap},
the calculation of $\K[d]$ amounts to the calculation of the subgroup
$\ker\omap^{d}\subseteq K(0,0)$. If $H_0$ is a generator of the corresponding
zero-dimensional ten-fold way phase, the manifestly separable Hamiltonian
$\mbox{diag}\, (H_0,{\cal S}H_0)$ is the generator of $\ker\omap$. Similarly,
$H_1$, a generator of one-dimensional ten-fold way phase, defines a
one-dimensional separable Hamiltonian $H_1^\prime=[H_1,{\cal S}H_1]$. If
$H_1^\prime$ is topologically non-trivial ({\em i.e.}, has no additional mass
terms) and has at least one crystalline-symmetry-breaking mass term $\rM_1$,
then the zero-dimensional Hamiltonian $\bar\omega(H_1^\prime,\rM_1)$ exists and
is an element the group $K(0,0)$. The kernel $\ker\omap^2$ is the subgroup of
$K(0,0)$ generated by $\ker\omap$ and $\bar\omega(H_1^\prime,\rM_1)$. (Note
that the subgroup $\ker\omap^2$ defined this way is uniquely defined, in spite
of the non-uniqueness of $\bar\omap$.) This procedure can be continued until
$\ker\omega^{d}$ is obtained, which then gives $\K[d]=K(0,0)/\ker\omap^d$.

\subsection{Classification anomalous boundary of codimension $n$}

The calculation of the anomalous boundary classification $\ki{n}$ starts from
the extrinsic boundary classification group $\ks[0]{n}$ and the decoration
subgroups $\ks[k]{n}$, see Sec.~\ref{sec:omap}. For $n=d_\parallel+1$, the
extrinsic classifying group $\ks[0]{n}=K(d-d_\parallel,0)$ is the
classification of $(d-d_\parallel)$-dimensional ten-fold way phases protected
with an on-site (internal) symmetry. The decoration groups are subgroups of the
extrinsic classifying group, $\ks[k]{n} = \ker \omega^{n-k} \subseteq
\ks[0]{n}$. Their calculations proceeds along the same line as in the
calculation of $\K[d]$ discussed above.

To calculate $\ki{n}$ for $n\le d_\parallel$, we need the classification of the
ten-fold way phases $K_{{\rm TF},{\cal S}}(d,d_{\parallel}) \subseteq K_{\rm
TF}(d)$ that are compatible with the crystalline (anti)symmetry $\cal S$. To
this end,  we notice that a canonical-form ten-fold way Hamiltonian $H$ has the
most symmetric form --- its low-energy expansion has the full rotational
symmetry, with generators that are pairwise products of the kinetic terms.
Accordingly, it is sufficient to check if a canonical-form Hamiltonian $H$ is
compatible with the (anti)symmetry ${\cal S}$, which is matter of algebra. Once
the extrinsic boundary classification group $\ks[0]{n}$ is known, the
decoration group $\ks[n-1]{n}$ is generated by the Hamiltonian
$\mbox{diag}\,(H_{d+1-n},{\cal S}H_{d+1-n})$, where $H_{d+1-n}$ is a generator
Hamiltonian of $(d+1-n)$-dimensional ten-fold way phase without additional
crystalline symmetry. Furthermore, if the Hamiltonian $H_{d+2-n}^\prime=
\mbox{diag}\, (H_{d+2-n},{\cal S}H_{d+2-n})$ is topologically non-trivial and
has one crystalline-symmetry-breaking mass term $\rM_1$, then the Hamiltonian
$\bar\omap(H_{d+2-n},\rM_1)$ is a $(d+1-n)$-dimensional Hamiltonian
representing an element of $K_{{\rm TF},{\cal S}}$. This way we obtain
$\ks[n-2]{n}$ as the subgroup spanned by $\ks[n-1]{n}$ and
$\bar\omap(H_{d+2-n},\rM_1)$. This procedure is repeated until the group
$\ks[1]{n}$ is reached.

\subsection{Calculating $\K[n]$ from the groups $\K[d]$ and $\ki{k}$}\label{sec:KnKd}

The bulk-boundary correspondence~(\ref{eq:bb}) can be rewritten in form of an exact sequence
\begin{align}
  0\rightarrow \K[n+1]\rightarrow&\K[n]\rightarrow \ki{n+1}\rightarrow 0,\ \
  n=0,1,\ldots,d.
	\label{eq:ext}
\end{align}
If the boundary classification group $\ki{n+1}$ is a free abelian group ({\em
i.e.}, it is of the form $\ZZ^k$), then the above exact sequence splits and has
the unique solution $\K[n]=\K[n+1]\oplus\ki{n+1}$. On the other hand, if the
boundary classification has a torsion subgroup ({\em e.g.}, it contains
$\ZZ_2$) then the above exact sequence has in general more than one solution
and knowledge of the groups $\ki{n+1}$ and $\K[n+1]$ is not sufficient to
determine $\K[n]$. Such an extension problem can be formulated as an algebraic
problem: Assume $g$ is a torsion element from $\ki{n+1}$, {\em i.e.}, $\oplus^k
g = e$ is the trivial element for some $k$. (For the order-two symmetries one
always has $k=2$.) Let $H$ be a bulk Hamiltonian that generates the state $g$
on its boundary. The Hamiltonian $H$ represents an element of $\K[n]$, but not
of $\K[n+1]$. The $k$-fold direct sum $\oplus^k H$ is either trivial or it is a
generator of the subgroup $\K[n+1] \subseteq \K[n]$. With this additional
knowledge, which can be determined by checking for additional mass terms of the
Hamiltonian $\oplus^k H$, the group $\K[n]$ can be uniquely determined from
$\K[n+1]$ and $\ki{n+1}$.

\subsection{Example: classification of inversion symmetric 3d topological insulator in class A}

As an example, we now show how the above procedure can be used to obtain the
full classifying subgroup sequence of an inversion-symmetric topological
insulator in three dimensions. We start with the group $\K[d]$ classifying
atomic-limit insulators. To obtain $\K[d]$, we need the classification $K(0,0)$
of zero-dimensional Hamiltonians protected by an on-site symmetry $\cal O$. Such
zero-dimensional Hamiltonians can be block-diagonalized where each block has
the classification of the zero-dimensional Hamiltonian of symmetry class A,
\begin{align}
  K(0,0)=\{(n_1,n_2),n_1,n_2\in\ZZ\}=\ZZ^2,
\end{align}
where the integers $n_1$ and $n_2$ count the numbers of occupied states of even
and odd parity, respectively. Furthermore, taking $H_0=\sigma_3$ as a generator
of the ten-fold way class A, the separable Hamiltonian $\mbox{diag}\,
(H_0,H_0)=\tau_0\sigma_3$ has on-site symmetry $U_{\cal O}=\tau_1$ and has
topological number $(1,1)$, so that
\begin{align}
  \ker\omap=\{(n,n),n\in\ZZ\}=\ZZ.
\end{align}
Further, since there are no topologically non-trivial one-dimensional ten-fold
way phases in class A, it follows that $\ker\omap^2 = \ker\omap$, and since the
separable nontrivial two-dimensional phase is a first-order phase (it has two
chiral modes on its boundary) one even has $\ker\omap^3=\ker\omap$. This way we
arrive at the classification of atomic limits
\begin{align}
  \K[3](3,3)&=K(0,0)/\ker\omap^3=\ZZ.
\end{align}

For the classification of anomalous boundary states, we can immediately
conclude that the groups $\ki{3}$ and $\ki{1}$ are trivial since there are no
topologically non-trivial ten-fold way phases in class A in one and three
dimensions and since inversion leaves no points on the boundary invariant. To
obtain the second-order boundary classification group $\ki{2}$ we first need to
check if a (two-dimensional) Chern insulator is compatible with two-fold
rotation symmetry. This is indeed the case, as demonstrated by the
canonical-form Hamiltonian
\begin{align}
  \Gamma_0=\sigma_1,\quad\bm\Gamma=(\sigma_2,\sigma_3),
\end{align}
which is compatible with the two-fold rotation symmetry represented by $U_{\cal
R}=\sigma_1$. Thus,
\begin{align}
  \ks[0]{2}=K_{{\rm TF},{\cal R}}(2,2)=\ZZ.
\end{align}
Moreover, since the diagonal sum $\mbox{diag}\,(H_2,{\cal R}H_2)$ has Chern
number equal to two ({\em i.e.}, it has two co-propagating chiral boundary
modes), one finds
\begin{align}
  \ks[1]{2}=2\ZZ\subseteq K_{\rm TF}(2,2),
\end{align}
so that 
\begin{equation}
  \ki{2} = \ZZ_2.
\end{equation}

To obtain the bulk subgroup sequence~(\ref{eq:subgroup}) from the
classification results $\K[3](3,3) = \ZZ$, $\ki{3} = 0$, $\ki{2} = \ZZ_2$, and
$\ki{1}=0$, we need to solve the exact sequence (\ref{eq:ext}) to obtain
$\K[2](3,3)$, $\K[1](3,3)$, and $\K[0](3,3)$. Here, only the case $n=1$ is
nontrivial,
\begin{align}
	0\rightarrow\mathbb{Z}\rightarrow \K[1](3,3)\rightarrow\mathbb{Z}_2\rightarrow 0.
\end{align}
In order to resolve the above sequence we need to know if the sum of two
second-order phases yields a trivial phase, in which case one has
$\K[1](3,3)=\mathbb{Z}_2\oplus\mathbb{Z}$, or if it yields a topologically
non-trivial atomic insulator, in which case the bulk classification is
$\K[1](3,3)=\mathbb{Z}$. We answer this question by considering the direct sum
of two second-order phases. Such a direct sum is given by the sum of two
second-order Chern insulators with co-propagating chiral modes along hinges.
Following the procedure outlined in Sec.\ \ref{sec:omegainverse}, these
three-dimensional inversion-symmetric second-order Chern insulators can be seen
as two-dimensional (first-order) Chern insulators embedded in an otherwise
trivial three-dimensional crystal. By rotating one of the two embedded Chern
insulators with respect to the other, the system may be deformed such that the
two chiral hinge modes are counter-propagating, corresponding to the
two-dimensional Hamiltonian
\begin{align}
	\Gamma_0=\tau_0\sigma_1,\,\bm\Gamma=(\tau_0\sigma_2,\tau_3\sigma_3).
\end{align}
The inversion symmetry of the three-dimensional host crystal becomes a twofold
rotation symmetry for the two-dimensional Chern insulator, represented by
$U_\mathcal{R}=\tau_0\sigma_1$. The above Hamiltonian has two mass terms
$\tau_1\sigma_3$ and $\tau_2\sigma_3$ which break the twofold rotation symmetry
${\cal R}$. We therefore conclude that the resulting
Hamiltonian is a topologically non-trivial atomic limit, so that $\K[1](3,3) =
\ZZ$. The resulting bulk subgroup sequence then reads
\begin{align}
  2\ZZ\subseteq2\ZZ\subseteq\ZZ\subseteq\ZZ,
\end{align}
which agrees with the corresponding entry in Table~\ref{tab:subgroupR}. (Note
that although we originally identified $\K[3]$ with $\ZZ$, this identification
must be reconsidered in view of the fact that $\K[3]=\K[2]$ is a subgroup of
$\K[1]=\K[0]$ and that $\K[1]/\K[2] =\ZZ_2$.)

\section{Conclusions}\label{sec:conclusions}

Topological crystalline insulators and superconductors have a more subtle
boundary signature of a nontrivial bulk topology than topological phases that
do not rely on the protection by a crystalline symmetry. Whereas the latter
have a bulk-boundary correspondence involving the crystal's full boundary,
such that a nontrivial topology is uniquely associated by a gapless boundary
state, topological crystalline
insulators or superconductors may also have protected gapless boundary states
of codimension larger than one or they may have no boundary signatures at all.
In this work we provide the formal framework for a classification of
topological crystalline phases that fully accounts for these different
scenarios and provide such a classification for topological crystalline phases
with an order-two crystalline symmetry or antisymmetry. This classification of
bulk crystalline phases consists of a subgroup sequence $\K[d] \subseteq
\K[d-1] \subseteq \ldots \subseteq K$, where the subgroup $\K[n]$ classifies
bulk phases with boundary states of codimension larger than $n$. 
The first group in the sequence, $\K[d]$
classifies those bulk phases for which no boundary signature exists. 
Our classification identifies such phases as $d$-dimensional ``stacks''
of disconnected (zero-dimensional) objects, {\em i.e.}, as an ``atomic-limit''
insulator. We
contrast the subgroup sequence describing the bulk topology with a
classification of codimension-$n$ boundary states. After dividing out
codimension-$n$ boundary states which can also be obtained as boundary
states of topological phases residing on the boundary --- {\em i.e.}, after dividing out boundary states that can be fully attributed to the
crystal's termination ---, the resulting anomalous boundary
classifying group $\ki{n} = \K[n-1]/\K[n]$. This is the bulk-boundary
correspondence for topological crystalline insulators.

A central role in our construction is played by an ``order-raising
homomorphism'', which simultaneously raises the dimensionality $d$ of the
Hamiltonian, the number of inverted dimensions $d_{\parallel}$ of the order-two
crystalline symmetry or antisymmetry, and the codimension $n$ of the boundary
states (if any). For order-two symmetries, we find that the layer stacking
construction used in
Refs.~\onlinecite{isobe2015,fulga2016,song2017b,khalaf2018b} is a realization
of the order-raising homomorphism. This is an important observation, since we
found the explicit expression for the order-raising homomorphism $\omap$ only
for order-two crystalline (anti)symmetries, whereas the layer stacking
construction can be applied to arbitrary crystalline (anti)symmetry, which
makes it a valuable tool in obtaining the anomalous boundary classification of
higher-order topological phases.~\cite{khalaf2018b} Finding anomalous boundary
classifying groups is simpler task~\cite{khalaf2018} compared to finding the
bulk classifying groups.~\cite{shiozaki2017}

Our algebraic approach allowed us to obtain a full classification of
higher-order phases of topological crystalline phases with an order-two
crystalline symmetry without having to analyze each symmetry class in detail.
This ``efficiency'' of the method also has a disadvantage, as it does not
provide explicit expressions for topological invariants. Nevertheless, since
our approach allows one to construct canonical-form Hamiltonians for the
generators of the bulk classifying groups, the combined knowledge of the full
classification and of the generators can be used to estimate to what extent
topology can be described by ``proxies'', such as the symmetry-based indicators
of Refs.\ \onlinecite{po2017,zhang2018}. (For example, a single $\ZZ_2$
indicator will provide a full description of a bulk topology if the classifying
group is $\ZZ_2$, but not if it is $\ZZ_2^2$ or $\ZZ$.) Examples of such a
procedure are given in Sec.\ \ref{sec:examples}. The relation of our algebraic
approach to other classification approaches, such as the momentum space
Atiyah-Hirzebruch spectral sequence~\cite{shiozaki2018} is still an open
question.

The first element in the group sequence, $\K[d]$, is zero for crystalline
(anti)symmetries with $d_{\parallel} < d$. These include mirror (anti)symmetry
in dimensions $d \ge 2$ and twofold rotation (anti)symmetry in dimensions $d
\ge 3$. On the other hand, for mirror symmetry with $d=1$, twofold rotation
symmetry with $d=2$, and inversion symmetry with $d=3$, $\K[d]$ may be nonzero.
A nonzero $\K[d]$ indicates that there topological phases with a nontrivial
bulk topology but without topologically protected boundary states. In some
cases, such topologically nontrivial phases without protected boundary states
are characterized by other observable signatures, such as the presence of
boundary charges (not states!),~\cite{lau2016,vanmiert2018} or quantized
electric~\cite{benalcazar2017,benalcazar2017b,shitade2018,gao2018,gao2018b} or
magnetic moments. Such signatures of a nontrivial bulk topology are not part
of the higher-order bulk boundary correspondence that we establish here, and it
is an interesting open problem how they can be incorporated.

We hope the results of this work not only bear theoretical relevance, but will
also help experimental efforts~\cite{peterson2018,serra-garcia2018,imhof2018}
to observe some of the rich boundary phenomenology of crystalline topological
insulators and superconductors in solid-state systems. Currently the list of
candidate materials for a second-order topological insulators consists of
tin-telluride,\cite{schindler2018} bismuth,\cite{schindler2018b} magnetically
doped bismuth selenide\cite{yue2018} and certain transition metal
dichalcogenides.\cite{wang2018c} Our complete classification may facilitate
the search for other material candidates. Finally, we note that in this work
only strong crystalline invariants were considered. We leave it for future
works the study of HOTPs originating from weak crystalline topological
invariants,\cite{ezawa2018c} which would further expand the list of potential
solid-state material candidates.

\acknowledgements

We thank Andrei Bernevig, Akira Furusaki, Max Geier, Eslam Khalaf, Felix von
Oppen, Ken Shiozaki, Simon Trebst, and Haruki Watanabe for stimulating
discussions. We acknowledge support by project A03 of the CRC-TR 183 and by the
priority programme SPP 1666 of the German Science Foundation (DFG). 

\appendix
\section{Bulk-boundary correspondence}\label{app:bb}

In this Appendix we show that a bulk-boundary correspondence for general topological crystalline phases follows immediately from the following statement, which has been proven for the phases with an order-two symmetry ${\cal S}$ in the main text:
\begin{itemize}
  \item The topological classification of atomic limits $K_A$ is the same as
the topological classification of the bulk phases with no boundary states $\K[d]$.
\end{itemize}
{}From the above statement it follows that every topologically non-trivial
non-atomic limit bulk, classified by $K/\K[d]$, needs to have anomalous
boundary states of a certain codimension. Thus, to prove the bulk-boundary
correspondence we need to show that for every anomalous boundary state of
codimension $n$ there is a bulk that is generating such boundary state. The
existence of such bulk readily follows since each boundary state of given
codimension can be generated by embedding a ten-fold way bulk phase (assuming
bulk-boundary correspondence for the ten-fold way phases) in a topologically
trivial bulk. (The stacking map $\sigma$ of Sec.~\ref{sec:omapstacking} or the
order-raising homomorphism $\omap$ of Sec.\ \ref{sec:omapraising} are nothing
but implementations of such an embedding procedure.)

To prove the statement that $K_A=\K[d]$ for an arbitrary symmetry group, it is enough to show
that $\K[d]\subseteq K_A$, since atomic limits have no boundary states, so that automatically $K_A\subseteq\K[d]$. Since a bulk Hamiltonian with topological invariants from
$\K[d]$ has no boundary states, we can perform the cutting procedure of
Sec.~\ref{sec:omegainverse} of the main text, to reduce the system to a phase that
consists of a zero-dimensional topologically non-trivial Hamiltonian embedded
in a topologically trivial bulk---clearly, an atomic limit phase, thus
$K_A=\K[d]$.

\section{Dimension-raising isomorphisms}\label{app:dmaps}

The construction of the order-raising homomorphism $\omap$ requires us to
include ``defect Hamiltonians'' $H(\vk,\vr)$ into our classification. Defect
Hamiltonians were introduced for the ten-fold way classes by Teo and
Kane,~\cite{teo2010} and considered for crystalline topological phases with an
order-two symmetry or antisymmetry by Shiozaki and Sato.~\cite{shiozaki2014} Defect Hamiltonians with a one-dimensional defect variable $\varphi$ appear in the algebraic construction of the order-raising map, see Sec.\ \ref{sec:omapraising}. Following Ref.\ \onlinecite{shiozaki2014}, in this Section we introduce defect Hamiltonians of canonical form in a slightly more general setting, using defect variables of arbitrary dimension $\vr$. We proceed with a discussion the associated dimension-raising isomorphisms $\kappa_{\parallel}$, $\kappa_{\perp}$, $\rho_{\parallel}$, and $\rho_{\perp}$, as well as the boundary map $\delta$.

{\em Defect Hamiltonians.---} We consider families of Hamiltonians $H(\bm k,\bm \varphi,\mm)$, where the
$D$-dimensional ``defect coordinate'' $\bm \varphi=(\bm \varphi_\parallel,\bm \varphi_\perp)$ is
defined on a torus.~\footnote{In this work we assume $\bm \varphi$ to be defined on a
torus rather than on a sphere around the defect, as in
Refs.~\onlinecite{teo2010,shiozaki2014}. Defining $\bm \varphi$ to be on a torus
introduces weak invariants which are inessential for the present work, since we
consider strong invariants only.} Denoting the number of ``inverted''
defect coordinates as $D_{\parallel}$, the family of Hamiltonians $H(\bm k,\bm
\varphi,\mm)$ transforms under unitary order-two (anti)symmetry $\mathcal{S}$ as
\begin{align}
  H(\bm k,\bm \varphi,\mm)&= {\cal S} H(\vk,\vr,\mm) \nonumber \\ &\equiv
  \sigma_{\cal S} U_{\cal S} H(\mathcal{S}\bm k,\mathcal{S}\vr,\mm) U_{\cal S}^{-1},\\
  \mathcal{S}\bm k&=(-\bm k_\parallel,\bm k_\perp),\,\mathcal{S}\vr=(-\vr_\parallel,\vr_\perp)\nonumber,
\end{align}
where $\bm k_\parallel=(k_1,\dots,k_{d_\parallel})$, $\bm
k_\perp=(k_{d_\parallel+1},\dots,k_{d})$,
$\vr_\parallel=(\varphi_1,\dots,\varphi_{D_\parallel})$,
$\vr_\perp=(\varphi_{D_\parallel+1},\dots,\varphi_D)$, and we used the notation of
Sec.~\ref{sec:ShiozakiSato}.
Similarly, antiunitary symmetry and antisymmetry operations are
represented by unitary matrices $U_{\cal S}$,
\begin{align}
  H(\bm k,\vr,\mm) &= {\cal S} H(\vk,\vr,\mm) \nonumber \\ &\equiv
  \sigma_{\cal S} U_{\cal S} H^*(-\mathcal{S}\bm k,\mathcal{S}\vr,\mm) U_{\cal S}^{-1}.
  \label{eq:krA}
\end{align}

{\em Dimension-raising isomorphisms.---}
The dimension-raising isomorphisms $\kappa_{\parallel}$ and $\kappa_{\perp}$, which
increase the dimension $d$ by one, were introduced in the main text. For defect
Hamiltonians, two additional dimension-raising isomorphisms can be defined: The
isomorphism $\rho_\parallel$, which increases by one both the defect dimension
$D$ and the number of inverted defect coordinates $D_\parallel$, and the map
$\rho_\perp$, which changes only the defect dimension $D$, such
that~\cite{shiozaki2014}
\begin{align}
  \K[0](s,t|d,d_{\parallel},D,D_\parallel) &\overset{\rho_\parallel}=\K[0](s-1,t-1|d,d_{\parallel},D+1,D_\parallel+1),\nonumber\\
  &\overset{\rho_\perp}=\K[0](s-1,t|d,d_{\parallel},D+1,D_\parallel),
\end{align}
for complex and real ten-fold way classes with a crystalline unitary
order-two (anti)symmetry, and
\begin{align}
  \K[0](s|d,d_{\parallel},D,D_\parallel) &\overset{\rho_\parallel}= \K[0](s+1|d,d_{\parallel},D+1,D_\parallel+1) \nonumber \\
  &\overset{\rho_\perp}= \K[0](s-1|d,d_{\parallel},D+1,D_\parallel),
\end{align}
for complex ten-fold way classes with a crystalline antiunitary order-two
(anti)symmetry. 

The action of these isomorphisms is defined~\cite{shiozaki2015,shiozaki2016}
analogously to Eq.~(\ref{eq:omapH}),
\begin{align}
	\label{eq:kappaH}
	\kappa(H(\bm k,\vr,\mm))&=\, H_\kappa(\bm k,\vr,\mm+1-\cos k^\prime)+\Gamma_\kappa\sin k^\prime. \\
	\rho(H(\bm k,\vr,\mm)) &=\, H_\rho(\bm k,\vr,\mm+1-\cos
	\varphi^\prime)+\Gamma_\rho\sin \varphi^\prime.
        \label{eq:rhoH}
\end{align}
If the defect coordinate $\varphi^\prime$ is flipped under the resulting
crystalline symmetry then the $(d+1)$-dimensional defect coordinate takes the form
$(\varphi^\prime,\vr)$, otherwise it is $(\vr,\varphi^\prime)$. The form of the mapped
Hamiltonian is listed in Tables~\ref{tab:rhoAZC}-\ref{tab:kappaAZ}. The
additional unitary (anti)symmetry ${\cal S}$ transforms as summarized in
Tables~\ref{tab:SSdim_raising} and \ref{tab:SSdim_raisingC}. 

\begin{table}
\begin{tabular*}{\columnwidth}{c @{\extracolsep{\fill}} cccc}
\hline\hline 
TF class   & $(H_{\rho},\Gamma_{\rho})$           & $\rho(U_\mathcal{C})$   \\ \hline
A          & $(\tau_3H,\tau_2)$  & $\tau_1$     \\
AIII       & $(H,U_\mathcal{C})$ & - \\
\hline\hline
\end{tabular*}
\caption{The mapped Hamiltonian~(\ref{eq:rhoH}) and the representation of the
chiral symmetry $\mathcal{C}$ under application of the dimension-raising
isomorphism $\rho$ for the complex ten-fold way classes.}
\label{tab:rhoAZC}
\end{table}
\begin{table}
\begin{tabular*}{\columnwidth}{c @{\extracolsep{\fill}} ccccc}
\hline\hline 
TF classes & $(H_{\rho},\Gamma_{\rho})$           & $\rho(U_\mathcal{T})$ & $\rho(U_\mathcal{P})$   \\ \hline
AI, AII    & $(\tau_3H,\tau_2)$  & $\tau_3U_\mathcal{T}$ & $\tau_2U_\mathcal{T}$     \\
BDI, CII   & $(H,U_\mathcal{C})$ & $U_\mathcal{T}$       & - \\
D, C       & $(\tau_3H,\tau_2)$  & $\tau_1U_\mathcal{P}$ & $\tau_0U_\mathcal{P}$ \\
DIII, CI   & $(H,U_\mathcal{C})$ & -                     & $U_\mathcal{P}$ \\
\hline\hline
\end{tabular*}
\caption{The mapped Hamiltonian~(\ref{eq:rhoH}) and the representation of
the antiunitary (anti)symmetries $\mathcal{T}$ and $\mathcal{P}$ under application
of the dimension-raising isomorphism $\rho$ for the real ten-fold way
classes.}
\label{tab:rhoAZ}
\end{table}

\begin{table}
\begin{tabular*}{\columnwidth}{c @{\extracolsep{\fill}} cc}
\hline\hline 
TF class & $(H_{\kappa},\Gamma_{\kappa})$         & $\kappa(U_\mathcal{C})$   \\ \hline
A        & $(\tau_3H,\tau_2)$  & $\tau_1$     \\
AIII     & $(H,U_\mathcal{C})$ & - \\
\hline\hline
\end{tabular*}
\caption{The mapped Hamiltonian~(\ref{eq:kappaH}) and the representation of the
chiral symmetry $\mathcal{C}$ under application of the dimension-raising
isomorphism $\kappa$ for the complex ten-fold way classes.}
\label{tab:kappaAZC}
\end{table}
\begin{table}
\begin{tabular*}{\columnwidth}{c @{\extracolsep{\fill}} ccc}
\hline\hline 
TF classes & $(H_{\kappa},\Gamma_{\kappa})$          & $\kappa(U_\mathcal{T})$ & $\kappa(U_\mathcal{P})$   \\ \hline
AI, AII    & $(\tau_3H,\tau_2)$   & $\tau_0U_\mathcal{T}$   & $\tau_1U_\mathcal{T}$  \\
BDI, CII   & $(H,U_\mathcal{C})$  & -                       & $U_\mathcal{P}$       \\
D, C       & $(\tau_3H,\tau_2)$   & $\tau_2U_\mathcal{P}$   & $\tau_3U_\mathcal{P}$ \\
DIII, CI   & $(H,U_\mathcal{C})$  & $U_\mathcal{T}$         & -                     \\
\hline\hline
\end{tabular*}
\caption{The mapped Hamiltonian~(\ref{eq:kappaH}) and the representation the
antiunitary (anti)symmetries $\mathcal{T}$ and $\mathcal{P}$ under application
of the dimension-raising isomorphism $\kappa$ for the real ten-fold way
classes.}
\label{tab:kappaAZ}
\end{table}

\begin{table}
\begin{tabular*}{\columnwidth}{c @{\extracolsep{\fill}} cccc}
\hline\hline 
TF classes & $\mathcal{S}$ symmetry &
$\kappa_\parallel,\rho_\parallel(U_\mathcal{S})$ &
$\kappa_\perp,\rho_\perp(U_\mathcal{S})$ \\
\hline
A          & $\mathcal{S}$            & $\tau_3U_\mathcal{S}$     & $\tau_0U_\mathcal{S}$              \\
AIII       & ${\mathcal{S}_+}$        & $U_\mathcal{C}U_\mathcal{S}$     & $U_\mathcal{S}$ \\
A          & $\mathcal{CS}$           & $\tau_1U_\mathcal{S}$     & $\tau_2U_\mathcal{S}$              \\
AIII       & ${\mathcal{S}_-}$        & $U_\mathcal{S}$     & $iU_\mathcal{C}U_\mathcal{S}$ \\
\hline\hline
\end{tabular*}
\caption{The mapped representation of the unitary order-two (anti)symmetry
  $\mathcal{S}$ under application of the dimension-raising isomorphisms
  $\kappa_\parallel$, $\kappa_\perp$, $\rho_\parallel$ and $\rho_\perp$ for the
  complex ten-fold way classes.  The mapping of the Hamiltonian and the ten-fold way symmetries is given in
  Table~\ref{tab:kappaAZC}.}
\label{tab:SSdim_raisingC}
\end{table}
\begin{table}
\begin{tabular*}{\columnwidth}{c @{\extracolsep{\fill}} ccc}
\hline\hline 
TF classes         & $\mathcal{S}$ symmetry                &
$\kappa_\parallel,\rho_\parallel(U_\mathcal{S})$    &
$\kappa_\perp,\rho_\perp(U_\mathcal{S})$ \\
\hline
AI, AII, D, C      & ${\mathcal{S}_+,\mathcal{S}_-}$       & $\tau_3U_\mathcal{S}$              & $\tau_0U_\mathcal{S}$              \\
AI, AII, D, C      & ${\mathcal{CS}_+,\mathcal{CS}_-}$     & $\tau_1U_\mathcal{S}$              & $\tau_2U_\mathcal{S}$              \\
BDI, CII, DIII, CI & ${\mathcal{S}_{++},\mathcal{S}_{--}}$ & $U_\mathcal{C}U_\mathcal{S}$ & $U_\mathcal{S}$              \\
BDI, CII, DIII, CI & ${\mathcal{S}_{+-},\mathcal{S}_{-+}}$ & $U_\mathcal{S}$              & $U_\mathcal{C}U_\mathcal{S}$ \\
\hline\hline
\end{tabular*}
\caption{The mapped representation of the unitary order-two (anti)symmetry
  $\mathcal{S}$ under application of the dimension-raising isomorphisms
  $\kappa_\parallel$, $\kappa_\perp$, $\rho_\parallel$ and $\rho_\perp$ for the
  real ten-fold way classes. The mapping of the Hamiltonian and the ten-fold way
  symmetries is given in Table~\ref{tab:kappaAZ}.}
\label{tab:SSdim_raising}
\end{table}

As explained in Ref.~\onlinecite{shiozaki2014}, the introduction of defect
dimensions can be used to define the dimension-raising isomorphisms for
ten-fold way classes with an order-two antiunitary (anti)symmetry, making
use of the fact that the complex Shiozaki-Sato classes with antiunitary
(anti)symmetry are isomorphic to real ten-fold way classes. Such an
isomorphism is most easily constructed~\cite{shiozaki2014} by noticing that
renaming the coordinates $(\bm k_\perp,\vr_\parallel)\rightarrow\tilde{\bm
k}$ and $(\vr_\perp,\bm k_\parallel)\rightarrow\tilde{\vr}$ gives a Hamiltonian
in the corresponding ten-fold way class, see the transformation
law~(\ref{eq:krA}). Such a transformation defines the isomorphism
\begin{align}
	K(s|d,d_{\parallel},D,D_\parallel)&=K_{\rm TF}(s\vert d-d_\parallel+D_\parallel,D-D_\parallel+d).
  \label{eq:AtoR}
\end{align}
Correspondingly, for the complex Shiozaki-Sato classes with an antiunitary
symmetry the dimension-raising isomorphisms are defined by first applying the above
isomorphism to a real ten-fold way class, then using Teo and Kane's
dimension-raising isomorphisms $\kappa$ and $\rho$,~\cite{teo2010} and then using the
inverse of the isomorphism (\ref{eq:AtoR}). From this procedure it is readily
seen that for complex Shiozaki-Sato classes with an antiunitary order-two
symmetry one has, up to the isomorphism~(\ref{eq:AtoR}),
\begin{align}
  \rho_\parallel=\kappa_\perp=\kappa,\nonumber\\
  \kappa_\parallel=\rho_\perp=\rho.
  \label{eq:kappaA}
\end{align}

{\em Boundary homomorphism $\delta$.---}
From the definition~(\ref{eq:delta}) of the boundary map $\delta$, we can write the action of the homomorphism $\delta\circ\rho$
\begin{align}
  \delta\circ\rho(H(\vk,\vr,\mm))&=H_\rho(\vk,\vr,\mm).
\end{align}
Together with the definition of $H_{\rho}$ given above this fully specified the product $\delta \circ \rho$.

\section{Properties of the order-raising homomorphism $\omap$}\label{app:omap}

The explicit expression (\ref{eq:omap}) for the order-raising homomorphism $\omap$ arises naturally in the context of an exact
sequence containing ten-fold way classifying groups $K_{\rm TF}$ and Shiozaki-Sato groups $K$. This exact sequence is a
variant of an exact sequence considered by Turner~\textit{et
al.}~\cite{turner2012} and by us~\cite{trifunovic2017} for the classification
of inversion-symmetric and mirror-symmetric topological insulators and
superconductors,
\begin{align}
  K(d,d_\parallel,D,D_\parallel-1) &\overset{i}{\rightarrow}\,
  K_{\rm TF}(d,D) \nonumber \\ &\overset{c_{\cal S}}{\rightarrow}\, K(d,d_\parallel,D,D_\parallel)\nonumber\\
  \label{eq:exactsec}
  &\overset{\omap}{\rightarrow}\,K(d+1,d_\parallel+1,D,D_\parallel)
  \nonumber\\ &\overset{i}{\rightarrow}\, K_{\rm TF}(d+1,D).
\end{align}
Here $i$ is the natural homomorphism, in the literature~\cite{shiozaki2016}
also called a ``symmetry forgetting functor'', that identifies a member of
Shiozaki-Sato group as a member of the corresponding ten-fold way group,
and $c_{\cal S}$ is the homomorphism that constructs
separable Hamiltonians
\begin{align}
	c_{\cal S}[H]&=
	\begin{pmatrix}
		H & 0\\
		0 & \mathcal{S}H
	\end{pmatrix},
	\label{eq:cmap}
\end{align}
where the symmetry ${\cal S}$ has $d_{\parallel}$ inverted spatial dimensions and $D_{\parallel}$ inverted defect dimensions. The homomorphism $\omap$ is defined by Eq.\ (\ref{eq:omap}), where --- using the more general definitions of the maps $\rho_{\parallel}$, $\delta$, and $\kappa_{\parallel}$ given in the previous appendix --- the homomorphism $\omap$ appearing here is a map between defect Hamiltonians. 

We first show that exactness of the sequence (\ref{eq:exactsec}) leads to the three properties of the order-raising homomorphism listed in Sec.\ \ref{sec:omap}. Exactness of the sequence (\ref{eq:exactsec}) will then be shown at the end of this appendix.

{\em Properties 1--3 of the order-raising homomorphism.---}
The maps in the exact sequence~(\ref{eq:exactsec}) all preserve the group
operations (\textit{i.e.}, they are homomorphisms), and the image of every map is the
same as the kernel of the subsequent one. Thus exactness at $K_{\rm TF}(d,D)$
immediately gives that $\omap(H)$ is trivial if and only if $H$ is separable,
\textit{i.e.}, $H\in c_{\cal S}[K]$.  This proves the first property
of the order-raising homomorphism $\omap$ listed in Sec.~\ref{sec:omap} of the
main text.

To prove the second property, we first notice that the natural homomorphism $i$
commutes with the dimension-raising isomorphisms, since the latter act the same way on
the Hamiltonians from the ten-fold way and Shiozaki-Sato classes, see
Sec.~\ref{sec:examples},
\begin{align}
  i\circ\chi_\parallel&=\chi\circ i,\nonumber\\
  i\circ\chi_\perp&=\chi\circ i,
  \label{eq:commute}
\end{align}
with $\chi=\rho,\kappa$. Exactness of the sequence~(\ref{eq:exactsec}) at
$K(d,d_\parallel,D,D_\parallel)$ and $K_{\rm TF}(d,D)$ 
yields the isomorphism
\begin{align}
  \ker\omap&=\, \img c_{\cal S} \nonumber \\ &=\,
  K_{\rm TF}(d,D)/K_{{\rm TF},{\cal S}}(d,d_\parallel,D,D_\parallel-1),
	\label{eq:KdKs}
\end{align}
with $\ker\omap\subseteq\K[0](d,d_\parallel,D,D_\parallel)$ and
$i[K]=K_{{\rm TF},{\cal S}}=\K[0]/\K[1]$. Due to commutation relations~(\ref{eq:commute}), we
conclude that the dimension-raising isomorphisms preserve the subgroups $K_{{\rm TF},{\cal S}}$,
and from Eq.~(\ref{eq:KdKs}) the same applies to the subgroups $\ker\omap$.
Furthermore, the exactness at $\K[0](d,d_\parallel,D,D_\parallel)$ gives,

\begin{align}
  \img\omap=\ker i,
\end{align}
thus the dimension-raising isomorphisms also preserve the subgroups $\img\omap$. We
conclude that the homomorphism $\omap$ commutes with the dimension-raising isomorphisms
up to an automorphism of $\img\omap$. Since the groups $\img\omap=\K[1]$ are at
most $\ZZ$ and $\mathrm{Aut}(\ZZ)=\ZZ_2$, the mentioned automorphism changes at
most the sign of the topological invariants. Such sign change is inessential 
and therefore the dimension-raising isomorphisms preserve the bulk classifying groups
of HOTPs $\K[n]$. This proves the second property of the order-raising
homomorphism $\omap$.

We prove the third property using the explicit expression~(\ref{eq:omap}) for
the $\omap$ homomorphism. Firstly, by comparing the dimension of a nontrivial
$\omap^n(H)$, where $H$ is a minimal canonical
model,~\cite{chiu2014,morimoto2013,shiozaki2014} a representative of
$\K[0]/\K[1]\ker\omap$, to the minimal dimension of the representative of
$\K[n]/\K[n+1]$ we find that $\omap^n(H)$ is also a minimal canonical model. We
therefore conclude that for a minimal canonical model $H$, representative of
either $\K[n]/\K[n+1]$ or $\K[0]/\K[1]\ker\omap$, $\omap(H)$ is also a minimal
canonical model.

Next we show that under the assumption that a minimal canonical model with
$n-1$ crystalline-symmetry-breaking mass terms $H^{(n)}$ (for a fixed $n$) is a
representative of $\K[n-1]/\K[n]$ for $n>1$ and $\K[0]/\K[1]\ker\omap$ for
$n=1$, $\omap(H)$ has $n$ boundary mass terms.  Since under these assumptions,
$\omap(H)$ is a minimal canonical model, the number of its
$\mathcal{S}$-symmetry breaking mass terms does not change under the continuous
Hamiltonian deformations. It is now a matter of simple algebra to show that
there are no additional $\mathcal{S}$-symmetry breaking mass terms beyond the
ones given in Tables~\ref{tab:omap}-\ref{tab:omapA}; We illustrate how this
proof works for classes BDI$^{\mathcal{S}_{++}}$, BDI$^{\mathcal{S}_{--}}$,
CII$^{\mathcal{S}_{++}}$ and CII$^{\mathcal{S}_{--}}$. Proofs for the other symmetry classes are analogous. In order to satisfy
chiral symmetry, the additional mass term needs to be of the form
$\tau_3\rM_{n+1}$ which has to anticommute with $\rM_n=\tau_3U_\mathcal{C}$.
Thus $\rM_{n+1}$ anticommutes with $U_\mathcal{C}$, which makes it a valid
$\mathcal{S}$-symmetry breaking mass term of the $H^{(n)}$ Hamiltonian,
contradicting the initial assumption on the number of the crystalline symmetry
breaking mass terms.  This proves the first statement of the third property of the $\omap$ homomorphism. The second statement of the third property directly follows from exactneses of the sequence (\ref{eq:exactsec}) at $K(d+1,d_{\parallel}+1,D,D_{\parallel})$ since $\ker i$ consists of those Hamiltonians that have at least one crystalline-symmetry-breaking mass term.

{\em Exactness of the sequence (\ref{eq:exactsec}).---}
The exactness of the sequence (\ref{eq:exactsec}) can be proven as follows:
Consider a one-parameter family $H(\varphi)$ of a Hamiltonian $H$ from
$\K[0](d,d_\parallel,D,D_\parallel-1)$, with the order-two symmetry
(antisymmetry) $U_\mathcal{S}$ acting locally as
$\sigma_\mathcal{S}U_\mathcal{S}H(\varphi)U_\mathcal{S}^\dagger=H(\varphi)$.
This one-parameter family is mapped via the homomorphism $c_{\cal S}\circ i$ to $H^\prime$,
\begin{align}
	H^\prime(\varphi)=H(\varphi)\oplus H(-\varphi),
	\label{eq:Hloop}
\end{align}
that is the $\mathcal{S}$ symmetry now acts non-locally on the coordinate
$\varphi$. The loop~(\ref{eq:Hloop}) is a topologically trivial loop.
Alternatively, each topologically trivial loop can be deformed to the above
form with an arbitrary $H(\varphi)$ proving that $\img i=\ker
c_{\cal S}$.

\begin{figure}
	\includegraphics[width=0.8\columnwidth]{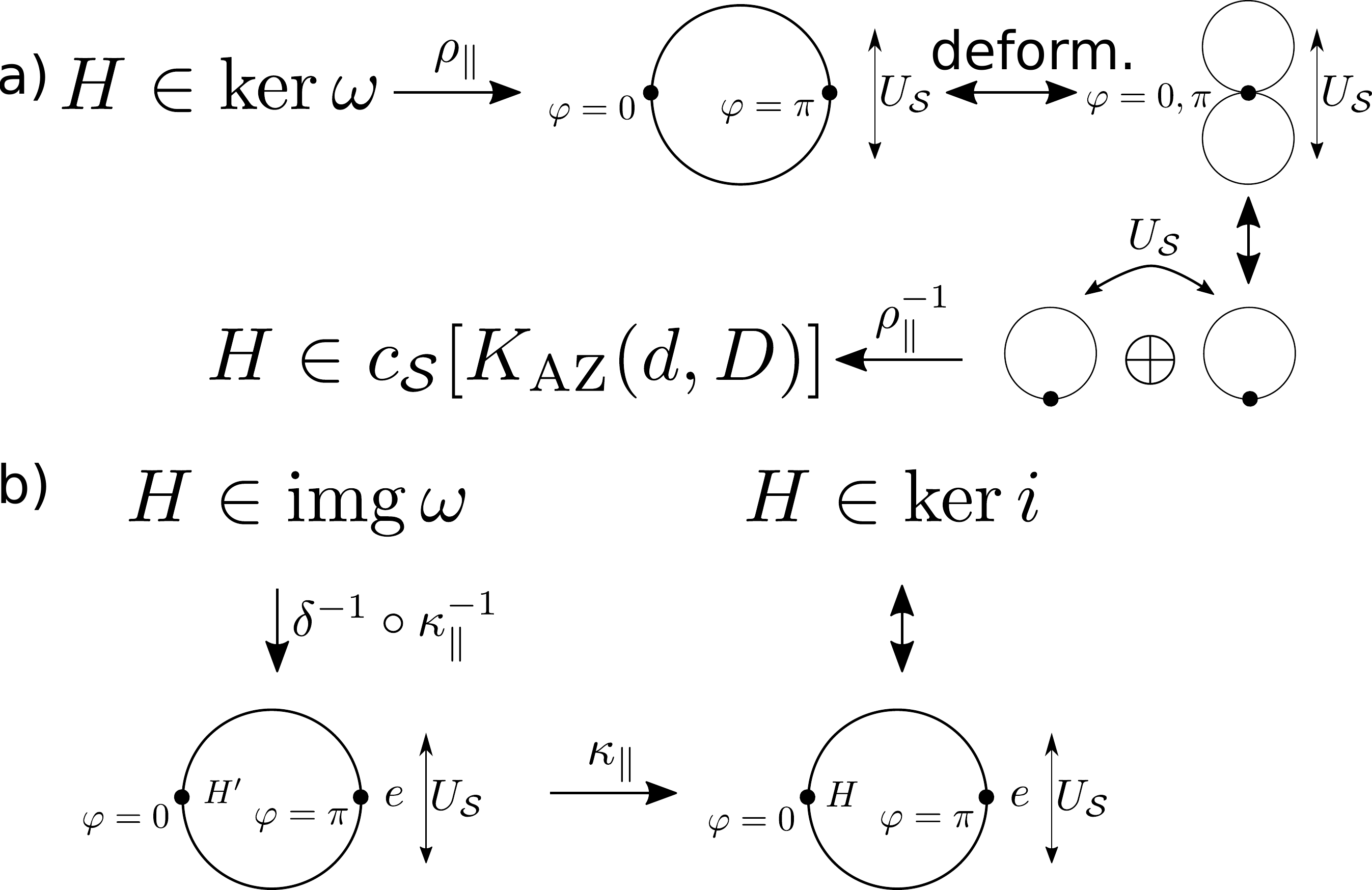}
	\caption{\label{fig:exactness} Hamiltonians from $\ker\omap$ can be
	deformed to the form~(\ref{eq:cmap}) and vice versa (a). For Hamiltonians
from $\img\omap$, a path can be constructed that connects them to the trivial
element (b). Similarly for every Hamiltonian in $\ker i$, the loop in canonical
form can be constructed that yields $H$ from $\img\omap$.}
\end{figure}

We next show that every Hamiltonian in $\ker\omap$ can be continuously
deformed to the diagonal form~(\ref{eq:cmap}). Since $\kappa_\parallel$ and
$\rho_\parallel$ are isomorphisms that preserve a diagonal form, it is
sufficient to show that every Hamiltonian in $\ker\delta$ can be deformed into
the diagonal form. Hereto we note that $\delta(H)=0$ implies that $H(0)$ and
$H(\pi)$ are both in the trivial equivalence class (nontrivial $H(0)=H(\pi)$
would correspond to a weak topological phase, which we do not consider here),
for which after continuous deformation, we may set $H(0)=H(\pi)=e$, $e$ being
the trivial element. Under stable equivalence we may replace $H(\varphi)$ by
$H(\varphi)\oplus e$ which may be smoothly deformed into

\begin{align}
  H(\varphi)\equiv
  \begin{cases}
    H(\varphi)\oplus e & \text{for } 0\le\varphi<\pi\\
    e\oplus H(\varphi) & \text{for } \pi\le\varphi<2\pi,
  \end{cases}
\end{align}
and subsequently , into a Hamiltonian of the form~(\ref{eq:cmap}), since
$\rho_\parallel \mathcal{S}\rho_\parallel^{-1}H(2\pi-\varphi)=H(\varphi)$. As
the procedure can be run backwards we conclude $\ker\omap=\img
c_{\cal S}$ giving the exactness of the
sequence~(\ref{eq:exactsec}) at $\K[0](d,d_\parallel,D,D_\parallel)$.
Figure.~\ref{fig:exactness}a illustrates the above steps of the proof.

Similarly, because $\kappa_\parallel$ is an isomorphism, to show exactness at
the second stage of the sequence~(\ref{eq:exactsec}) it is sufficient to show
that any element of $\img\delta$ can be smoothly deformed to the trivial
element $e$ if the crystalline symmetry $\rho_\parallel
\mathcal{S}\rho_\parallel^{-1}$ is no longer imposed, and vice versa, see
Fig.~\ref{fig:exactness}b. Again we may assume that $H(0)=e$, and the
continuous deformation linking $H(\pi)\ominus H(0)$ to $e\ominus e$ is
$H(\varphi)\ominus H(0)$ with $0<\varphi<\pi$.  Similarly, if such a
transformation exists, \textit{i.e.}, if there exists a continuous function
$\tilde{H}(\varphi)=H(\varphi)\ominus H(0)$ interpolating between $H(0)\ominus
H(0)$ and $H(\pi)\ominus H(0)$, then there also exists a family of
$\rho_\parallel \mathcal{S}\rho_\parallel^{-1}$-symmetric Hamiltonians 

\begin{align}
  H(\varphi)\equiv
  \begin{cases}
    H(\varphi) & \text{for } 0\le\varphi<\pi\\
    H(2\pi-\varphi) & \text{for } \pi\le\varphi<2\pi,
  \end{cases}
\end{align}
such that $\tilde{H}(\varphi)=H(\varphi)\ominus H(0)$.

\section{Calculation of $\ker\omap^k$ and $K_{{\rm TF},{\cal S}}$}\label{app:kerk}

Because the order-raising map $\omap$ commutes with the isomorphisms
$\kappa_{\parallel}$ and $\kappa_{\perp}$ it is sufficient to calculate the
groups $\ker \omap$ and $K_{{\rm TF},{\cal O}}$ for the case $d=d_{\parallel} =
0$. The results for $d=0$ can be lifted to $d > 0$ by suitable application of
the dimension-raising isomorphisms $\kappa_{\parallel}$ and $\kappa_{\perp}$.
To obtain $\ker \omap$ for $d=0$ we note that the bulk-boundary
correspondence~(\ref{eq:bb}) then gives
\begin{align}
  \K[1](1,1)&=\ki{2}=\K[0](0,0)/\ker\omap,
\end{align}
whereas $K_{{\rm TF},{\cal S}}$ can be obtained from the isomorphism
(\ref{eq:KdKs}). For most symmetry classes these two relations are sufficient
to determine $\ker\omap$ and $K_{{\rm TF},{\cal S}}$ from the known
groups~\cite{shiozaki2014} $\K[0]$ and~\cite{langbehn2017,geier2018} $\K[1]$,
owing to the simple structure of these groups. There are a few cases for which
both $\K[1]$ and $\ker \omap$ are nontrivial and non-unique subgroups of
$\K[0]$. For these an explicit calculation is needed. These special cases are
considered below. The results for the groups $\ker \omap$ and $K_{{\rm
TF},{\cal O}}$ for $d=0$ are given in
Tables~\ref{tab:Kd_complex}-\ref{tab:Kd_complexA} and
\ref{tab:Ks_complex}-\ref{tab:Ks_real}, respectively.

\begin{table}
	\begin{tabular*}{\columnwidth}{c@{\extracolsep{\fill}}ccccc}
\hline\hline 
$s$ & TF class & $t=0$ &  $t=1$ \\
\hline
    $0$ & A    & $\ZZ$ & $0$  \\
    $1$ & AIII & $0$   & $0$  \\
    \hline\hline
  \end{tabular*}
  \caption{The subgroups $\ker\omap^k\subseteq\K[0](s,t\vert0,0,0,0)$ for
  complex Shiozaki-Sato classes with a unitary order-two (anti)symmetry.} 
\label{tab:Kd_complex}
\end{table}
\begin{table}
	\begin{tabular*}{\columnwidth}{c@{\extracolsep{\fill}}cccc}
\hline\hline 
$s$    & Shiozaki-Sato class            & $\ker\omap$ \\
\hline
$0$    & A$^{{\cal T}^+{\cal O}}$       & $2\ZZ^{(q=2)}$\\
$1$    & AIII$^{{\cal P}^+{\cal O}_+}$  & $0^{(q=1)}$   \\
$2$    & A$^{{\cal P}^+{\cal O}}$       & $\ZZ_2$       \\
$3$    & AIII$^{{\cal T}^-{\cal O}_-}$  & $0$           \\
$4$    & A$^{{\cal T}^-{\cal O}}$       & $2\ZZ$        \\
$5$    & AIII$^{{\cal P}^-{\cal O}_+}$  & $0$           \\
$6$    & A$^{{\cal P}^-{\cal O}}$       & $0$           \\
$7$    & AIII$^{{\cal T}^+{\cal O}_-}$  & $0$           \\
    \hline\hline
  \end{tabular*}
  \caption{The subgroups $\ker\omap\subseteq\K[0](s\vert0,0,0,0)$ for complex
    Shiozaki-Sato classes with an antiunitary order-two (anti)symmetry. The
  integer in the superscript gives the $q$ so that $\ker\omap^k=K$ for
$k>q$.} 
\label{tab:Kd_complexA}
\end{table}
\begin{table}
\begin{tabular*}{\columnwidth}{c @{\extracolsep{\fill}} ccccc}
\hline\hline 
$s$ & TF class & $t=0$                          & $t=1$ & $t=2$  & $t=3$\\
\hline
$0$ & AI   & $\ZZ$           & $0$         & $2\ZZ$         & $\ZZ_2$ \\
$1$ & BDI  & $\ZZ_2^{(q=1)}$ & $0$         & $0$            & $\ZZ_2$ \\
$2$ & D    & $\ZZ_2^{(q=1)}$ & $0^{(q=1)}$ & $0$            & $0$ \\
$3$ & DIII & $0$             & $0^{(q=1)}$ & $0$            & $0$ \\
$4$ & AII  & $2\ZZ$          & $0$         & $4\ZZ^{(q=2)}$ & $0$ \\
$5$ & CII  & $0$             & $0$         & $0$            & $0$ \\
$6$ & C    & $0$             & $0$         & $0$            & $0$ \\
$7$ & CI   & $0$             & $0$         & $0$            & $0$ \\
\hline\hline
\end{tabular*}
\caption{The subgroups $\ker\omap\subseteq\K[0](s,t\vert0,0,0,0)$ for real Shiozaki-Sato
  classes with a unitary order-two (anti)symmetry. The integer in the superscript gives the integer $q$
  so that $\ker\omap^k=K$ for $k>q$.} 
\label{tab:Kd_real}
\end{table}
\begin{table}
	\begin{tabular*}{\columnwidth}{c@{\extracolsep{\fill}}cccccc}
\hline\hline 
$s$ & TF class & $K_{\rm TF}(s)$ & $K_{{\rm TF},{\cal O}}(s,0)$ &  $K_{{\rm TF},{\cal O}}(s,1)$ \\
\hline
    $0$ & A & $\ZZ$ & $\ZZ$ & $0$ \\
    $1$ & AIII & $0$   & $0$   & $0$ \\
    \hline\hline
  \end{tabular*}
  \caption{The groups $K_{\rm TF}(s\vert0,0)$ and $K_{{\rm TF},{\cal O}}(s,t\vert0,0,0,0)$
for complex Shiozaki-Sato classes with a unitary order-two (anti)symmetry.} 
\label{tab:Ks_complex}
\end{table}
\begin{table}
	\begin{tabular*}{\columnwidth}{c@{\extracolsep{\fill}}ccccc}
\hline\hline 
$s$ & Shiozaki-Sato class & $K_{\rm TF}(s)$ & $K_{{\rm TF},{\cal O}}(s)$ \\
\hline
$0$ & A$^{{\cal T}^+{\cal O}}$     & $\ZZ$ & $\ZZ$  \\
$1$ & AIII$^{{\cal P}^+{\cal O}_+}$& $0$ & $0$    \\
$2$ & A$^{{\cal P}^+{\cal O}}$     & $\ZZ$ & $0$    \\
$3$ & AIII$^{{\cal T}^-{\cal O}_-}$& $0$ & $0$    \\
$4$ & A$^{{\cal T}^-{\cal O}}$     & $\ZZ$ & $2\ZZ$ \\
$5$ & AIII$^{{\cal P}^-{\cal O}_+}$& $0$ & $0$    \\
$6$ & A$^{{\cal P}^-{\cal O}}$     & $\ZZ$ & $0$    \\
$7$ & AIII$^{{\cal T}^+{\cal O}_-}$& $0$ & $0$    \\
    \hline\hline
  \end{tabular*}
  \caption{The subgroups $K_{{\rm TF},{\cal O}}(s\vert0,0,0,0)$ for complex Shiozaki-Sato classes with an antiunitary
  order-two (anti)symmetry.} 
\label{tab:Ks_complexA}
\end{table}
\begin{table}
	\begin{tabular*}{\columnwidth}{c@{\extracolsep{\fill}}cccccc}
\hline\hline 
\multirow{2}{*}{$s$} & \multirow{2}{*}{TF class} & \multirow{2}{*}{$K_{\rm TF}(s)$} & \multicolumn{4}{c}{$K_{{\rm TF},{\cal O}}(s,t)$} \\ & & & $t=0$ & $t=1$ & $t=2$ & $t=3$ \\
\hline
    $0$ & AI & $\ZZ$   & $\ZZ$   & $0$     & $2\ZZ$   & $0$ \\
    $1$ & BDI & $\ZZ_2$ & $\ZZ_2$ & $\ZZ_2$ & $0$     & $0$    \\
    $2$ & D & $\ZZ_2$ & $\ZZ_2$ & $\ZZ_2$     & $\ZZ_2$ & $0$    \\
    $3$ & DIII & $0$     & $0$     & $0$     & $0$     & $0$    \\
    $4$ & AII & $2\ZZ$  & $2\ZZ$  & $0$     & $2\ZZ$  & $0$ \\
    $5$ & CII & $0$     & $0$     & $0$     & $0$     & $0$    \\
    $6$ & C & $0$     & $0$     & $0$     & $0$     & $0$    \\
    $7$ & CI & $0$     & $0$     & $0$     & $0$     & $0$    \\
    \hline\hline
  \end{tabular*}
  \caption{The groups $K_{\rm TF}(s\vert0,0)$ and $K_{{\rm TF},{\cal O}}(s,t\vert0,0,0,0)$
for the real Shiozaki-Sato classes with a unitary order-two (anti)symmetry.} 
\label{tab:Ks_real}
\end{table}

\subsection{Classes A$^\cO$, $(s,t)=(0,0)$,
  AI$^{\cO_+}$, $(s,t)=(0,0)$ and AII$^{\cO_+}$, $(s,t)=(4,0)$}
A zero-dimensional Hamiltonian $H_0$ in classes A and AI with an order-two
on-site symmetry $\cO$ is classified by
\begin{align}
	\K[0]&=\{(n_+,n_-),n_+,n_-\in \ZZ\}=\ZZ^2,
	\label{eq:K00}
\end{align}
where $n_\pm$ is the difference between the number of positive and negative
energy levels of $H_0$ with $\pm$ parity under $\cO$ symmetry. In class AII,
due to Kramers degeneracy, the integers $n_\pm$ need to be even. Since the
local symmetry $\cO$ commutes with the time-reversal symmetry (class AI), the
subgroups $\ker i$ and $\ker\omap$ are easily obtained,
\begin{align}
	\ker i&=\{(n,-n),n\in\ZZ\}=\ZZ,\nonumber\\
	\ker\omap&=\{(n,n),n\in\ZZ\}=\ZZ,
	\label{eq:Kd00}
\end{align}
since Hamiltonians with $n_+ = n_- = n$ can be deformed into a separable
Hamiltonian, whereas Hamiltonians with $n_+ + n_- = 0$ are trivial when the
protection by the on-site symmetry $\cO$ is lifted.

\subsection{Classes BDI$^{\cO_{++}}$, $(s,t)=(1,0)$ and D$^{\cO_+}$,
$(s,t)=(2,0)$}

Hamiltonians $H$ from these classes are classified by
\begin{align}
	\K[0]&=\{(n_+,n_-),n_\pm\in\ZZ_2\}=\ZZ_2^2,
\end{align}
with $n_\pm=\sign[\pf(H_\pm)]$, where $H_\pm$ is the block of the Hamiltonian
$H$ with $\pm$ parity under $\cO$. The Hamiltonian $H$ is taken in a basis
where particle-hole antisymmetry is represented by $U_\mathcal{P}=1$. In this
class, the subgroups $\ker i$ and $\ker \omap$ are identical,
\begin{align}
	\ker i&=\{(n,n),n\in\ZZ_2\}=\ZZ_2,\nonumber\\
	\ker\omap&=\{(n,n),n\in\ZZ_2\}=\ZZ_2.
	\label{eq:st10r}
\end{align}
\bibliography{refs}

\begin{thebibliography}{78}%
\makeatletter
\providecommand \@ifxundefined [1]{%
 \@ifx{#1\undefined}
}%
\providecommand \@ifnum [1]{%
 \ifnum #1\expandafter \@firstoftwo
 \else \expandafter \@secondoftwo
 \fi
}%
\providecommand \@ifx [1]{%
 \ifx #1\expandafter \@firstoftwo
 \else \expandafter \@secondoftwo
 \fi
}%
\providecommand \natexlab [1]{#1}%
\providecommand \enquote  [1]{``#1''}%
\providecommand \bibnamefont  [1]{#1}%
\providecommand \bibfnamefont [1]{#1}%
\providecommand \citenamefont [1]{#1}%
\providecommand \href@noop [0]{\@secondoftwo}%
\providecommand \href [0]{\begingroup \@sanitize@url \@href}%
\providecommand \@href[1]{\@@startlink{#1}\@@href}%
\providecommand \@@href[1]{\endgroup#1\@@endlink}%
\providecommand \@sanitize@url [0]{\catcode `\\12\catcode `\$12\catcode
  `\&12\catcode `\#12\catcode `\^12\catcode `\_12\catcode `\%12\relax}%
\providecommand \@@startlink[1]{}%
\providecommand \@@endlink[0]{}%
\providecommand \url  [0]{\begingroup\@sanitize@url \@url }%
\providecommand \@url [1]{\endgroup\@href {#1}{\urlprefix }}%
\providecommand \urlprefix  [0]{URL }%
\providecommand \Eprint [0]{\href }%
\providecommand \doibase [0]{http://dx.doi.org/}%
\providecommand \selectlanguage [0]{\@gobble}%
\providecommand \bibinfo  [0]{\@secondoftwo}%
\providecommand \bibfield  [0]{\@secondoftwo}%
\providecommand \translation [1]{[#1]}%
\providecommand \BibitemOpen [0]{}%
\providecommand \bibitemStop [0]{}%
\providecommand \bibitemNoStop [0]{.\EOS\space}%
\providecommand \EOS [0]{\spacefactor3000\relax}%
\providecommand \BibitemShut  [1]{\csname bibitem#1\endcsname}%
\let\auto@bib@innerbib\@empty
%</preamble>
\bibitem [{\citenamefont {Hasan}\ and\ \citenamefont {Kane}(2010)}]{hasan2010}%
  \BibitemOpen
  \bibfield  {author} {\bibinfo {author} {\bibfnamefont {M.~Z.}\ \bibnamefont
  {Hasan}}\ and\ \bibinfo {author} {\bibfnamefont {C.~L.}\ \bibnamefont
  {Kane}},\ }\href {\doibase 10.1103/RevModPhys.82.3045} {\bibfield  {journal}
  {\bibinfo  {journal} {Rev. Mod. Phys.}\ }\textbf {\bibinfo {volume} {82}},\
  \bibinfo {pages} {3045} (\bibinfo {year} {2010})}\BibitemShut {NoStop}%
\bibitem [{\citenamefont {Bernevig}\ and\ \citenamefont
  {Hughes}(2013)}]{bernevig2013}%
  \BibitemOpen
  \bibfield  {author} {\bibinfo {author} {\bibfnamefont {B.~A.}\ \bibnamefont
  {Bernevig}}\ and\ \bibinfo {author} {\bibfnamefont {T.~L.}\ \bibnamefont
  {Hughes}},\ }\href@noop {} {\emph {\bibinfo {title} {Topological Insulators
  and Topological Superconductors}}}\ (\bibinfo  {publisher} {Princeton
  University Press},\ \bibinfo {year} {2013})\BibitemShut {NoStop}%
\bibitem [{\citenamefont {Qi}\ and\ \citenamefont {Zhang}(2011)}]{qi2011}%
  \BibitemOpen
  \bibfield  {author} {\bibinfo {author} {\bibfnamefont {X.-L.}\ \bibnamefont
  {Qi}}\ and\ \bibinfo {author} {\bibfnamefont {S.-C.}\ \bibnamefont {Zhang}},\
  }\href {\doibase 10.1103/RevModPhys.83.1057} {\bibfield  {journal} {\bibinfo
  {journal} {Rev. Mod. Phys.}\ }\textbf {\bibinfo {volume} {83}},\ \bibinfo
  {pages} {1057} (\bibinfo {year} {2011})}\BibitemShut {NoStop}%
\bibitem [{\citenamefont {Fu}(2011)}]{fu2011}%
  \BibitemOpen
  \bibfield  {author} {\bibinfo {author} {\bibfnamefont {L.}~\bibnamefont
  {Fu}},\ }\href {\doibase 10.1103/PhysRevLett.106.106802} {\bibfield
  {journal} {\bibinfo  {journal} {Phys. Rev. Lett.}\ }\textbf {\bibinfo
  {volume} {106}},\ \bibinfo {pages} {106802} (\bibinfo {year}
  {2011})}\BibitemShut {NoStop}%
\bibitem [{\citenamefont {Fang}\ \emph {et~al.}(2012)\citenamefont {Fang},
  \citenamefont {Gilbert},\ and\ \citenamefont {Bernevig}}]{fang2012}%
  \BibitemOpen
  \bibfield  {author} {\bibinfo {author} {\bibfnamefont {C.}~\bibnamefont
  {Fang}}, \bibinfo {author} {\bibfnamefont {M.~J.}\ \bibnamefont {Gilbert}}, \
  and\ \bibinfo {author} {\bibfnamefont {B.~A.}\ \bibnamefont {Bernevig}},\
  }\href@noop {} {\bibfield  {journal} {\bibinfo  {journal} {Phys. Rev. B}\
  }\textbf {\bibinfo {volume} {86}},\ \bibinfo {pages} {115112} (\bibinfo
  {year} {2012})}\BibitemShut {NoStop}%
\bibitem [{\citenamefont {Turner}\ \emph {et~al.}(2012)\citenamefont {Turner},
  \citenamefont {Zhang}, \citenamefont {Mong},\ and\ \citenamefont
  {Vishwanath}}]{turner2012}%
  \BibitemOpen
  \bibfield  {author} {\bibinfo {author} {\bibfnamefont {A.~M.}\ \bibnamefont
  {Turner}}, \bibinfo {author} {\bibfnamefont {Y.}~\bibnamefont {Zhang}},
  \bibinfo {author} {\bibfnamefont {R.~S.~K.}\ \bibnamefont {Mong}}, \ and\
  \bibinfo {author} {\bibfnamefont {A.}~\bibnamefont {Vishwanath}},\ }\href
  {\doibase 10.1103/PhysRevB.85.165120} {\bibfield  {journal} {\bibinfo
  {journal} {Phys. Rev. B}\ }\textbf {\bibinfo {volume} {85}},\ \bibinfo
  {pages} {165120} (\bibinfo {year} {2012})}\BibitemShut {NoStop}%
\bibitem [{\citenamefont {Fang}\ \emph {et~al.}(2013)\citenamefont {Fang},
  \citenamefont {Gilbert},\ and\ \citenamefont {Bernevig}}]{fang2013}%
  \BibitemOpen
  \bibfield  {author} {\bibinfo {author} {\bibfnamefont {C.}~\bibnamefont
  {Fang}}, \bibinfo {author} {\bibfnamefont {M.~J.}\ \bibnamefont {Gilbert}}, \
  and\ \bibinfo {author} {\bibfnamefont {B.~A.}\ \bibnamefont {Bernevig}},\
  }\href@noop {} {\bibfield  {journal} {\bibinfo  {journal} {Phys. Rev. B}\
  }\textbf {\bibinfo {volume} {87}},\ \bibinfo {pages} {035119} (\bibinfo
  {year} {2013})}\BibitemShut {NoStop}%
\bibitem [{\citenamefont {Chiu}\ \emph {et~al.}(2013)\citenamefont {Chiu},
  \citenamefont {Yao},\ and\ \citenamefont {Ryu}}]{chiu2013}%
  \BibitemOpen
  \bibfield  {author} {\bibinfo {author} {\bibfnamefont {C.-K.}\ \bibnamefont
  {Chiu}}, \bibinfo {author} {\bibfnamefont {H.}~\bibnamefont {Yao}}, \ and\
  \bibinfo {author} {\bibfnamefont {S.}~\bibnamefont {Ryu}},\ }\href {\doibase
  10.1103/PhysRevB.88.075142} {\bibfield  {journal} {\bibinfo  {journal} {Phys.
  Rev. B}\ }\textbf {\bibinfo {volume} {88}},\ \bibinfo {pages} {075142}
  (\bibinfo {year} {2013})}\BibitemShut {NoStop}%
\bibitem [{\citenamefont {Morimoto}\ and\ \citenamefont
  {Furusaki}(2013)}]{morimoto2013}%
  \BibitemOpen
  \bibfield  {author} {\bibinfo {author} {\bibfnamefont {T.}~\bibnamefont
  {Morimoto}}\ and\ \bibinfo {author} {\bibfnamefont {A.}~\bibnamefont
  {Furusaki}},\ }\href {\doibase 10.1103/PhysRevB.88.125129} {\bibfield
  {journal} {\bibinfo  {journal} {Phys. Rev. B}\ }\textbf {\bibinfo {volume}
  {88}},\ \bibinfo {pages} {125129} (\bibinfo {year} {2013})}\BibitemShut
  {NoStop}%
\bibitem [{\citenamefont {Slager}\ \emph {et~al.}(2013)\citenamefont {Slager},
  \citenamefont {Mesaros}, \citenamefont {Juricic},\ and\ \citenamefont
  {Zaanen}}]{slager2013}%
  \BibitemOpen
  \bibfield  {author} {\bibinfo {author} {\bibfnamefont {R.-J.}\ \bibnamefont
  {Slager}}, \bibinfo {author} {\bibfnamefont {A.}~\bibnamefont {Mesaros}},
  \bibinfo {author} {\bibfnamefont {V.}~\bibnamefont {Juricic}}, \ and\
  \bibinfo {author} {\bibfnamefont {J.}~\bibnamefont {Zaanen}},\ }\href
  {\doibase 10.1038/nphys2513} {\bibfield  {journal} {\bibinfo  {journal}
  {Nature Phys.}\ }\textbf {\bibinfo {volume} {9}},\ \bibinfo {pages} {98}
  (\bibinfo {year} {2013})}\BibitemShut {NoStop}%
\bibitem [{\citenamefont {Jadaun}\ \emph {et~al.}(2013)\citenamefont {Jadaun},
  \citenamefont {Xiao}, \citenamefont {Niu},\ and\ \citenamefont
  {Banerjee}}]{jadaun2013}%
  \BibitemOpen
  \bibfield  {author} {\bibinfo {author} {\bibfnamefont {P.}~\bibnamefont
  {Jadaun}}, \bibinfo {author} {\bibfnamefont {D.}~\bibnamefont {Xiao}},
  \bibinfo {author} {\bibfnamefont {Q.}~\bibnamefont {Niu}}, \ and\ \bibinfo
  {author} {\bibfnamefont {S.~K.}\ \bibnamefont {Banerjee}},\ }\href@noop {}
  {\bibfield  {journal} {\bibinfo  {journal} {Phys. Rev. B}\ }\textbf {\bibinfo
  {volume} {88}},\ \bibinfo {pages} {085110} (\bibinfo {year}
  {2013})}\BibitemShut {NoStop}%
\bibitem [{\citenamefont {Liu}\ \emph {et~al.}(2014)\citenamefont {Liu},
  \citenamefont {He},\ and\ \citenamefont {Law}}]{liu2014b}%
  \BibitemOpen
  \bibfield  {author} {\bibinfo {author} {\bibfnamefont {X.-J.}\ \bibnamefont
  {Liu}}, \bibinfo {author} {\bibfnamefont {J.~J.}\ \bibnamefont {He}}, \ and\
  \bibinfo {author} {\bibfnamefont {K.~T.}\ \bibnamefont {Law}},\ }\href@noop
  {} {\bibfield  {journal} {\bibinfo  {journal} {Phys. Rev. B}\ }\textbf
  {\bibinfo {volume} {90}},\ \bibinfo {pages} {235141} (\bibinfo {year}
  {2014})}\BibitemShut {NoStop}%
\bibitem [{\citenamefont {Alexandradinata}\ \emph {et~al.}(2014)\citenamefont
  {Alexandradinata}, \citenamefont {Fang}, \citenamefont {Gilbert},\ and\
  \citenamefont {Bernevig}}]{alexandradinata2014}%
  \BibitemOpen
  \bibfield  {author} {\bibinfo {author} {\bibfnamefont {A.}~\bibnamefont
  {Alexandradinata}}, \bibinfo {author} {\bibfnamefont {C.}~\bibnamefont
  {Fang}}, \bibinfo {author} {\bibfnamefont {M.~J.}\ \bibnamefont {Gilbert}}, \
  and\ \bibinfo {author} {\bibfnamefont {B.~A.}\ \bibnamefont {Bernevig}},\
  }\href@noop {} {\bibfield  {journal} {\bibinfo  {journal} {Phys. Rev. Lett.}\
  }\textbf {\bibinfo {volume} {113}},\ \bibinfo {pages} {116403} (\bibinfo
  {year} {2014})}\BibitemShut {NoStop}%
\bibitem [{\citenamefont {Dong}\ and\ \citenamefont {Liu}(2016)}]{dong2016}%
  \BibitemOpen
  \bibfield  {author} {\bibinfo {author} {\bibfnamefont {X.-Y.}\ \bibnamefont
  {Dong}}\ and\ \bibinfo {author} {\bibfnamefont {C.-X.}\ \bibnamefont {Liu}},\
  }\href@noop {} {\bibfield  {journal} {\bibinfo  {journal} {Phys. Rev. B}\
  }\textbf {\bibinfo {volume} {93}},\ \bibinfo {pages} {045429} (\bibinfo
  {year} {2016})}\BibitemShut {NoStop}%
\bibitem [{\citenamefont {Chiu}\ \emph {et~al.}(2016)\citenamefont {Chiu},
  \citenamefont {Teo}, \citenamefont {Schnyder},\ and\ \citenamefont
  {Ryu}}]{chiu2016}%
  \BibitemOpen
  \bibfield  {author} {\bibinfo {author} {\bibfnamefont {C.-K.}\ \bibnamefont
  {Chiu}}, \bibinfo {author} {\bibfnamefont {J.~C.~Y.}\ \bibnamefont {Teo}},
  \bibinfo {author} {\bibfnamefont {A.~P.}\ \bibnamefont {Schnyder}}, \ and\
  \bibinfo {author} {\bibfnamefont {S.}~\bibnamefont {Ryu}},\ }\href@noop {}
  {\bibfield  {journal} {\bibinfo  {journal} {Rev. Mod. Phys.}\ }\textbf
  {\bibinfo {volume} {88}},\ \bibinfo {pages} {035005} (\bibinfo {year}
  {2016})}\BibitemShut {NoStop}%
\bibitem [{\citenamefont {Trifunovic}\ and\ \citenamefont
  {Brouwer}(2017)}]{trifunovic2017}%
  \BibitemOpen
  \bibfield  {author} {\bibinfo {author} {\bibfnamefont {L.}~\bibnamefont
  {Trifunovic}}\ and\ \bibinfo {author} {\bibfnamefont {P.~W.}\ \bibnamefont
  {Brouwer}},\ }\href {\doibase 10.1103/PhysRevB.96.195109} {\bibfield
  {journal} {\bibinfo  {journal} {Phys. Rev. B}\ }\textbf {\bibinfo {volume}
  {96}},\ \bibinfo {pages} {195109} (\bibinfo {year} {2017})}\BibitemShut
  {NoStop}%
\bibitem [{\citenamefont {Kruthoff}\ \emph {et~al.}(2017)\citenamefont
  {Kruthoff}, \citenamefont {de~Boer}, \citenamefont {van Wezel}, \citenamefont
  {Kane},\ and\ \citenamefont {Slager}}]{kruthoff2017}%
  \BibitemOpen
  \bibfield  {author} {\bibinfo {author} {\bibfnamefont {J.}~\bibnamefont
  {Kruthoff}}, \bibinfo {author} {\bibfnamefont {J.}~\bibnamefont {de~Boer}},
  \bibinfo {author} {\bibfnamefont {J.}~\bibnamefont {van Wezel}}, \bibinfo
  {author} {\bibfnamefont {C.~L.}\ \bibnamefont {Kane}}, \ and\ \bibinfo
  {author} {\bibfnamefont {R.-J.}\ \bibnamefont {Slager}},\ }\href {\doibase
  10.1103/PhysRevX.7.041069} {\bibfield  {journal} {\bibinfo  {journal} {Phys.
  Rev. X}\ }\textbf {\bibinfo {volume} {7}},\ \bibinfo {pages} {041069}
  (\bibinfo {year} {2017})}\BibitemShut {NoStop}%
\bibitem [{\citenamefont {Po}\ \emph {et~al.}(2017)\citenamefont {Po},
  \citenamefont {Vishwanath},\ and\ \citenamefont {Watanabe}}]{po2017}%
  \BibitemOpen
  \bibfield  {author} {\bibinfo {author} {\bibfnamefont {H.~C.}\ \bibnamefont
  {Po}}, \bibinfo {author} {\bibfnamefont {A.}~\bibnamefont {Vishwanath}}, \
  and\ \bibinfo {author} {\bibfnamefont {H.}~\bibnamefont {Watanabe}},\
  }\href@noop {} {\bibfield  {journal} {\bibinfo  {journal} {Nature Comm.}\
  }\textbf {\bibinfo {volume} {8}},\ \bibinfo {pages} {50} (\bibinfo {year}
  {2017})}\BibitemShut {NoStop}%
\bibitem [{\citenamefont {Bradlyn}\ \emph {et~al.}(2017)\citenamefont
  {Bradlyn}, \citenamefont {Elcoro}, \citenamefont {Cano}, \citenamefont
  {Vergniory}, \citenamefont {Wang}, \citenamefont {Felser}, \citenamefont
  {Aroyo},\ and\ \citenamefont {Bernevig}}]{bradlyn2017}%
  \BibitemOpen
  \bibfield  {author} {\bibinfo {author} {\bibfnamefont {B.}~\bibnamefont
  {Bradlyn}}, \bibinfo {author} {\bibfnamefont {L.}~\bibnamefont {Elcoro}},
  \bibinfo {author} {\bibfnamefont {J.}~\bibnamefont {Cano}}, \bibinfo {author}
  {\bibfnamefont {M.~G.}\ \bibnamefont {Vergniory}}, \bibinfo {author}
  {\bibfnamefont {Z.}~\bibnamefont {Wang}}, \bibinfo {author} {\bibfnamefont
  {C.}~\bibnamefont {Felser}}, \bibinfo {author} {\bibfnamefont {M.~I.}\
  \bibnamefont {Aroyo}}, \ and\ \bibinfo {author} {\bibfnamefont {B.~A.}\
  \bibnamefont {Bernevig}},\ }\href@noop {} {\bibfield  {journal} {\bibinfo
  {journal} {Nature}\ }\textbf {\bibinfo {volume} {547}},\ \bibinfo {pages}
  {298} (\bibinfo {year} {2017})}\BibitemShut {NoStop}%
\bibitem [{\citenamefont {Khalaf}\ \emph {et~al.}(2018)\citenamefont {Khalaf},
  \citenamefont {Po}, \citenamefont {Vishwanath},\ and\ \citenamefont
  {Watanabe}}]{khalaf2018}%
  \BibitemOpen
  \bibfield  {author} {\bibinfo {author} {\bibfnamefont {E.}~\bibnamefont
  {Khalaf}}, \bibinfo {author} {\bibfnamefont {H.~C.}\ \bibnamefont {Po}},
  \bibinfo {author} {\bibfnamefont {A.}~\bibnamefont {Vishwanath}}, \ and\
  \bibinfo {author} {\bibfnamefont {H.}~\bibnamefont {Watanabe}},\ }\href
  {\doibase 10.1103/PhysRevX.8.031070} {\bibfield  {journal} {\bibinfo
  {journal} {Phys. Rev. X}\ }\textbf {\bibinfo {volume} {8}},\ \bibinfo {pages}
  {031070} (\bibinfo {year} {2018})}\BibitemShut {NoStop}%
\bibitem [{\citenamefont {Shiozaki}\ and\ \citenamefont
  {Sato}(2014)}]{shiozaki2014}%
  \BibitemOpen
  \bibfield  {author} {\bibinfo {author} {\bibfnamefont {K.}~\bibnamefont
  {Shiozaki}}\ and\ \bibinfo {author} {\bibfnamefont {M.}~\bibnamefont
  {Sato}},\ }\href@noop {} {\bibfield  {journal} {\bibinfo  {journal} {Phys.
  Rev. B}\ }\textbf {\bibinfo {volume} {90}},\ \bibinfo {pages} {165114}
  (\bibinfo {year} {2014})}\BibitemShut {NoStop}%
\bibitem [{\citenamefont {Shiozaki}\ \emph {et~al.}(2016)\citenamefont
  {Shiozaki}, \citenamefont {Sato},\ and\ \citenamefont {Gomi}}]{shiozaki2016}%
  \BibitemOpen
  \bibfield  {author} {\bibinfo {author} {\bibfnamefont {K.}~\bibnamefont
  {Shiozaki}}, \bibinfo {author} {\bibfnamefont {M.}~\bibnamefont {Sato}}, \
  and\ \bibinfo {author} {\bibfnamefont {K.}~\bibnamefont {Gomi}},\ }\href
  {\doibase 10.1103/PhysRevB.93.195413} {\bibfield  {journal} {\bibinfo
  {journal} {Phys. Rev. B}\ }\textbf {\bibinfo {volume} {93}},\ \bibinfo
  {pages} {195413} (\bibinfo {year} {2016})}\BibitemShut {NoStop}%
\bibitem [{\citenamefont {Shiozaki}\ \emph {et~al.}(2017)\citenamefont
  {Shiozaki}, \citenamefont {Sato},\ and\ \citenamefont {Gomi}}]{shiozaki2017}%
  \BibitemOpen
  \bibfield  {author} {\bibinfo {author} {\bibfnamefont {K.}~\bibnamefont
  {Shiozaki}}, \bibinfo {author} {\bibfnamefont {M.}~\bibnamefont {Sato}}, \
  and\ \bibinfo {author} {\bibfnamefont {K.}~\bibnamefont {Gomi}},\ }\href
  {\doibase 10.1103/PhysRevB.95.235425} {\bibfield  {journal} {\bibinfo
  {journal} {Phys. Rev. B}\ }\textbf {\bibinfo {volume} {95}},\ \bibinfo
  {pages} {235425} (\bibinfo {year} {2017})}\BibitemShut {NoStop}%
\bibitem [{\citenamefont {{Shiozaki}}\ \emph {et~al.}(2018)\citenamefont
  {{Shiozaki}}, \citenamefont {{Sato}},\ and\ \citenamefont
  {{Gomi}}}]{shiozaki2018}%
  \BibitemOpen
  \bibfield  {author} {\bibinfo {author} {\bibfnamefont {K.}~\bibnamefont
  {{Shiozaki}}}, \bibinfo {author} {\bibfnamefont {M.}~\bibnamefont {{Sato}}},
  \ and\ \bibinfo {author} {\bibfnamefont {K.}~\bibnamefont {{Gomi}}},\
  }\href@noop {} {\bibfield  {journal} {\bibinfo  {journal} {ArXiv e-prints}\ }
  (\bibinfo {year} {2018})},\ \Eprint {http://arxiv.org/abs/1802.06694}
  {arXiv:1802.06694 [cond-mat.str-el]} \BibitemShut {NoStop}%
\bibitem [{\citenamefont {Thorngren}\ and\ \citenamefont
  {Else}(2018)}]{thorngren2018}%
  \BibitemOpen
  \bibfield  {author} {\bibinfo {author} {\bibfnamefont {R.}~\bibnamefont
  {Thorngren}}\ and\ \bibinfo {author} {\bibfnamefont {D.~V.}\ \bibnamefont
  {Else}},\ }\href {\doibase 10.1103/PhysRevX.8.011040} {\bibfield  {journal}
  {\bibinfo  {journal} {Phys. Rev. X}\ }\textbf {\bibinfo {volume} {8}},\
  \bibinfo {pages} {011040} (\bibinfo {year} {2018})}\BibitemShut {NoStop}%
\bibitem [{\citenamefont {Rhim}\ \emph {et~al.}(2018)\citenamefont {Rhim},
  \citenamefont {Bardarson},\ and\ \citenamefont {Slager}}]{rhim2018}%
  \BibitemOpen
  \bibfield  {author} {\bibinfo {author} {\bibfnamefont {J.-W.}\ \bibnamefont
  {Rhim}}, \bibinfo {author} {\bibfnamefont {J.~H.}\ \bibnamefont {Bardarson}},
  \ and\ \bibinfo {author} {\bibfnamefont {R.-J.}\ \bibnamefont {Slager}},\
  }\href {\doibase 10.1103/PhysRevB.97.115143} {\bibfield  {journal} {\bibinfo
  {journal} {Phys. Rev. B}\ }\textbf {\bibinfo {volume} {97}},\ \bibinfo
  {pages} {115143} (\bibinfo {year} {2018})}\BibitemShut {NoStop}%
\bibitem [{\citenamefont {Parameswaran}\ and\ \citenamefont
  {Wan}(2017)}]{parameswaran2017}%
  \BibitemOpen
  \bibfield  {author} {\bibinfo {author} {\bibfnamefont {S.~A.}\ \bibnamefont
  {Parameswaran}}\ and\ \bibinfo {author} {\bibfnamefont {Y.}~\bibnamefont
  {Wan}},\ }\href {\doibase 10.1103/Physics.10.132} {\bibfield  {journal}
  {\bibinfo  {journal} {Physics}\ }\textbf {\bibinfo {volume} {10}},\ \bibinfo
  {pages} {132} (\bibinfo {year} {2017})}\BibitemShut {NoStop}%
\bibitem [{\citenamefont {Schindler}\ \emph
  {et~al.}(2018{\natexlab{a}})\citenamefont {Schindler}, \citenamefont {Cook},
  \citenamefont {Vergniory}, \citenamefont {Wang}, \citenamefont {Parkin},
  \citenamefont {Bernevig},\ and\ \citenamefont {Neupert}}]{schindler2018}%
  \BibitemOpen
  \bibfield  {author} {\bibinfo {author} {\bibfnamefont {F.}~\bibnamefont
  {Schindler}}, \bibinfo {author} {\bibfnamefont {A.~M.}\ \bibnamefont {Cook}},
  \bibinfo {author} {\bibfnamefont {M.~G.}\ \bibnamefont {Vergniory}}, \bibinfo
  {author} {\bibfnamefont {Z.}~\bibnamefont {Wang}}, \bibinfo {author}
  {\bibfnamefont {S.~S.~P.}\ \bibnamefont {Parkin}}, \bibinfo {author}
  {\bibfnamefont {B.~A.}\ \bibnamefont {Bernevig}}, \ and\ \bibinfo {author}
  {\bibfnamefont {T.}~\bibnamefont {Neupert}},\ }\href {\doibase
  10.1126/sciadv.aat0346} {\bibfield  {journal} {\bibinfo  {journal} {Science
  Advances}\ }\textbf {\bibinfo {volume} {4}} (\bibinfo {year}
  {2018}{\natexlab{a}}),\ 10.1126/sciadv.aat0346}\BibitemShut {NoStop}%
\bibitem [{\citenamefont {Peng}\ \emph {et~al.}(2017)\citenamefont {Peng},
  \citenamefont {Bao},\ and\ \citenamefont {von Oppen}}]{peng2017}%
  \BibitemOpen
  \bibfield  {author} {\bibinfo {author} {\bibfnamefont {Y.}~\bibnamefont
  {Peng}}, \bibinfo {author} {\bibfnamefont {Y.}~\bibnamefont {Bao}}, \ and\
  \bibinfo {author} {\bibfnamefont {F.}~\bibnamefont {von Oppen}},\ }\href@noop
  {} {\bibfield  {journal} {\bibinfo  {journal} {Phys. Rev. B}\ }\textbf
  {\bibinfo {volume} {95}},\ \bibinfo {pages} {235143} (\bibinfo {year}
  {2017})}\BibitemShut {NoStop}%
\bibitem [{\citenamefont {Langbehn}\ \emph {et~al.}(2017)\citenamefont
  {Langbehn}, \citenamefont {Peng}, \citenamefont {Trifunovic}, \citenamefont
  {von Oppen},\ and\ \citenamefont {Brouwer}}]{langbehn2017}%
  \BibitemOpen
  \bibfield  {author} {\bibinfo {author} {\bibfnamefont {J.}~\bibnamefont
  {Langbehn}}, \bibinfo {author} {\bibfnamefont {Y.}~\bibnamefont {Peng}},
  \bibinfo {author} {\bibfnamefont {L.}~\bibnamefont {Trifunovic}}, \bibinfo
  {author} {\bibfnamefont {F.}~\bibnamefont {von Oppen}}, \ and\ \bibinfo
  {author} {\bibfnamefont {P.~W.}\ \bibnamefont {Brouwer}},\ }\href {\doibase
  10.1103/PhysRevLett.119.246401} {\bibfield  {journal} {\bibinfo  {journal}
  {Phys. Rev. Lett.}\ }\textbf {\bibinfo {volume} {119}},\ \bibinfo {pages}
  {246401} (\bibinfo {year} {2017})}\BibitemShut {NoStop}%
\bibitem [{\citenamefont {Song}\ \emph
  {et~al.}(2017{\natexlab{a}})\citenamefont {Song}, \citenamefont {Fang},\ and\
  \citenamefont {Fang}}]{song2017}%
  \BibitemOpen
  \bibfield  {author} {\bibinfo {author} {\bibfnamefont {Z.}~\bibnamefont
  {Song}}, \bibinfo {author} {\bibfnamefont {Z.}~\bibnamefont {Fang}}, \ and\
  \bibinfo {author} {\bibfnamefont {C.}~\bibnamefont {Fang}},\ }\href {\doibase
  10.1103/PhysRevLett.119.246402} {\bibfield  {journal} {\bibinfo  {journal}
  {Phys. Rev. Lett.}\ }\textbf {\bibinfo {volume} {119}},\ \bibinfo {pages}
  {246402} (\bibinfo {year} {2017}{\natexlab{a}})}\BibitemShut {NoStop}%
\bibitem [{\citenamefont {Benalcazar}\ \emph
  {et~al.}(2017{\natexlab{a}})\citenamefont {Benalcazar}, \citenamefont
  {Bernevig},\ and\ \citenamefont {Hughes}}]{benalcazar2017}%
  \BibitemOpen
  \bibfield  {author} {\bibinfo {author} {\bibfnamefont {W.~A.}\ \bibnamefont
  {Benalcazar}}, \bibinfo {author} {\bibfnamefont {B.~A.}\ \bibnamefont
  {Bernevig}}, \ and\ \bibinfo {author} {\bibfnamefont {T.~L.}\ \bibnamefont
  {Hughes}},\ }\href@noop {} {\bibfield  {journal} {\bibinfo  {journal}
  {Science}\ }\textbf {\bibinfo {volume} {357}},\ \bibinfo {pages} {61}
  (\bibinfo {year} {2017}{\natexlab{a}})}\BibitemShut {NoStop}%
\bibitem [{\citenamefont {Benalcazar}\ \emph
  {et~al.}(2017{\natexlab{b}})\citenamefont {Benalcazar}, \citenamefont
  {Bernevig},\ and\ \citenamefont {Hughes}}]{benalcazar2017b}%
  \BibitemOpen
  \bibfield  {author} {\bibinfo {author} {\bibfnamefont {W.~A.}\ \bibnamefont
  {Benalcazar}}, \bibinfo {author} {\bibfnamefont {B.~A.}\ \bibnamefont
  {Bernevig}}, \ and\ \bibinfo {author} {\bibfnamefont {T.~L.}\ \bibnamefont
  {Hughes}},\ }\href {\doibase 10.1103/PhysRevB.96.245115} {\bibfield
  {journal} {\bibinfo  {journal} {Phys. Rev. B}\ }\textbf {\bibinfo {volume}
  {96}},\ \bibinfo {pages} {245115} (\bibinfo {year}
  {2017}{\natexlab{b}})}\BibitemShut {NoStop}%
\bibitem [{\citenamefont {Fang}\ and\ \citenamefont {Fu}(2017)}]{fang2018}%
  \BibitemOpen
  \bibfield  {author} {\bibinfo {author} {\bibfnamefont {C.}~\bibnamefont
  {Fang}}\ and\ \bibinfo {author} {\bibfnamefont {L.}~\bibnamefont {Fu}},\
  }\href@noop {} {\bibfield  {journal} {\bibinfo  {journal} {arXiv:1709.01929}\
  } (\bibinfo {year} {2017})}\BibitemShut {NoStop}%
\bibitem [{\citenamefont {Ezawa}(2018{\natexlab{a}})}]{ezawa2018}%
  \BibitemOpen
  \bibfield  {author} {\bibinfo {author} {\bibfnamefont {M.}~\bibnamefont
  {Ezawa}},\ }\href {\doibase 10.1103/PhysRevLett.120.026801} {\bibfield
  {journal} {\bibinfo  {journal} {Phys. Rev. Lett.}\ }\textbf {\bibinfo
  {volume} {120}},\ \bibinfo {pages} {026801} (\bibinfo {year}
  {2018}{\natexlab{a}})}\BibitemShut {NoStop}%
\bibitem [{\citenamefont {Shapourian}\ \emph {et~al.}(2018)\citenamefont
  {Shapourian}, \citenamefont {Wang},\ and\ \citenamefont
  {Ryu}}]{shapourian2017}%
  \BibitemOpen
  \bibfield  {author} {\bibinfo {author} {\bibfnamefont {H.}~\bibnamefont
  {Shapourian}}, \bibinfo {author} {\bibfnamefont {Y.}~\bibnamefont {Wang}}, \
  and\ \bibinfo {author} {\bibfnamefont {S.}~\bibnamefont {Ryu}},\ }\href
  {\doibase 10.1103/PhysRevB.97.094508} {\bibfield  {journal} {\bibinfo
  {journal} {Phys. Rev. B}\ }\textbf {\bibinfo {volume} {97}},\ \bibinfo
  {pages} {094508} (\bibinfo {year} {2018})}\BibitemShut {NoStop}%
\bibitem [{\citenamefont {Zhu}(2018)}]{zhu2018}%
  \BibitemOpen
  \bibfield  {author} {\bibinfo {author} {\bibfnamefont {X.}~\bibnamefont
  {Zhu}},\ }\href {\doibase 10.1103/PhysRevB.97.205134} {\bibfield  {journal}
  {\bibinfo  {journal} {Phys. Rev. B}\ }\textbf {\bibinfo {volume} {97}},\
  \bibinfo {pages} {205134} (\bibinfo {year} {2018})}\BibitemShut {NoStop}%
\bibitem [{\citenamefont {Yan}\ \emph {et~al.}(2018)\citenamefont {Yan},
  \citenamefont {Song},\ and\ \citenamefont {Wang}}]{yan2018}%
  \BibitemOpen
  \bibfield  {author} {\bibinfo {author} {\bibfnamefont {Z.}~\bibnamefont
  {Yan}}, \bibinfo {author} {\bibfnamefont {F.}~\bibnamefont {Song}}, \ and\
  \bibinfo {author} {\bibfnamefont {Z.}~\bibnamefont {Wang}},\ }\href {\doibase
  10.1103/PhysRevLett.121.096803} {\bibfield  {journal} {\bibinfo  {journal}
  {Phys. Rev. Lett.}\ }\textbf {\bibinfo {volume} {121}},\ \bibinfo {pages}
  {096803} (\bibinfo {year} {2018})}\BibitemShut {NoStop}%
\bibitem [{\citenamefont {{Wang}}\ \emph
  {et~al.}(2018{\natexlab{a}})\citenamefont {{Wang}}, \citenamefont {{Lin}},\
  and\ \citenamefont {{Hughes}}}]{wang2018}%
  \BibitemOpen
  \bibfield  {author} {\bibinfo {author} {\bibfnamefont {Y.}~\bibnamefont
  {{Wang}}}, \bibinfo {author} {\bibfnamefont {M.}~\bibnamefont {{Lin}}}, \
  and\ \bibinfo {author} {\bibfnamefont {T.~L.}\ \bibnamefont {{Hughes}}},\
  }\href@noop {} {\bibfield  {journal} {\bibinfo  {journal} {ArXiv e-prints}\ }
  (\bibinfo {year} {2018}{\natexlab{a}})},\ \Eprint
  {http://arxiv.org/abs/1804.01531} {arXiv:1804.01531 [cond-mat.supr-con]}
  \BibitemShut {NoStop}%
\bibitem [{\citenamefont {{Wang}}\ \emph
  {et~al.}(2018{\natexlab{b}})\citenamefont {{Wang}}, \citenamefont {{Liu}},
  \citenamefont {{Lu}},\ and\ \citenamefont {{Zhang}}}]{wang2018b}%
  \BibitemOpen
  \bibfield  {author} {\bibinfo {author} {\bibfnamefont {Q.}~\bibnamefont
  {{Wang}}}, \bibinfo {author} {\bibfnamefont {C.-C.}\ \bibnamefont {{Liu}}},
  \bibinfo {author} {\bibfnamefont {Y.-M.}\ \bibnamefont {{Lu}}}, \ and\
  \bibinfo {author} {\bibfnamefont {F.}~\bibnamefont {{Zhang}}},\ }\href@noop
  {} {\bibfield  {journal} {\bibinfo  {journal} {ArXiv e-prints}\ } (\bibinfo
  {year} {2018}{\natexlab{b}})},\ \Eprint {http://arxiv.org/abs/1804.04711}
  {arXiv:1804.04711 [cond-mat.mes-hall]} \BibitemShut {NoStop}%
\bibitem [{\citenamefont {Geier}\ \emph {et~al.}(2018)\citenamefont {Geier},
  \citenamefont {Trifunovic}, \citenamefont {Hoskam},\ and\ \citenamefont
  {Brouwer}}]{geier2018}%
  \BibitemOpen
  \bibfield  {author} {\bibinfo {author} {\bibfnamefont {M.}~\bibnamefont
  {Geier}}, \bibinfo {author} {\bibfnamefont {L.}~\bibnamefont {Trifunovic}},
  \bibinfo {author} {\bibfnamefont {M.}~\bibnamefont {Hoskam}}, \ and\ \bibinfo
  {author} {\bibfnamefont {P.~W.}\ \bibnamefont {Brouwer}},\ }\href {\doibase
  10.1103/PhysRevB.97.205135} {\bibfield  {journal} {\bibinfo  {journal} {Phys.
  Rev. B}\ }\textbf {\bibinfo {volume} {97}},\ \bibinfo {pages} {205135}
  (\bibinfo {year} {2018})}\BibitemShut {NoStop}%
\bibitem [{\citenamefont {van Miert}\ and\ \citenamefont
  {Ortix}(2018{\natexlab{a}})}]{vanmiert2018b}%
  \BibitemOpen
  \bibfield  {author} {\bibinfo {author} {\bibfnamefont {G.}~\bibnamefont {van
  Miert}}\ and\ \bibinfo {author} {\bibfnamefont {C.}~\bibnamefont {Ortix}},\
  }\href {\doibase 10.1103/PhysRevB.98.081110} {\bibfield  {journal} {\bibinfo
  {journal} {Phys. Rev. B}\ }\textbf {\bibinfo {volume} {98}},\ \bibinfo
  {pages} {081110} (\bibinfo {year} {2018}{\natexlab{a}})}\BibitemShut
  {NoStop}%
\bibitem [{\citenamefont {{Calugaru}}\ \emph {et~al.}(2018)\citenamefont
  {{Calugaru}}, \citenamefont {{Juricic}},\ and\ \citenamefont
  {{Roy}}}]{calugaru2018}%
  \BibitemOpen
  \bibfield  {author} {\bibinfo {author} {\bibfnamefont {D.}~\bibnamefont
  {{Calugaru}}}, \bibinfo {author} {\bibfnamefont {V.}~\bibnamefont
  {{Juricic}}}, \ and\ \bibinfo {author} {\bibfnamefont {B.}~\bibnamefont
  {{Roy}}},\ }\href@noop {} {\bibfield  {journal} {\bibinfo  {journal} {ArXiv
  e-prints}\ } (\bibinfo {year} {2018})},\ \Eprint
  {http://arxiv.org/abs/1808.08965} {arXiv:1808.08965 [cond-mat.mes-hall]}
  \BibitemShut {NoStop}%
\bibitem [{\citenamefont {Schindler}\ \emph
  {et~al.}(2018{\natexlab{b}})\citenamefont {Schindler}, \citenamefont {Wang},
  \citenamefont {Vergniory}, \citenamefont {Cook}, \citenamefont {Murani},
  \citenamefont {Sengupta}, \citenamefont {Kasumov}, \citenamefont {Deblock},
  \citenamefont {Jeon}, \citenamefont {Drozdov}, \citenamefont {Bouchiat},
  \citenamefont {Gu{\'e}ron}, \citenamefont {Yazdani}, \citenamefont
  {Bernevig},\ and\ \citenamefont {Neupert}}]{schindler2018b}%
  \BibitemOpen
  \bibfield  {author} {\bibinfo {author} {\bibfnamefont {F.}~\bibnamefont
  {Schindler}}, \bibinfo {author} {\bibfnamefont {Z.}~\bibnamefont {Wang}},
  \bibinfo {author} {\bibfnamefont {M.~G.}\ \bibnamefont {Vergniory}}, \bibinfo
  {author} {\bibfnamefont {A.~M.}\ \bibnamefont {Cook}}, \bibinfo {author}
  {\bibfnamefont {A.}~\bibnamefont {Murani}}, \bibinfo {author} {\bibfnamefont
  {S.}~\bibnamefont {Sengupta}}, \bibinfo {author} {\bibfnamefont {A.~Y.}\
  \bibnamefont {Kasumov}}, \bibinfo {author} {\bibfnamefont {R.}~\bibnamefont
  {Deblock}}, \bibinfo {author} {\bibfnamefont {S.}~\bibnamefont {Jeon}},
  \bibinfo {author} {\bibfnamefont {I.}~\bibnamefont {Drozdov}}, \bibinfo
  {author} {\bibfnamefont {H.}~\bibnamefont {Bouchiat}}, \bibinfo {author}
  {\bibfnamefont {S.}~\bibnamefont {Gu{\'e}ron}}, \bibinfo {author}
  {\bibfnamefont {A.}~\bibnamefont {Yazdani}}, \bibinfo {author} {\bibfnamefont
  {B.~A.}\ \bibnamefont {Bernevig}}, \ and\ \bibinfo {author} {\bibfnamefont
  {T.}~\bibnamefont {Neupert}},\ }\href {\doibase 10.1038/s41567-018-0224-7}
  {\bibfield  {journal} {\bibinfo  {journal} {Nature Physics}\ }\textbf
  {\bibinfo {volume} {14}},\ \bibinfo {pages} {918} (\bibinfo {year}
  {2018}{\natexlab{b}})}\BibitemShut {NoStop}%
\bibitem [{\citenamefont {Volovik}(2010)}]{volovik2010}%
  \BibitemOpen
  \bibfield  {author} {\bibinfo {author} {\bibfnamefont {G.~E.}\ \bibnamefont
  {Volovik}},\ }\href@noop {} {\bibfield  {journal} {\bibinfo  {journal} {JETP
  Letters}\ }\textbf {\bibinfo {volume} {91}},\ \bibinfo {pages} {201}
  (\bibinfo {year} {2010})}\BibitemShut {NoStop}%
\bibitem [{\citenamefont {Sitte}\ \emph {et~al.}(2012)\citenamefont {Sitte},
  \citenamefont {Rosch}, \citenamefont {Altman},\ and\ \citenamefont
  {Fritz}}]{sitte2012}%
  \BibitemOpen
  \bibfield  {author} {\bibinfo {author} {\bibfnamefont {M.}~\bibnamefont
  {Sitte}}, \bibinfo {author} {\bibfnamefont {A.}~\bibnamefont {Rosch}},
  \bibinfo {author} {\bibfnamefont {E.}~\bibnamefont {Altman}}, \ and\ \bibinfo
  {author} {\bibfnamefont {L.}~\bibnamefont {Fritz}},\ }\href {\doibase
  10.1103/PhysRevLett.108.126807} {\bibfield  {journal} {\bibinfo  {journal}
  {Phys. Rev. Lett.}\ }\textbf {\bibinfo {volume} {108}},\ \bibinfo {pages}
  {126807} (\bibinfo {year} {2012})}\BibitemShut {NoStop}%
\bibitem [{\citenamefont {Zhang}\ \emph {et~al.}(2013)\citenamefont {Zhang},
  \citenamefont {Kane},\ and\ \citenamefont {Mele}}]{zhang2013}%
  \BibitemOpen
  \bibfield  {author} {\bibinfo {author} {\bibfnamefont {F.}~\bibnamefont
  {Zhang}}, \bibinfo {author} {\bibfnamefont {C.~L.}\ \bibnamefont {Kane}}, \
  and\ \bibinfo {author} {\bibfnamefont {E.~J.}\ \bibnamefont {Mele}},\ }\href
  {\doibase 10.1103/PhysRevLett.110.046404} {\bibfield  {journal} {\bibinfo
  {journal} {Phys. Rev. Lett.}\ }\textbf {\bibinfo {volume} {110}},\ \bibinfo
  {pages} {046404} (\bibinfo {year} {2013})}\BibitemShut {NoStop}%
\bibitem [{\citenamefont {Khalaf}(2018)}]{khalaf2018b}%
  \BibitemOpen
  \bibfield  {author} {\bibinfo {author} {\bibfnamefont {E.}~\bibnamefont
  {Khalaf}},\ }\href {\doibase 10.1103/PhysRevB.97.205136} {\bibfield
  {journal} {\bibinfo  {journal} {Phys. Rev. B}\ }\textbf {\bibinfo {volume}
  {97}},\ \bibinfo {pages} {205136} (\bibinfo {year} {2018})}\BibitemShut
  {NoStop}%
\bibitem [{\citenamefont {Altland}\ and\ \citenamefont
  {Zirnbauer}(1997)}]{altland1997}%
  \BibitemOpen
  \bibfield  {author} {\bibinfo {author} {\bibfnamefont {A.}~\bibnamefont
  {Altland}}\ and\ \bibinfo {author} {\bibfnamefont {M.~R.}\ \bibnamefont
  {Zirnbauer}},\ }\href {\doibase 10.1103/PhysRevB.55.1142} {\bibfield
  {journal} {\bibinfo  {journal} {Phys. Rev. B}\ }\textbf {\bibinfo {volume}
  {55}},\ \bibinfo {pages} {1142} (\bibinfo {year} {1997})}\BibitemShut
  {NoStop}%
\bibitem [{\citenamefont {Hughes}\ \emph {et~al.}(2011)\citenamefont {Hughes},
  \citenamefont {Prodan},\ and\ \citenamefont {Bernevig}}]{hughes2011}%
  \BibitemOpen
  \bibfield  {author} {\bibinfo {author} {\bibfnamefont {T.~L.}\ \bibnamefont
  {Hughes}}, \bibinfo {author} {\bibfnamefont {E.}~\bibnamefont {Prodan}}, \
  and\ \bibinfo {author} {\bibfnamefont {B.~A.}\ \bibnamefont {Bernevig}},\
  }\href {\doibase 10.1103/PhysRevB.83.245132} {\bibfield  {journal} {\bibinfo
  {journal} {Phys. Rev. B}\ }\textbf {\bibinfo {volume} {83}},\ \bibinfo
  {pages} {245132} (\bibinfo {year} {2011})}\BibitemShut {NoStop}%
\bibitem [{\citenamefont {Lau}\ \emph {et~al.}(2016)\citenamefont {Lau},
  \citenamefont {van~den Brink},\ and\ \citenamefont {Ortix}}]{lau2016}%
  \BibitemOpen
  \bibfield  {author} {\bibinfo {author} {\bibfnamefont {A.}~\bibnamefont
  {Lau}}, \bibinfo {author} {\bibfnamefont {J.}~\bibnamefont {van~den Brink}},
  \ and\ \bibinfo {author} {\bibfnamefont {C.}~\bibnamefont {Ortix}},\ }\href
  {\doibase 10.1103/PhysRevB.94.165164} {\bibfield  {journal} {\bibinfo
  {journal} {Phys. Rev. B}\ }\textbf {\bibinfo {volume} {94}},\ \bibinfo
  {pages} {165164} (\bibinfo {year} {2016})}\BibitemShut {NoStop}%
\bibitem [{\citenamefont {Teo}\ and\ \citenamefont {Kane}(2010)}]{teo2010}%
  \BibitemOpen
  \bibfield  {author} {\bibinfo {author} {\bibfnamefont {J.~C.~Y.}\
  \bibnamefont {Teo}}\ and\ \bibinfo {author} {\bibfnamefont {C.~L.}\
  \bibnamefont {Kane}},\ }\href {\doibase 10.1103/PhysRevB.82.115120}
  {\bibfield  {journal} {\bibinfo  {journal} {Phys. Rev. B}\ }\textbf {\bibinfo
  {volume} {82}},\ \bibinfo {pages} {115120} (\bibinfo {year}
  {2010})}\BibitemShut {NoStop}%
\bibitem [{\citenamefont {Isobe}\ and\ \citenamefont {Fu}(2015)}]{isobe2015}%
  \BibitemOpen
  \bibfield  {author} {\bibinfo {author} {\bibfnamefont {H.}~\bibnamefont
  {Isobe}}\ and\ \bibinfo {author} {\bibfnamefont {L.}~\bibnamefont {Fu}},\
  }\href@noop {} {\bibfield  {journal} {\bibinfo  {journal} {Phys. Rev. B}\
  }\textbf {\bibinfo {volume} {92}},\ \bibinfo {pages} {081304} (\bibinfo
  {year} {2015})}\BibitemShut {NoStop}%
\bibitem [{\citenamefont {Fulga}\ \emph {et~al.}(2016)\citenamefont {Fulga},
  \citenamefont {Avraham}, \citenamefont {Beidenkopf},\ and\ \citenamefont
  {Stern}}]{fulga2016}%
  \BibitemOpen
  \bibfield  {author} {\bibinfo {author} {\bibfnamefont {I.~C.}\ \bibnamefont
  {Fulga}}, \bibinfo {author} {\bibfnamefont {N.}~\bibnamefont {Avraham}},
  \bibinfo {author} {\bibfnamefont {H.}~\bibnamefont {Beidenkopf}}, \ and\
  \bibinfo {author} {\bibfnamefont {A.}~\bibnamefont {Stern}},\ }\href@noop {}
  {\bibfield  {journal} {\bibinfo  {journal} {Phys. Rev. B}\ }\textbf {\bibinfo
  {volume} {94}},\ \bibinfo {pages} {125405} (\bibinfo {year}
  {2016})}\BibitemShut {NoStop}%
\bibitem [{\citenamefont {Huang}\ \emph {et~al.}(2017)\citenamefont {Huang},
  \citenamefont {Song}, \citenamefont {Huang},\ and\ \citenamefont
  {Hermele}}]{huang2017}%
  \BibitemOpen
  \bibfield  {author} {\bibinfo {author} {\bibfnamefont {S.-J.}\ \bibnamefont
  {Huang}}, \bibinfo {author} {\bibfnamefont {H.}~\bibnamefont {Song}},
  \bibinfo {author} {\bibfnamefont {Y.-P.}\ \bibnamefont {Huang}}, \ and\
  \bibinfo {author} {\bibfnamefont {M.}~\bibnamefont {Hermele}},\ }\href
  {\doibase 10.1103/PhysRevB.96.205106} {\bibfield  {journal} {\bibinfo
  {journal} {Phys. Rev. B}\ }\textbf {\bibinfo {volume} {96}},\ \bibinfo
  {pages} {205106} (\bibinfo {year} {2017})}\BibitemShut {NoStop}%
\bibitem [{\citenamefont {{Matsugatani}}\ and\ \citenamefont
  {{Watanabe}}(2018)}]{matsugatani2018}%
  \BibitemOpen
  \bibfield  {author} {\bibinfo {author} {\bibfnamefont {A.}~\bibnamefont
  {{Matsugatani}}}\ and\ \bibinfo {author} {\bibfnamefont {H.}~\bibnamefont
  {{Watanabe}}},\ }\href@noop {} {\bibfield  {journal} {\bibinfo  {journal}
  {ArXiv e-prints}\ } (\bibinfo {year} {2018})},\ \Eprint
  {http://arxiv.org/abs/1804.02794} {arXiv:1804.02794 [cond-mat.str-el]}
  \BibitemShut {NoStop}%
\bibitem [{\citenamefont {Nakahara}(2003)}]{nakahara2003}%
  \BibitemOpen
  \bibfield  {author} {\bibinfo {author} {\bibfnamefont {M.}~\bibnamefont
  {Nakahara}},\ }\href {https://books.google.de/books?id=cH-XQB0Ex5wC} {\emph
  {\bibinfo {title} {Geometry, Topology and Physics, Second Edition}}},\
  Graduate student series in physics\ (\bibinfo  {publisher} {Taylor \&
  Francis},\ \bibinfo {year} {2003})\BibitemShut {NoStop}%
\bibitem [{\citenamefont {Shiozaki}\ \emph {et~al.}(2015)\citenamefont
  {Shiozaki}, \citenamefont {Sato},\ and\ \citenamefont {Gomi}}]{shiozaki2015}%
  \BibitemOpen
  \bibfield  {author} {\bibinfo {author} {\bibfnamefont {K.}~\bibnamefont
  {Shiozaki}}, \bibinfo {author} {\bibfnamefont {M.}~\bibnamefont {Sato}}, \
  and\ \bibinfo {author} {\bibfnamefont {K.}~\bibnamefont {Gomi}},\ }\href
  {\doibase 10.1103/PhysRevB.91.155120} {\bibfield  {journal} {\bibinfo
  {journal} {Phys. Rev. B}\ }\textbf {\bibinfo {volume} {91}},\ \bibinfo
  {pages} {155120} (\bibinfo {year} {2015})}\BibitemShut {NoStop}%
\bibitem [{\citenamefont {Kells}\ \emph {et~al.}(2012)\citenamefont {Kells},
  \citenamefont {Meidan},\ and\ \citenamefont {Brouwer}}]{kells2012}%
  \BibitemOpen
  \bibfield  {author} {\bibinfo {author} {\bibfnamefont {G.}~\bibnamefont
  {Kells}}, \bibinfo {author} {\bibfnamefont {D.}~\bibnamefont {Meidan}}, \
  and\ \bibinfo {author} {\bibfnamefont {P.~W.}\ \bibnamefont {Brouwer}},\
  }\href {\doibase 10.1103/PhysRevB.85.060507} {\bibfield  {journal} {\bibinfo
  {journal} {Phys. Rev. B}\ }\textbf {\bibinfo {volume} {85}},\ \bibinfo
  {pages} {060507} (\bibinfo {year} {2012})}\BibitemShut {NoStop}%
\bibitem [{\citenamefont {Tewari}\ and\ \citenamefont
  {Sau}(2012)}]{tewari2012}%
  \BibitemOpen
  \bibfield  {author} {\bibinfo {author} {\bibfnamefont {S.}~\bibnamefont
  {Tewari}}\ and\ \bibinfo {author} {\bibfnamefont {J.~D.}\ \bibnamefont
  {Sau}},\ }\href {\doibase 10.1103/PhysRevLett.109.150408} {\bibfield
  {journal} {\bibinfo  {journal} {Phys. Rev. Lett.}\ }\textbf {\bibinfo
  {volume} {109}},\ \bibinfo {pages} {150408} (\bibinfo {year}
  {2012})}\BibitemShut {NoStop}%
\bibitem [{\citenamefont {Dwivedi}\ \emph {et~al.}(2018)\citenamefont
  {Dwivedi}, \citenamefont {Hickey}, \citenamefont {Eschmann},\ and\
  \citenamefont {Trebst}}]{dwivedi2018}%
  \BibitemOpen
  \bibfield  {author} {\bibinfo {author} {\bibfnamefont {V.}~\bibnamefont
  {Dwivedi}}, \bibinfo {author} {\bibfnamefont {C.}~\bibnamefont {Hickey}},
  \bibinfo {author} {\bibfnamefont {T.}~\bibnamefont {Eschmann}}, \ and\
  \bibinfo {author} {\bibfnamefont {S.}~\bibnamefont {Trebst}},\ }\href
  {\doibase 10.1103/PhysRevB.98.054432} {\bibfield  {journal} {\bibinfo
  {journal} {Phys. Rev. B}\ }\textbf {\bibinfo {volume} {98}},\ \bibinfo
  {pages} {054432} (\bibinfo {year} {2018})}\BibitemShut {NoStop}%
\bibitem [{Note1()}]{Note1}%
  \BibitemOpen
  \bibinfo {note} {In a non-minimal model (which can be obtained, {\protect \em
  e.g.}, by adding trivial bands to a minimal model of a topological phase),
  the number of mutually anticommuting ${\protect \cal S}$-breaking mass terms
  $M_l$ may be less than $n-1$ or the $M_l$ do not all change sign under
  ${\protect \cal S}$. However, in that case a Hamiltonian with $n-1$ mutually
  anticommuting ${\protect \cal S}$-breaking mass terms may be recovered by
  removing or adding trivial bands.}\BibitemShut {Stop}%
\bibitem [{\citenamefont {Song}\ \emph
  {et~al.}(2017{\natexlab{b}})\citenamefont {Song}, \citenamefont {Huang},
  \citenamefont {Fu},\ and\ \citenamefont {Hermele}}]{song2017b}%
  \BibitemOpen
  \bibfield  {author} {\bibinfo {author} {\bibfnamefont {H.}~\bibnamefont
  {Song}}, \bibinfo {author} {\bibfnamefont {S.-J.}\ \bibnamefont {Huang}},
  \bibinfo {author} {\bibfnamefont {L.}~\bibnamefont {Fu}}, \ and\ \bibinfo
  {author} {\bibfnamefont {M.}~\bibnamefont {Hermele}},\ }\href {\doibase
  10.1103/PhysRevX.7.011020} {\bibfield  {journal} {\bibinfo  {journal} {Phys.
  Rev. X}\ }\textbf {\bibinfo {volume} {7}},\ \bibinfo {pages} {011020}
  (\bibinfo {year} {2017}{\natexlab{b}})}\BibitemShut {NoStop}%
\bibitem [{Note2()}]{Note2}%
  \BibitemOpen
  \bibinfo {note} {This property follows directly from the observation that the
  kernel of the natural quotient map $\protect \mathaccentV {tilde}07E\alpha :
  K \to \alpha [K]/\alpha [G]$ is $G \protect \qopname \relax o{ker}\alpha
  $.}\BibitemShut {Stop}%
\bibitem [{\citenamefont {{Tuegel}}\ \emph {et~al.}(2018)\citenamefont
  {{Tuegel}}, \citenamefont {{Chua}},\ and\ \citenamefont
  {{Hughes}}}]{tuegel2018}%
  \BibitemOpen
  \bibfield  {author} {\bibinfo {author} {\bibfnamefont {T.~I.}\ \bibnamefont
  {{Tuegel}}}, \bibinfo {author} {\bibfnamefont {V.}~\bibnamefont {{Chua}}}, \
  and\ \bibinfo {author} {\bibfnamefont {T.~L.}\ \bibnamefont {{Hughes}}},\
  }\href@noop {} {\bibfield  {journal} {\bibinfo  {journal} {ArXiv e-prints}\ }
  (\bibinfo {year} {2018})},\ \Eprint {http://arxiv.org/abs/1802.06790}
  {arXiv:1802.06790} \BibitemShut {NoStop}%
\bibitem [{\citenamefont {{Zhang}}\ \emph {et~al.}(2018)\citenamefont
  {{Zhang}}, \citenamefont {{Jiang}}, \citenamefont {{Song}}, \citenamefont
  {{Huang}}, \citenamefont {{He}}, \citenamefont {{Fang}}, \citenamefont
  {{Weng}},\ and\ \citenamefont {{Fang}}}]{zhang2018}%
  \BibitemOpen
  \bibfield  {author} {\bibinfo {author} {\bibfnamefont {T.}~\bibnamefont
  {{Zhang}}}, \bibinfo {author} {\bibfnamefont {Y.}~\bibnamefont {{Jiang}}},
  \bibinfo {author} {\bibfnamefont {Z.}~\bibnamefont {{Song}}}, \bibinfo
  {author} {\bibfnamefont {H.}~\bibnamefont {{Huang}}}, \bibinfo {author}
  {\bibfnamefont {Y.}~\bibnamefont {{He}}}, \bibinfo {author} {\bibfnamefont
  {Z.}~\bibnamefont {{Fang}}}, \bibinfo {author} {\bibfnamefont
  {H.}~\bibnamefont {{Weng}}}, \ and\ \bibinfo {author} {\bibfnamefont
  {C.}~\bibnamefont {{Fang}}},\ }\href@noop {} {\bibfield  {journal} {\bibinfo
  {journal} {ArXiv e-prints}\ } (\bibinfo {year} {2018})},\ \Eprint
  {http://arxiv.org/abs/1807.08756} {arXiv:1807.08756 [cond-mat.mtrl-sci]}
  \BibitemShut {NoStop}%
\bibitem [{\citenamefont {van Miert}\ and\ \citenamefont
  {Ortix}(2018{\natexlab{b}})}]{vanmiert2018}%
  \BibitemOpen
  \bibfield  {author} {\bibinfo {author} {\bibfnamefont {G.}~\bibnamefont {van
  Miert}}\ and\ \bibinfo {author} {\bibfnamefont {C.}~\bibnamefont {Ortix}},\
  }\href {\doibase 10.1103/PhysRevB.97.201111} {\bibfield  {journal} {\bibinfo
  {journal} {Phys. Rev. B}\ }\textbf {\bibinfo {volume} {97}},\ \bibinfo
  {pages} {201111} (\bibinfo {year} {2018}{\natexlab{b}})}\BibitemShut
  {NoStop}%
\bibitem [{\citenamefont {Shitade}\ \emph {et~al.}(2018)\citenamefont
  {Shitade}, \citenamefont {Watanabe},\ and\ \citenamefont
  {Yanase}}]{shitade2018}%
  \BibitemOpen
  \bibfield  {author} {\bibinfo {author} {\bibfnamefont {A.}~\bibnamefont
  {Shitade}}, \bibinfo {author} {\bibfnamefont {H.}~\bibnamefont {Watanabe}}, \
  and\ \bibinfo {author} {\bibfnamefont {Y.}~\bibnamefont {Yanase}},\ }\href
  {\doibase 10.1103/PhysRevB.98.020407} {\bibfield  {journal} {\bibinfo
  {journal} {Phys. Rev. B}\ }\textbf {\bibinfo {volume} {98}},\ \bibinfo
  {pages} {020407} (\bibinfo {year} {2018})}\BibitemShut {NoStop}%
\bibitem [{\citenamefont {Gao}\ \emph {et~al.}(2018)\citenamefont {Gao},
  \citenamefont {Vanderbilt},\ and\ \citenamefont {Xiao}}]{gao2018}%
  \BibitemOpen
  \bibfield  {author} {\bibinfo {author} {\bibfnamefont {Y.}~\bibnamefont
  {Gao}}, \bibinfo {author} {\bibfnamefont {D.}~\bibnamefont {Vanderbilt}}, \
  and\ \bibinfo {author} {\bibfnamefont {D.}~\bibnamefont {Xiao}},\ }\href
  {\doibase 10.1103/PhysRevB.97.134423} {\bibfield  {journal} {\bibinfo
  {journal} {Phys. Rev. B}\ }\textbf {\bibinfo {volume} {97}},\ \bibinfo
  {pages} {134423} (\bibinfo {year} {2018})}\BibitemShut {NoStop}%
\bibitem [{\citenamefont {Gao}\ and\ \citenamefont {Xiao}(2018)}]{gao2018b}%
  \BibitemOpen
  \bibfield  {author} {\bibinfo {author} {\bibfnamefont {Y.}~\bibnamefont
  {Gao}}\ and\ \bibinfo {author} {\bibfnamefont {D.}~\bibnamefont {Xiao}},\
  }\href {\doibase 10.1103/PhysRevB.98.060402} {\bibfield  {journal} {\bibinfo
  {journal} {Phys. Rev. B}\ }\textbf {\bibinfo {volume} {98}},\ \bibinfo
  {pages} {060402} (\bibinfo {year} {2018})}\BibitemShut {NoStop}%
\bibitem [{\citenamefont {Peterson}\ \emph {et~al.}(2018)\citenamefont
  {Peterson}, \citenamefont {Benalcazar}, \citenamefont {Hughes},\ and\
  \citenamefont {Bahl}}]{peterson2018}%
  \BibitemOpen
  \bibfield  {author} {\bibinfo {author} {\bibfnamefont {C.~W.}\ \bibnamefont
  {Peterson}}, \bibinfo {author} {\bibfnamefont {W.~A.}\ \bibnamefont
  {Benalcazar}}, \bibinfo {author} {\bibfnamefont {T.~L.}\ \bibnamefont
  {Hughes}}, \ and\ \bibinfo {author} {\bibfnamefont {G.}~\bibnamefont
  {Bahl}},\ }\href {http://dx.doi.org/10.1038/nature25777} {\bibfield
  {journal} {\bibinfo  {journal} {Nature}\ }\textbf {\bibinfo {volume} {555}},\
  \bibinfo {pages} {346} (\bibinfo {year} {2018})}\BibitemShut {NoStop}%
\bibitem [{\citenamefont {Serra-Garcia}\ \emph {et~al.}(2018)\citenamefont
  {Serra-Garcia}, \citenamefont {Peri}, \citenamefont {S{\"u}sstrunk},
  \citenamefont {Bilal}, \citenamefont {Larsen}, \citenamefont {Villanueva},\
  and\ \citenamefont {Huber}}]{serra-garcia2018}%
  \BibitemOpen
  \bibfield  {author} {\bibinfo {author} {\bibfnamefont {M.}~\bibnamefont
  {Serra-Garcia}}, \bibinfo {author} {\bibfnamefont {V.}~\bibnamefont {Peri}},
  \bibinfo {author} {\bibfnamefont {R.}~\bibnamefont {S{\"u}sstrunk}}, \bibinfo
  {author} {\bibfnamefont {O.~R.}\ \bibnamefont {Bilal}}, \bibinfo {author}
  {\bibfnamefont {T.}~\bibnamefont {Larsen}}, \bibinfo {author} {\bibfnamefont
  {L.~G.}\ \bibnamefont {Villanueva}}, \ and\ \bibinfo {author} {\bibfnamefont
  {S.~D.}\ \bibnamefont {Huber}},\ }\href
  {http://dx.doi.org/10.1038/nature25156} {\bibfield  {journal} {\bibinfo
  {journal} {Nature}\ }\textbf {\bibinfo {volume} {555}},\ \bibinfo {pages}
  {342} (\bibinfo {year} {2018})}\BibitemShut {NoStop}%
\bibitem [{\citenamefont {Imhof}\ \emph {et~al.}(2018)\citenamefont {Imhof},
  \citenamefont {Berger}, \citenamefont {Bayer}, \citenamefont {Brehm},
  \citenamefont {Molenkamp}, \citenamefont {Kiessling}, \citenamefont
  {Schindler}, \citenamefont {Lee}, \citenamefont {Greiter}, \citenamefont
  {Neupert},\ and\ \citenamefont {Thomale}}]{imhof2018}%
  \BibitemOpen
  \bibfield  {author} {\bibinfo {author} {\bibfnamefont {S.}~\bibnamefont
  {Imhof}}, \bibinfo {author} {\bibfnamefont {C.}~\bibnamefont {Berger}},
  \bibinfo {author} {\bibfnamefont {F.}~\bibnamefont {Bayer}}, \bibinfo
  {author} {\bibfnamefont {J.}~\bibnamefont {Brehm}}, \bibinfo {author}
  {\bibfnamefont {L.~W.}\ \bibnamefont {Molenkamp}}, \bibinfo {author}
  {\bibfnamefont {T.}~\bibnamefont {Kiessling}}, \bibinfo {author}
  {\bibfnamefont {F.}~\bibnamefont {Schindler}}, \bibinfo {author}
  {\bibfnamefont {C.~H.}\ \bibnamefont {Lee}}, \bibinfo {author} {\bibfnamefont
  {M.}~\bibnamefont {Greiter}}, \bibinfo {author} {\bibfnamefont
  {T.}~\bibnamefont {Neupert}}, \ and\ \bibinfo {author} {\bibfnamefont
  {R.}~\bibnamefont {Thomale}},\ }\href {\doibase 10.1038/s41567-018-0246-1}
  {\bibfield  {journal} {\bibinfo  {journal} {Nature Physics}\ }\textbf
  {\bibinfo {volume} {14}},\ \bibinfo {pages} {925} (\bibinfo {year}
  {2018})}\BibitemShut {NoStop}%
\bibitem [{\citenamefont {{Yue}}\ \emph {et~al.}(2018)\citenamefont {{Yue}},
  \citenamefont {{Xu}}, \citenamefont {{Song}}, \citenamefont {{Lu}},
  \citenamefont {{Weng}}, \citenamefont {{Fang}},\ and\ \citenamefont
  {{Dai}}}]{yue2018}%
  \BibitemOpen
  \bibfield  {author} {\bibinfo {author} {\bibfnamefont {C.}~\bibnamefont
  {{Yue}}}, \bibinfo {author} {\bibfnamefont {Y.}~\bibnamefont {{Xu}}},
  \bibinfo {author} {\bibfnamefont {Z.}~\bibnamefont {{Song}}}, \bibinfo
  {author} {\bibfnamefont {Y.-M.}\ \bibnamefont {{Lu}}}, \bibinfo {author}
  {\bibfnamefont {H.}~\bibnamefont {{Weng}}}, \bibinfo {author} {\bibfnamefont
  {C.}~\bibnamefont {{Fang}}}, \ and\ \bibinfo {author} {\bibfnamefont
  {X.}~\bibnamefont {{Dai}}},\ }\href@noop {} {\bibfield  {journal} {\bibinfo
  {journal} {ArXiv e-prints}\ } (\bibinfo {year} {2018})},\ \Eprint
  {http://arxiv.org/abs/1807.01414} {arXiv:1807.01414 [cond-mat.mes-hall]}
  \BibitemShut {NoStop}%
\bibitem [{\citenamefont {{Wang}}\ \emph
  {et~al.}(2018{\natexlab{c}})\citenamefont {{Wang}}, \citenamefont {{Wieder}},
  \citenamefont {{Li}}, \citenamefont {{Yan}},\ and\ \citenamefont
  {{Bernevig}}}]{wang2018c}%
  \BibitemOpen
  \bibfield  {author} {\bibinfo {author} {\bibfnamefont {Z.}~\bibnamefont
  {{Wang}}}, \bibinfo {author} {\bibfnamefont {B.~J.}\ \bibnamefont
  {{Wieder}}}, \bibinfo {author} {\bibfnamefont {J.}~\bibnamefont {{Li}}},
  \bibinfo {author} {\bibfnamefont {B.}~\bibnamefont {{Yan}}}, \ and\ \bibinfo
  {author} {\bibfnamefont {B.~A.}\ \bibnamefont {{Bernevig}}},\ }\href@noop {}
  {\bibfield  {journal} {\bibinfo  {journal} {ArXiv e-prints}\ } (\bibinfo
  {year} {2018}{\natexlab{c}})},\ \Eprint {http://arxiv.org/abs/1806.11116}
  {arXiv:1806.11116 [cond-mat.mtrl-sci]} \BibitemShut {NoStop}%
\bibitem [{\citenamefont {Ezawa}(2018{\natexlab{b}})}]{ezawa2018c}%
  \BibitemOpen
  \bibfield  {author} {\bibinfo {author} {\bibfnamefont {M.}~\bibnamefont
  {Ezawa}},\ }\href {\doibase 10.1103/PhysRevB.97.241402} {\bibfield  {journal}
  {\bibinfo  {journal} {Phys. Rev. B}\ }\textbf {\bibinfo {volume} {97}},\
  \bibinfo {pages} {241402} (\bibinfo {year} {2018}{\natexlab{b}})}\BibitemShut
  {NoStop}%
\bibitem [{Note3()}]{Note3}%
  \BibitemOpen
  \bibinfo {note} {In this work we assume $\protect \bm {\varphi }$ to be
  defined on a torus rather than on a sphere around the defect, as in
  Refs.~\protect \rev@citealpnum {teo2010,shiozaki2014}. Defining $\protect \bm
  {\varphi }$ to be on a torus introduces weak invariants which are inessential
  for the present work, since we consider strong invariants only.}\BibitemShut
  {Stop}%
\bibitem [{\citenamefont {{Chiu}}(2014)}]{chiu2014}%
  \BibitemOpen
  \bibfield  {author} {\bibinfo {author} {\bibfnamefont {C.-K.}\ \bibnamefont
  {{Chiu}}},\ }\href@noop {} {\bibfield  {journal} {\bibinfo  {journal} {ArXiv
  e-prints}\ } (\bibinfo {year} {2014})},\ \Eprint
  {http://arxiv.org/abs/1410.1117} {arXiv:1410.1117} \BibitemShut {NoStop}%
\end{thebibliography}%
\end{document}